\documentclass{aa}  
\usepackage{graphicx}
\usepackage{txfonts}
\usepackage{xcolor}
\usepackage{comment}
\usepackage{multirow}
\usepackage{hhline}
\usepackage[shortlabels]{enumitem}
\usepackage{amsmath}
\usepackage{placeins}
\usepackage{multicol}
\usepackage{lineno}
\usepackage{hyperref} %% to avoid \citeads line fills
\makeatletter
  \renewcommand*\aa@pageof{, page \thepage{} of \pageref*{LastPage}}
\makeatother

\usepackage{booktabs}
\newcommand\doubleRule{\toprule\toprule}

\usepackage{tikz}
\usepackage{scalerel}

\usepackage{orcidlink}

\newcommand \msun{\mathrm{M}_{\odot}}

\newcommand{\rr}[1]{{{#1}}}

\begin{document}

\title{The dynamical evolution of the stellar clumps in the Sparkler galaxy}
%\subtitle{}
   \author{E. Giunchi\thanks{E-mail: eric.giunchi2@unibo.it}
          \orcidlink{0000-0002-3818-1746}\,
          \inst{1}
          \and
          F. Marinacci
          \orcidlink{0000-0003-3816-7028}\,
          \inst{1}
          \and
          C. Nipoti
          \orcidlink{0000-0003-3121-6616}\,
          \inst{1}
          \and
          A. Claeyssens
          \orcidlink{0000-0001-7940-1816}\,
          \inst{2,3}
          \and
          R. Pascale
          \orcidlink{0000-0002-6389-6268}\,
          \inst{4}
          \and
          F. Calura
          \orcidlink{0000-0002-6175-0871}\,
          \inst{4}
          \and
          A. Ragagnin
          \orcidlink{0000-0002-8106-2742}\,
          \inst{4,5}}

   \institute{Dipartimento di Fisica e Astronomia "Augusto Righi", Universit\`a di Bologna, via Piero Gobetti 93/2, I-40129 Bologna, Italy
   \and
    The Oskar Klein Centre, Department of Astronomy, Stockholm University, AlbaNova, SE-10691 Stockholm, Sweden
    \and 
    Univ Lyon, Univ Lyon1, ENS de Lyon, CNRS, Centre de Recherche Astrophysique de Lyon UMR5574, Saint-Genis-Laval, France
    \and 
    INAF - Osservatorio di Astrofisica e Scienza dello Spazio di Bologna, Via Gobetti 93/3, 40129 Bologna, Italy
    \and
    INAF - Osservatorio Astronomico di Trieste, via G.B. Tiepolo 11, I-34143
    Trieste, Italy
   }

   \date{\today}

% \abstract{}{}{}{}{} 
% 5 {} token are mandatory
 
  \abstract
   {Recent James Webb space telescope observations detected a system of stellar clumps around the $z\simeq1.4$ gravitationally lensed Sparkler galaxy (of stellar mass $M_*\sim10^9\,\msun$), with ages and metallicities compatible with globular cluster (GC) progenitors. However, most of their masses ($>10^6\,\msun$) and sizes ($>30$ pc) are about 10 times those of GCs in the local Universe.}
   {To assess whether these clumps can evolve into GC-like objects, we performed $N$-body simulations of their dynamical evolution from $z\simeq1.4$ to $z=0$ ($\sim9.23$ Gyr), under the effect of dynamical friction and tidal stripping.}
   {Dynamical friction is studied performing multiple runs of a system of clumps in a Sparkler-like spherical halo of mass $M_ {200}\simeq5\times10^{11}\,\msun$, inferred from the stellar-to-halo mass relation. For the tidal stripping, we simulated resolved clumps, orbiting in an external, static gravitational potential including the same halo as in the dynamical friction simulations and also a Sparkler-like stellar disk.}
   {Dynamical friction causes the clumps with mass $>10^7\,\msun$ to sink into the galaxy central regions, possibly contributing to the bulge growth. In absence of tidal stripping, the mass distribution of the surviving clumps ($\approx40\%$) peaks at $\approx5\times10^6\,\msun$, implying the presence of uncommonly over-massive clumps at $z=0$. Tidal shocks by the stellar disk strip considerable mass from low-mass clumps, even though their sizes remain larger than those of present-day GCs.
   When the surviving clumps are corrected for tidal stripping, their mass distribution peak shifts to $\sim2\times10^6\,\msun$, compatible with very massive GCs.}
   {Our simulations suggest that a fraction of the Sparkler clumps is expected to fall into the central regions, where they might become bulge fossil fragments or contribute to form a nuclear star cluster. The remaining clumps are too large in size to be progenitors of GCs.}


   \authorrunning{Eric Giunchi et al.}
   \maketitle
%
%-------------------------------------------------------------------

\section{Introduction}\label{sec:intro}
One of the earliest results of the James Webb Space Telescope (JWST) Early Release Observations was the set of NIRCam observations of the galaxy cluster SMACS J0723.3–7323 (hereafter SMACS0723, $z = 0.388$, \citealt{Pontoppidan2022}).
The high-resolution images gave to the community access to a large number of high-redshift galaxies gravitationally lensed by the galaxy cluster. Since gravitational lensing magnifies both the angular size and the flux of the lensed object, JWST unveiled the presence of many compact and bright stellar clumps located in high-$z$ galaxies \citep{Mowla2022,Claeyssens2023}, which could be studied with unparalleled high resolution and spectral coverage.

In particular, the Sparkler galaxy ($z\simeq 1.4$, \citealt{Mowla2022}) is one of the few high-$z$ objects known to host a system of stellar clumps located outside the observed main body of the galaxy (another example is the Relic galaxy at $z\simeq2.5$, \citealt{Whitaker2025}), following a spatial distribution that can resemble that of local globular clusters (GCs, \citealt{Forbes2023}).
Recent studies \citep{Mowla2022,Claeyssens2023,Adamo2023,Tomasetti2024} performed spectral energy distribution (SED) fitting and, although differing in some of the quantitative results, have consistently found that the Sparkler galaxy is a star-forming galaxy with a stellar mass of a few times $10^8\,\msun$, likely to be a progenitor of a galaxy of a few times $10^9\,\msun$ \citep{Adamo2023}, between the Large Magellanic Cloud (LMC) and the Milky Way (MW).
Furthermore, these works have confirmed the presence of about 10 clumps located far away from the main body of the galaxy, at about $2-4$ kpc from the centre of the system \citep{Forbes2023}, with a sub-sample of clumps having age and metallicity compatible with those predicted from simulations for GC systems of galaxies with mass and redshift comparable with those of the Sparkler \citep{Adamo2023}. Therefore, the distribution of the Sparkler clumps is consistent with that of disk/old halo GCs for a galaxy slightly less massive than the MW.

The clumps span a mass range $\log (M_*/\msun)=6-7.5$, and are, on average, considerably more massive than the local GCs observed in the MW -- for instance, \rr{NGC 6715 and NGC 6441, two of the most massive GCs hosted by the MW, have masses of $\log (M_*/\msun)=6.15-6.2$ \citep{Baumgardt2023}. Even $\omega$ Centauri, a GC-like object which is thought to be the remnant nucleus of a stripped dwarf galaxy \citep{Lee1999,Bekki2003,Bekki2019}, has mass $\log (M_*/\msun)=6.7$, less massive than the majority of the Sparkler clumps}.
It is remarkable to notice that the majority of these clumps have effective radii of $30-50$ pc, larger than those of local GCs, which are typically of a few parsecs \citep{Jordan2005,Spitler2006,Baumgardt2023}.
In conclusion, these objects, whilst compatible with GCs in terms of age and spatial distribution (and metallicity, \citealt{Adamo2023}), possess masses and sizes which are not.
However, dynamical processes such as dynamical friction and tidal stripping their dynamical evolution could drastically change their mass, size and spatial distributions between $z\approx 1.4$ and $z=0$.

Theoretical studies and simulations of dynamical friction and tidal stripping, carried out by means of $N$-body simulations, are made difficult by the dependence on various factors (such as the density profiles of the host system and of the satellite, and the orbital parameters of the latter) and by the different spatial scales at which these processes act.
To tackle the problem of including the effects of both processes, a number of approaches have been developed.
A family of methods is based on simulating the clumps as single particles, with a host represented as an $N$-body system that self-consistently produces dynamical friction. The tidal stripping is accounted for by means of analytic \citep{Kruijssen2015} or semi-analytic methods \citep{Vesperini1997,Baumgardt2003,Kravtsov2005}. For instance, making assumptions on the structure of the clumps and computing at each time-step the tidal tensor produced by the host $N$-body system, one can evaluate the amount of stripped material out of the particle-clump, as done in the E-MOSAICS project \citep{Pfeffer2018,ReinaCampos2019,Horta2021}.
A possible alternative is to resolve the clump in an external, static gravitational potential, in order to directly simulate the tidal stripping. To account for the effects of dynamical friction on the orbit of the clumps, at each time-step one can rely on analytic models (such as the dynamical friction formula by \citealt{Chandrasekhar1943}; see also \citealt{Binney2008}) which, under a certain assumptions, give the net acceleration \citep{Petts2016,DiCintio2024}.

In this work, we opted for performing two different sets of simulations that take into account one of the two dynamical processes at a time. The first setup includes a Sparkler-like host system represented by $10^7$ particles, with single-particle clumps experiencing dynamical friction, but in the absence of tidal stripping. The second setup is based on resolved $N$-body clumps orbiting in an external potential, to focus on the tidal stripping effects.
These sets of simulations allow us to explore in a statistically robust way different configurations for the morphology of the Sparkler galaxy and the spatial distribution of its system of clumps, which are unconstrained by currently available data.

\rr{The simulations are performed using the code \textsc{AREPO}\footnote{\url{https://arepo-code.org/}} \citep{Springel2005}, as available in its publicly released version \citep{Weinberger2020}.
To compute the gravitational acceleration exerted on each particle, \textsc{AREPO} distinguishes between short-range and long-range forces.
Long-range components are calculated following a grid-based method, where the gravitational force is computed on a Cartesian grid using Fourier methods and then interpolated to the position of each particle, while short-range components are calculated employing an oct-tree algorithm \citep{Barnes1986}, which calculated the collective contribution of distant particles to the overall acceleration by grouping them.
The result is a \textsc{TreePM} algorithm \citep{Bagla2002} conceptually similar to the one implemented in \textsc{GADGET} \citep{Springel2005b}, which benefits from the advantages of both methods, minimizing the disadvantages. In our simulations only the tree algorithm is used, which is the standard choice in \textsc{AREPO} for non-cosmological runs.
When dealing with collisionless gravity-only systems, each particle has its own time step. The time step of the $i$th particle is $\Delta t_i\propto(2\epsilon_\mathrm{soft}/|\pmb{a}_i|)^{1/2}$, where $\epsilon_\mathrm{soft}$ is the softening length (defined in Sects. \ref{sec:spark_dyn_mass} and \ref{sec:resolved_clumps_setup}) and $\pmb{a}_i$ is the $i$th particle acceleration.
The time integration is performed employing a second-order accurate leapfrog scheme, which alternates drift (updating the positions) and kick (updating the velocities) operations.
We refer the interested reader to \citet{Weinberger2020} for a more detailed description of the code.}

This paper is structured as follows: in Sect. \ref{sec:spark_properties} we present the Sparkler dataset, compute the clump size-mass relation, the clump mass distribution and the clump distance distribution under different morphological assumptions; in Sect. \ref{sec:halo_properties} we define the parameters of the Sparkler dark matter halo; in Sects. \ref{sec:dyn_fric} and \ref{sec:resolved_clumps} we present the setup and results of the dynamical friction and tidal stripping simulations, respectively; in Sect. \ref{sec:discussion} we discuss our results, combine them and compare the predicted stellar clumps descendants with present-day GCs; in Sect. \ref{sec:summary} we summarize our results.

\section{Properties of the Sparkler system}\label{sec:spark_properties}
The Sparkler galaxy is at $z=1.378$, gravitationally lensed by the galaxy cluster SMACS0723 \citep{Pontoppidan2022}, with stellar mass $M_*=(0.5-1)\times 10^9\,\msun$ \citep{Adamo2023}. According to the EAGLE simulation \citep{Crain2015,Schaye2015}, such a galaxy is expected to reach a few times $10^9\,\msun$ in stellar mass at $z=0$, in between the LMC and the MW galaxy \citep{Adamo2023}.
Studying its photometric colour, \cite{Mowla2022} have found that this galaxy is star forming, even though the distortion caused by the lensing makes it difficult to understand whether the baryonic component is dominated by a disk.

By means of SED fitting, a few recent studies \citep{Mowla2022,Claeyssens2023,Adamo2023,Tomasetti2024} have investigated the age and metallicity of the clumps observed in the galaxy.
For this work we focus on the ten clumps that \cite{Adamo2023} classified as GC candidates.
Regarding the ages of the clumps, from three to five clumps (thus about half of the sample) are older than 1 Gyr, with some of these clumps consistent with ages of $\sim 4$ Gyr and therefore a formation redshift between 7 and 11. Such formation redshifts are remarkably consistent with those of GCs, further hinting that these clumps might be good candidate progenitors of the GCs that we observe in the local Universe \citep{Vanzella2017,Calura2019}. The remaining sub-sample is constituted by younger clumps, with ages ranging between 0.1 and 1 Gyr, which are thought to have been recently accreted by the Sparkler galaxy \citep{Forbes2023}.

\begin{table}[t!]
\fontsize{10pt}{10pt}\selectfont
\setlength{\tabcolsep}{4pt}
\renewcommand{\arraystretch}{1.4} % Default value: 1
\centering
\caption{Parameters of the Sparkler clumps selected as a reference for this work.}
\begin{tabular}{cccccc}
\doubleRule
C23(ID) & $\log (M_*/\msun)$ & $R_\mathrm{eff}/\mathrm{pc}$ & $R_\mathrm{obs}/\mathrm{kpc}$\\
\hline
C1 & $6.3^{+0.0}_{-0.0}$ & $<12$ & $2.98$\\
C2 & $7.4^{+0.0}_{-0.0}$ & $52_{-13}^{+21}$ & $2.24$\\
C4 & $7.1^{+0.0}_{-0.0}$ & $<11$ & $1.88$\\
C5 & $6.8^{+0.0}_{-0.1}$ & $52_{-15}^{+21}$ & $2.26$\\
C6 & $6.7^{+0.0}_{-0.1}$ & $30_{-9}^{+14}$ & $3.73$\\
C7 & $6.3^{+0.1}_{-0.2}$ & $32_{-8}^{+12}$ & $4.52$\\
C9 & $7.0^{+0.0}_{-0.0}$ & $<12$ & $2.77$\\
C10 & $7.1^{+0.1}_{-0.2}$ & $45_{-12}^{+16}$ & $3.09$\\
C11 & $7.0^{+0.0}_{-0.1}$ & $46_{-12}^{+16}$ & $2.20$\\
C12 & $6.9^{+0.1}_{-0.0}$ & $34_{-10}^{+12}$ & $2.33$\\
\toprule
\end{tabular}
\tablefoot{From the left: ID of the clump (as used by \citealt{Claeyssens2023}), stellar mass, effective radius, observed distance from the center of the galaxy. All values are taken from \cite{Claeyssens2023}, except the distance which is computed from the source-plane image of Sparkler, as described in Sect. \ref{sec:spark_properties}.}
\label{tab:clumps}
\end{table}

The properties of this sample of clumps that will be used throughout the paper, namely the stellar mass, the effective radius and the observed distance from the centre of the galaxy, are listed in Table \ref{tab:clumps} and taken from \cite{Claeyssens2023}.
The masses of these clumps are $6.3<\log (M_*/\msun)<7.4$, even though a large source of uncertainty resides in the estimate of the magnification factor of the gravitational lensing.
The effective radii are defined as 2D half-mass radii, obtained fitting a 2D Gaussian function, convolved with the PSF of NIRCam, to the light profiles of the clumps, and then corrected for the magnification, as described in \cite{Claeyssens2023}. Apart from a few unresolved clumps, with an upper limit of $11-12$ pc, most of the clumps have larger effective radii ranging between 30 pc and 52 pc (third column in Table \ref{tab:clumps}).

The distance of these objects from the centre of the system is not trivial to evaluate, not only due to projection effects, but also because it must be corrected for the lensing effect.
\cite{Forbes2023} assess that most of the clumps are located (in projection) within $2-4$ kpc from the centre of the Sparkler. As a comparison, $60\%$ of the disk GCs of the MW are located within a radius of 3 kpc ($80\%$ within 5 kpc), with a distance above or below the galactic plane typically smaller than 2 kpc \citep{Mackey2004}. The old halo GCs are concentrated in the innermost regions as well, with $65\%$ of the clusters within 6 kpc \citep{Mackey2004}.
The distance distribution of the clumps is fundamental to determine their dynamical evolution since it affects their orbits. However, it is poorly constrained owing to the aforementioned observational caveats. Therefore, in this work we tested different clump spatial configurations that are described in detail in Sect. \ref{sec:clumps_distances_measurement}.

\subsection{Size-mass relation of the Sparkler clumps}\label{sec:size_mass}
We have computed the size-mass relation of the Sparkler clumps, considering the 2D half-mass radii $R_\mathrm{eff}$ and stellar masses $M_*$ listed in Table \ref{tab:clumps}. Considering that three clumps have only upper limits on their sizes, we fitted just the seven resolved ones.
We modelled the relation as

\begin{equation}\label{eq:size_mass}
    \log( R_\mathrm{eff}/\mathrm{pc})=m\log (M_*/\msun)+q.
\end{equation}

\begin{figure}[t!]
\centering
 \includegraphics[width=0.48\textwidth]{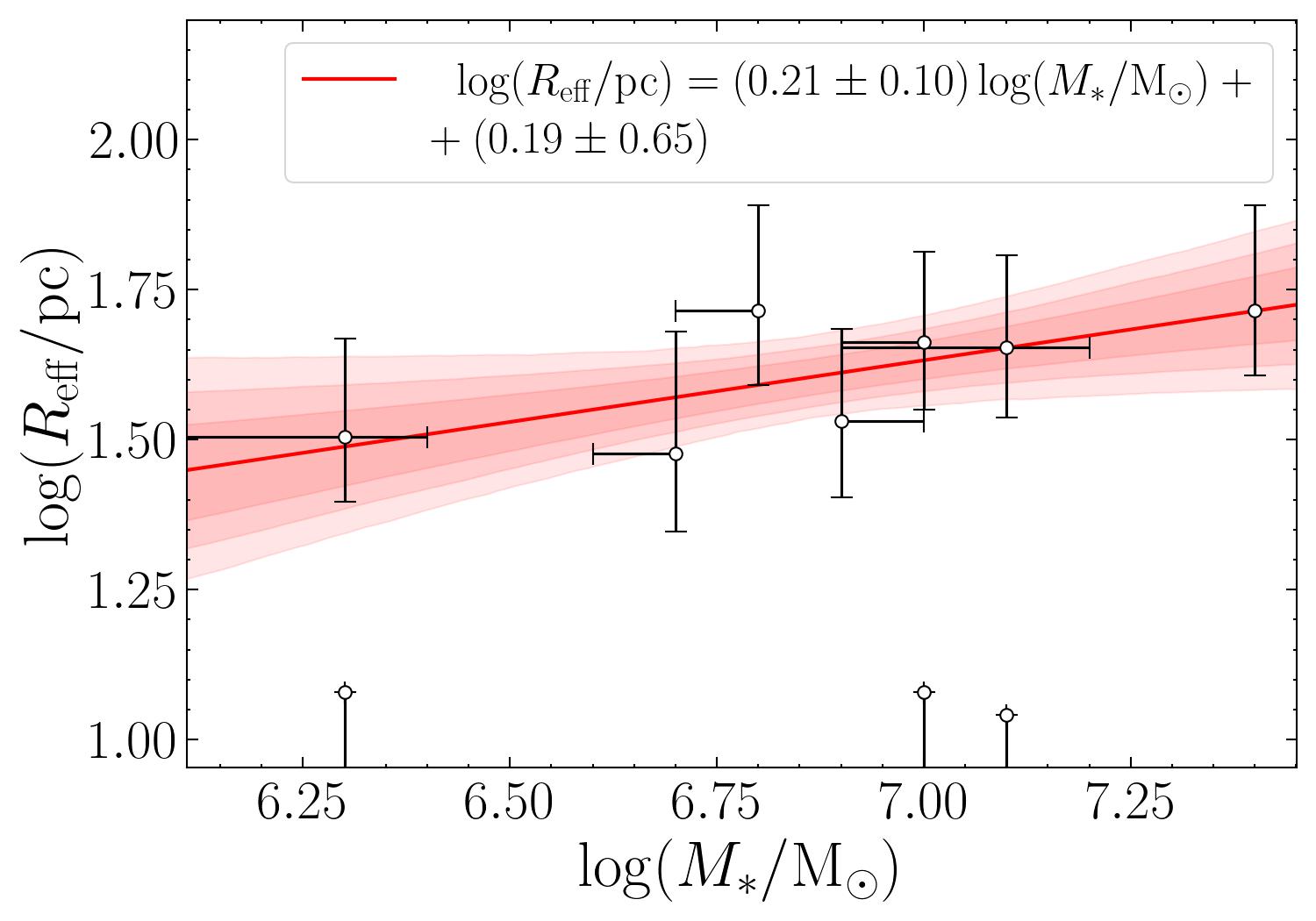}
 \caption{Size-mass relation for the extraplanar clumps of the Sparkler galaxy (Sect. \ref{sec:spark_properties}). Clumps are shown as black empty circles, while the best-fitting line (obtained as described in Sect. \ref{sec:size_mass}) is plotted as the red solid line. The red shaded areas (from opaque to transparent red) show the $1\sigma$, $2\sigma$, $3\sigma$ intervals of the fit, respectively.}
 \label{fig:size_mass}
\end{figure}

\noindent The free parameters $(m,q)$ are fitted using the least squares method, implemented in the \textsc{numpy}\footnote{\url{https://numpy.org/doc/2.1/index.html}} function \textsc{polyfit}. The best-fitting parameters are $m=0.21\pm 0.10$ and $q=0.19 \pm 0.65$. The uncertainties are computed as the square-root of the diagonal terms of the co-variance matrix. \rr{Our slope is consistent with the one found by \cite{Brown2021}, for young stellar clusters of the LEGUS sample of local star-forming galaxies, and by \citep{Claeyssens2023}, for stellar clusters and clumps in lensed galaxies at redshift $1.3-8.5$, that both found $m=0.24$. We are also in agreement with the slopes found in simulations of the formation of high-$z$ clumps, like SERRA \citep{Pallottini2022} and SIEGE \citep{Calura2022,Calura2024,Pascale2025}.}
The best fit is shown in Fig. \ref{fig:size_mass}, together with the fitted data points and the unresolved clumps.

Such correlation implies that the massive clumps tend to be denser. The average density $\overline{\rho}$ scales as

\begin{equation}\label{eq:average_dens}
    \overline{\rho}\propto M_*/R_\mathrm{eff}^3\propto M_*/\left(M_*^{m}\right)^3\propto M_*^{1-3m}.
\end{equation}

For $m=0.21$, the dependency becomes $\overline{\rho}\propto M_*^{0.37}$, which demonstrates how the average density positively correlates with the clump stellar mass \citep{Mestric2022,Calura2022}.

\subsection{Mass distribution}\label{sec:clumps_mass_distribution}
The set of 10 stellar masses of the Sparkler clumps is fitted to a Gaussian distribution $\mathcal{G}(\mathcal{D}\,;\,\pmb{\theta})$, where $\mathcal{D}_i=\log M_{*,i}$ is the mass of the $i$th Sparkler clump and $\pmb{\theta}=(\log \widetilde{M}_*,\sigma)$ are the mean and the standard deviation of the Gaussian distribution, respectively.
In the case of a Gaussian function, the best-fitting parameters can be obtained analytically and they correspond to the statistical mean and standard deviation of the dataset and reported in the first row of Table \ref{tab:mdf}.

\begin{table}[t!]
\fontsize{10pt}{10pt}\selectfont
\setlength{\tabcolsep}{4pt}
\renewcommand{\arraystretch}{1.4} % Default value: 1
\centering
\caption{Best-fitting parameters of the mass distributions for the initial (first columns) and survived (second-to-fourth rows) samples of clumps.}
\begin{tabular}{lc|c|ccc}
\bottomrule
\bottomrule
$p$ & $\pmb{\theta}$ & Initial & face-on & edge-on & spherical\\
\toprule
\multirow{2}{*}{$\mathcal{G}$} & $\log (\widetilde{M}_*/\mathrm{M_\odot})$ & $6.86$ & $6.76$ & $6.77$ & $6.69$\\
& $\sigma$ & $0.31$ & $0.28$ & $0.27$ & $0.23$\\
\bottomrule
\end{tabular}
\tablefoot{The distribution of $\log M_*$ is modelled as a Gaussian distribution with mean $\log \widetilde{M}_*$ and standard deviation $\sigma$ (first and second rows). The three configurations (face-on, edge-on and spherical) are defined as in Sect. \ref{sec:clumps_dist_distribution}.}
\label{tab:mdf}
\end{table}

\subsection{Clump distances measurement and distribution}\label{sec:clumps_distances_measurement}
The procedure used to obtain the physical distances of the clumps from the centre of the Sparkler galaxy follows the same method presented in \cite{Claeyssens2025}. In order to identify the central point of the galaxy, we use the lens model (presented in \citealt{Mahler2023} and used in \citealt{Claeyssens2023}) to project the image of the Sparkler in the source plane and recover its intrinsic morphology before the gravitational lensing effect. 
This source reconstruction method is based only on the direct inversion of the lensing equation for each pixel of an image of the Sparkler (NIRCam/F150W) and does not include PSF deconvolution and noise subtraction.
The observed clumps positions are projected to the source-plane with the same method and their physical distance from the centre is measured to obtain their galactocentric distance. 
We tested many possible centres, for instance in correspondence of the clumps C13 and C26 (see \citealt{Claeyssens2023}), which are in the centre either of the lensed image or of the source-plane image. The best-fit parameters of the distance distributions described in Sect. \ref{sec:clumps_dist_distribution} remain consistent within $1\sigma$, when changing the centre from which the fitted distances are computed, therefore we arbitrarily opted for locating the centre of the system in correspondence of the clump C13.

\subsubsection{Spatial distribution}\label{sec:clumps_dist_distribution}
Given the uncertainties on the spatial distribution of the Sparkler clumps, described in Sect. \ref{sec:spark_properties}, we studied three different configurations that can give a comprehensive view of how the clump spatial distribution affects their survival.
We define $r \equiv\sqrt{x^2+y^2+z^2}$ as the intrinsic, 3D distance of the cluster from the centre of the galaxy; $R \equiv\sqrt{x^2+y^2}$ is the distance of the cluster from the centre of the galaxy in the $x-y$ plane (which will coincide with the plane of the disk when assuming that the clusters are distributed in a disk-like configuration); $R_\mathrm{obs}$ is the position projected on the plane of the sky.
Each configuration is described by a density probability distribution $p(R_\mathrm{obs}\,;\,\pmb{\theta})$, where $\pmb{\theta}$ is the set of free parameters, which depends on the assumed spatial distribution.
The three configurations are defined as follows:

\begin{enumerate}
    \item \textit{face-on disk}: we assume that the Sparkler clumps are located in an infinitesimally thin disk plane seen face on. In this configuration, the observed distances $R_\mathrm{obs}$ (neglecting uncertainties and systematics) coincide with the intrinsic distances $R$, and no projection effects have to be taken into account.
    Motivated by the observed distribution of the disk GCs in the MW \citep{Binney2017,Posti2019}, we assume that the clumps are distributed following the stellar disk surface density, which we modelled as an exponential disk.
    Therefore, the clump spatial distribution is modelled as a disk on the $x-y$ plane, with $z$ as symmetry axis. This implies that all the clumps have $z=0$.
    The radial distribution on the $x-y$ plane is $p_\mathrm{exp}(R\,;R_\mathrm{e})=K\,\mathrm{exp}(-R/R_\mathrm{e})$, where $K$ is a normalization factor and $R_\mathrm{e}$ the scale radius.
    \item \textit{edge-on disk}: in this case, the assumptions we made are very similar to those of the face-on disk, with the Sparkler clumps lying in a disk seen edge-on. Therefore, the cluster spatial distribution in the disk is still modelled with an exponential function with effective radius $R_\mathrm{e}$.
    The probability of observing a clump at a projected distance $R_\mathrm{obs}$ is

    \begin{equation}\label{eq:edge_on_distr}
        p(R_\mathrm{obs}\,;R_\mathrm{e})=2\int_{R_\mathrm{obs}}^{\infty}\frac{p_\mathrm{exp}(R\,;R_\mathrm{e})\,R}{\sqrt{R^2-R_\mathrm{obs}^2}}\,dR\,.
    \end{equation}

    \item \textit{spherical}: in this configuration, we assume the Sparkler has a spherically symmetric system of clumps. The observed distance distribution $p_\mathrm{Sersic}(R_\mathrm{obs}\,;R_\mathrm{e},n)$ is modelled with a Sérsic profile \citep{Sersic1968}

    \begin{equation}\label{eq:sersic}
        p_\mathrm{Sersic}(R_\mathrm{obs}\,;R_\mathrm{e},n)=\Sigma_0 \exp\left[-b\left(\frac{R_\mathrm{obs}}{R_\mathrm{e}}\right)^{1/n}\right]\,,
    \end{equation}
    
    \noindent where $R_\mathrm{e}$ and $n$ are the effective radius and the Sérsic index, respectively, and
    
    \begin{equation}\label{eq:bn}
    \begin{split}
        b(n)\simeq\, & 2n-\dfrac{1}{3}+\dfrac{4}{405n}+\dfrac{46}{25515n^2}+\\
        & +\dfrac{131}{1148175n^3}-\dfrac{2194697}{30690717750n^4}+O\left(n^{-5}\right)
    \end{split}
    \end{equation}
    
    \noindent \citep{Ciotti1999}. The central density is
    
    \begin{equation}\label{eq:sigma0}
        \Sigma_0=\frac{N_\mathrm{tot}\,b^{2n}}{2\pi n\,R_\mathrm{e}^2\,\Gamma(2n)},
    \end{equation}
    
    \noindent where $N_\mathrm{tot}$ is the normalization factor ($N_\mathrm{tot}=1$ for a probability density function) and $\Gamma(x)$ is the complete gamma function.
    The intrinsic distance distribution is obtained de-projecting Eq. \ref{eq:sersic}, for which we followed the approximation by \cite{LimaNeto1999}, which is implemented in the code \textsc{OpOpGadget}, that we used to generate the initial conditions of our simulations (see Sect. \ref{sec:res_spherical}).
    In this approximation, the volume density $p_\mathrm{Sersic3D}(r\,;R_\mathrm{e},n)$ of clusters is given by

    \begin{equation}\label{eq:sersic_3D}
    p_\mathrm{Sersic3D}(r\,;R_\mathrm{e},n)=\rho_0\, \left(\frac{r\,b^n}{R_\mathrm{e}}\right)^{-p} \exp\left[-b\left(\frac{r}{R_\mathrm{e}}\right)^{1/n}\right]\,,
    \end{equation}

    \noindent where $p=1.0-0.6097/n+0.05463/n^2$ and the central density is

    \begin{equation}\label{eq:rho0}
    \rho_0=\frac{N_\mathrm{tot}\,b^{3n}}{4\pi n\,R_\mathrm{e}^3\,\Gamma[(3-p)n]}.
    \end{equation}

    We explore the ranges $0.56\leq n \leq 10$ and $10^{-2}\leq R_\mathrm{obs}/R_\mathrm{e} \leq 10^3$, for which Eq. \ref{eq:sersic_3D} deviates from the exact numerical de-projection by less than 5\%.

\end{enumerate}

Each configuration is fitted to the same distances of the 10 extraplanar Sparkler clumps (Table \ref{tab:clumps}) and the best fit is obtained by sampling the parameter space through the likelihood function

\begin{equation}\label{eq:likelihood}
    \ln \mathcal{L}(\pmb{\theta})=\sum_{i=1}^{N_\mathrm{c}}\ln p(R_{\mathrm{obs},i}\,;\,\pmb{\theta}),
\end{equation}

\noindent where $N_\mathrm{c}$ is the number of clumps. The sampling is performed by means of an MCMC method, using the \textsc{emcee} package\footnote{\url{https://emcee.readthedocs.io/en/stable/}}, with $50$ walkers for $400$ steps ($200$ of burn-in). We assumed flat priors, with the scale radius in the range $[0.1,15]$ kpc and the Sérsic index (only in the spherical configuration) in the range $[0.56,10]$. The reference parameters are defined as the medians of the posterior parameter distribution and listed in the first row of Table \ref{tab:spatial_df}.

\begin{table}[t!]
\fontsize{10pt}{10pt}\selectfont
\setlength{\tabcolsep}{4pt}
\renewcommand{\arraystretch}{1.4} % Default value: 1
\centering
\caption{Best-fitting parameters of the distance distributions for the initial (first row) and survived (second row) samples of clumps.}
\begin{tabular}{l|c|c|c}
\bottomrule
\bottomrule
Config. & face-on & edge-on & spherical\\\hline
$p(R\,;\,\pmb{\theta})$ & $p_\mathrm{exp}(R\,;R_\mathrm{e})$ & $p_\mathrm{exp}(R\,;R_\mathrm{e})$ & $p_\mathrm{Sersic}(R\,;R_\mathrm{e},n)$\\
$\pmb{\theta}$ & $R_\mathrm{e}/$kpc & $R_\mathrm{e}/$kpc & $R_\mathrm{e}/\mathrm{kpc},n$\\
\hline
Initial & $2.52^{+0.68}_{-0.5}$ & $4.06^{+1.55}_{-1.02}$ & $2.41^{+0.69}_{-0.75},0.70^{+0.25}_{-0.11}$\\
Survived & $3.30^{+0.23}_{-0.21}$ & $3.92^{+0.22}_{-0.21}$ & $2.13^{+0.12}_{-0.16},0.59^{+0.04}_{-0.02}$\\
\bottomrule
\end{tabular}
\vspace{0.15cm}
\tablefoot{As described in Sect. \ref{sec:clumps_dist_distribution}, the distance distribution is modelled as an exponential function for the face-on and edge-on disk cases, and as a de-projected Sérsic profile for the spherical case. Uncertainties are given at $1\sigma$.}
\label{tab:spatial_df}
\end{table}

\section{Dark-matter halo of the Sparkler system}\label{sec:halo_properties}
We modelled the dark matter (DM) halo with a Navarro-Frenk-White (NFW) density profile \citep{Navarro1996}

\begin{equation}\label{eq:nfw}
    \rho_\mathrm{NFW}(r)=\dfrac{4\,\rho_{-2}}{\dfrac{r}{r_{-2}}\left(1+\dfrac{r}{r_{-2}}\right)^2},
\end{equation}

\noindent where $r_{-2}$ is the radius with log-slope $d\log\rho_\mathrm{NFW}/d\log r=-2$ and

\begin{equation}
    \rho_{-2}=\rho_\mathrm{NFW}(r_{-2})=\dfrac{M_\Delta}{16\pi r_{-2}^3\left[\ln(1+c_\Delta)-\dfrac{c_\Delta}{(1+c_\Delta)}\right]},
\end{equation}

\noindent where $c_\Delta=r_\Delta/r_{-2}$ is the halo concentration and $M_\Delta$ is the total mass within the radius $r_\Delta$, such that the average halo density within $r_\Delta$ is $\Delta \rho_\mathrm{crit}$, where $\rho_\mathrm{crit}(z) \equiv3H(z)^2/8\pi G$ is the critical density of the universe. Here $H(z)$ is the Hubble parameter at redshift $z$ and $G$ is the gravitational constant.
We consider both $r_\mathrm{200}$, with $\Delta=200$ independent of $z$, and $r_\Delta=r_\mathrm{vir}$, where $\Delta=\Delta(z)$ as defined by \cite{Bryan1998}.

We used the \textsc{universemachine} \citep{Behroozi2019} relationships, derived from the Bolshoi-Planck simulations and successfully compared to observations, to get the typical virial mass of the DM halo of the Sparkler galaxy. Given the interval of the Sparkler stellar mass (first row in Table \ref{tab:halo}), this value spans the range $M_\mathrm{vir}(z\approx1.4)=(1.91-2.65)\times 10^{11}\,\msun$ (assuming $h\equiv H_0/100 {\rm \,km\,s^{-1}\,Mpc^{-1}}=0.678$). In order to maximize the dynamical effects that affect the clump evolution, throughout this work we have fixed the halo mass to the highest possible value.
The concentration is given by the analytic relations by \cite{Dutton2014}, assuming an NFW profile, and spans the interval $c_\mathrm{vir}(z\approx1.4)=5.79$.
We converted the $M_\mathrm{vir}-c_\mathrm{vir}$ pairs to $M_{200}-c_{200}$ pairs by means of the \textsc{colossus} package \citep{Diemer2018}. The resulting values are $M_{200}(z\approx1.4)=2.50\times 10^{11}\,\msun$ and $c_{200}(z\approx1.4)=5.33$.

\begin{figure}[t!]
\centering
 \includegraphics[width=0.49\textwidth]{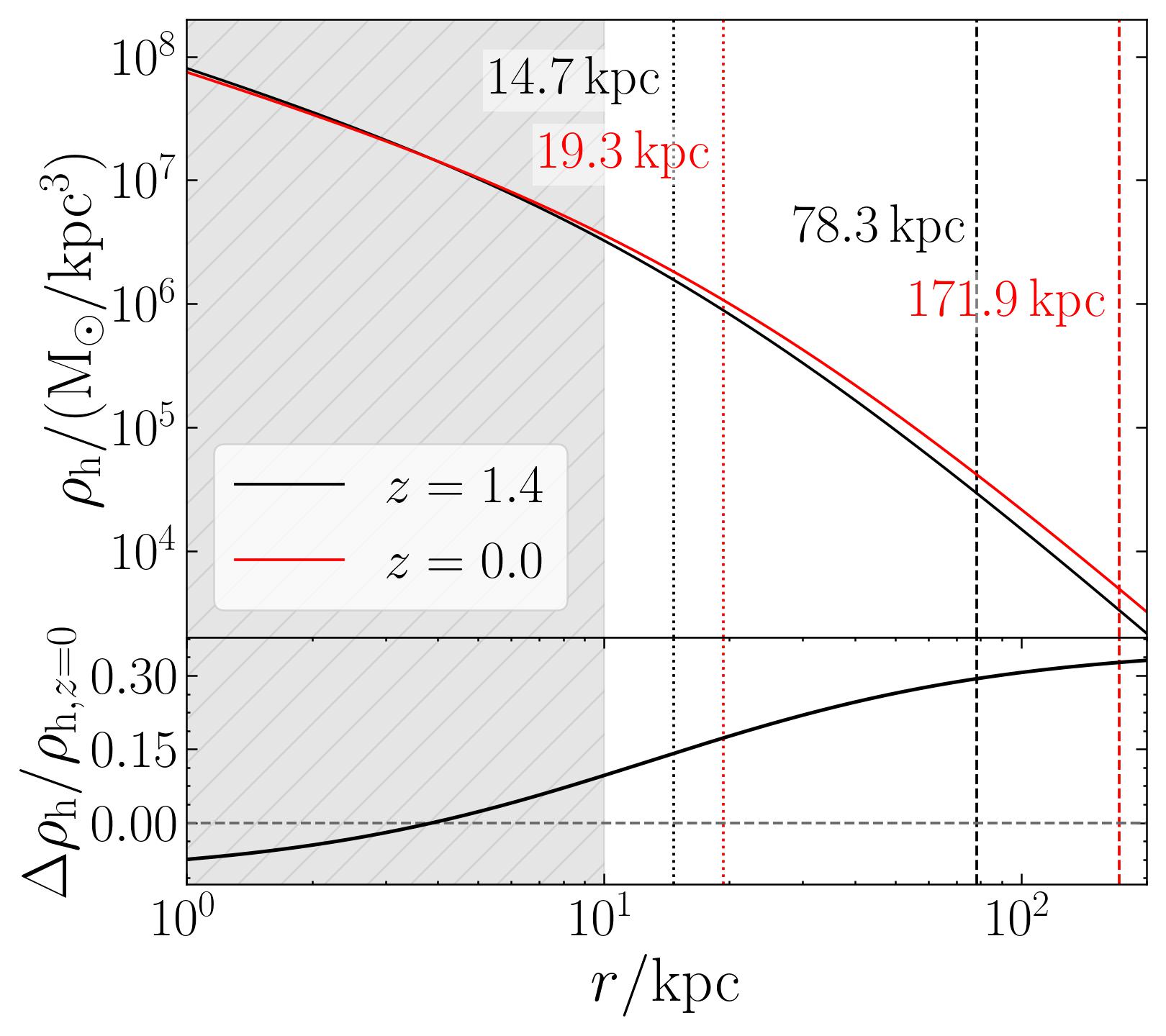}
 \caption{Top panel: density profile of the highest-mass dark matter halo of the Sparkler galaxy, as predicted by the models by \cite{Correa2015}. Black lines indicate the $z=0$ profile while red lines the $z=1.378$ one. The dotted lines mark $r_{-2}$, the dashed lines mark $r_{200}$. The grey shaded area marks the region within 1 and 10 kpc, which is the interval under study. Bottom panel: difference between the halo density profile at $z=0$ and at $z=1.378$, normalized to the density profile at $z=0$.}
 \label{fig:halo_profiles}
\end{figure}

In order to estimate the halo evolution from $z=1.378$ to $z=0$, we traced its mass through redshift using the mass accretion history recipes by \cite{Correa2015}, implemented in the \textsc{cummah} package. The resulting masses and concentrations are $M_{200}(z=0)=5.43\times 10^{11}\,\msun$ and $c_{200}(z=0)=8.89$.
Using \textsc{colossus} \citep{Diemer2018}, we have computed and compared the mass density profiles of the halo at $z\approx1.4$ and $z=0$, considering the case of the highest total mass. The resulting profiles are shown in the top panel of Fig. \ref{fig:halo_profiles}, with their difference (normalized to the density at $z=0$) shown in the bottom panel: we find that the change in mass and concentration is mainly driven by the pseudo-evolution of the halo and that the density profile does not change significantly, never exceeding the value of $0.1$ within 10 kpc (marked by the grey shaded area).
The virial radius $r_{200}$ increases from $78.34\,\mathrm{kpc}$ to $171.87\,\mathrm{kpc}$ (more than a factor $2$) and causes the increase in mass and concentration as well, while other structural parameters such as the scale radius $r_{-2}$ do not substantially change (from $14.71\,\mathrm{kpc}$ to $19.34\,\mathrm{kpc}$, about a factor $1.3$).

We conclude that the DM density profile of a Sparkler-like galaxy does not substantially change from $z\approx1.4$ to $z=0$, at least in the regions of interest for the present work. Therefore, for our purposes, the halo can be modelled as an isolated stationary system, neglecting its cosmological evolution and accretion history, by assuming that the galaxy did not undergo any major merger in this interval of time.
Finally, we computed the predicted stellar mass of the Sparkler galaxy at $z=0$, using the corresponding \textsc{universemachine} halo-to-stellar mass relation, finding $M_*(z=0)=8.79\times 10^9\,\msun$, in agreement with the prediction by \cite{Adamo2023} using results from the EAGLE simulation \citep{Crain2015,Schaye2015}. In Table \ref{tab:halo} we list the parameters of the halo of the Sparkler galaxy at $z=1.378$ and $z=0$, together with the Sparkler stellar mass. For completeness, we include also the parameters of the low-mass halo, obtained from the lowest bound of the estimate of the Sparkler stellar mass.

\begin{table}[t!]
\fontsize{10pt}{10pt}\selectfont
\setlength{\tabcolsep}{4pt}
\renewcommand{\arraystretch}{1.4} % Default value: 1
\centering
\caption{Parameters of the Sparkler galaxy and of its halo, modelled as an NFW following the prescription described in Sects. \ref{sec:spark_properties} and \ref{sec:halo_properties}, in the highest-mass case scenario, which is used throughout this work (in parentheses, we also report the values for the lowest-mass case scenario).}
\begin{tabular}{lc|cc}
\bottomrule\bottomrule
$z$ & & $1.378$ & $0$\\\hline
$M_*$ & $[10^9\,\msun]$ & $1.0\,(0.5)$ & $8.79\,(4.63)$\\
$M_{200}$ & $[10^{11}\,\msun]$ & $2.50\,(1.81)$ & $5.43\,(3.84)$\\
$c_{200}$ & & $5.33\,(5.44)$ & $8.89\,(9.21)$\\
$r_{-2}$ & [kpc] & $14.71\,(12.91)$ & $19.34\,(16.63)$\\
$r_{200}$ & [kpc] & $78.34\,(70.25)$ & $171.87\,(153.13)$\\
\toprule
\end{tabular}
\tablefoot{From top to bottom: redshift $z$, galactic stellar mass $M_*$, DM halo mass $M_{200}$, DM halo concentration $c_{200}$, DM halo scale radius $r_{-2}$, DM halo virial radius $r_{200}$.}
\label{tab:halo}
\end{table}

\section{Dynamical friction simulations with single-particle clumps}\label{sec:dyn_fric}

In this section, we study the effects of dynamical friction on the system of clumps orbiting around the Sparkler galaxy.
We first computed the dynamical friction timescale of the clumps in the Sparkler system, in order to assess if this process is expected to play a major role in the dynamical evolution of the clumps.
The dynamical friction timescale for a clump with mass $M_\mathrm{cl}$, in a circular orbit at a distance $r$ from the centre of a spherical halo can be estimated as \citep[][Eq. 8.13]{Binney2008}

\begin{equation}\label{eq:tfric}
    t_\mathrm{fric}=\frac{1.17}{\ln\Lambda}\frac{M_\mathrm{h}(r)}{M_\mathrm{cl}}t_\mathrm{cross},
\end{equation}

\noindent where $\ln\Lambda$ is the Coulomb logarithm, $M_\mathrm{h}(r)$ is the mass of the host system within radius $r$ and $t_\mathrm{cross}=r/v_\mathrm{c}$ is the crossing time, with $v_\mathrm{c}=[G\,M_\mathrm{h}(r)/r]^{1/2}$ the circular velocity at $r$. Throughout this work, we assume $\ln\Lambda=10$.

\begin{figure}[t!]
\centering
 \includegraphics[width=0.49\textwidth]{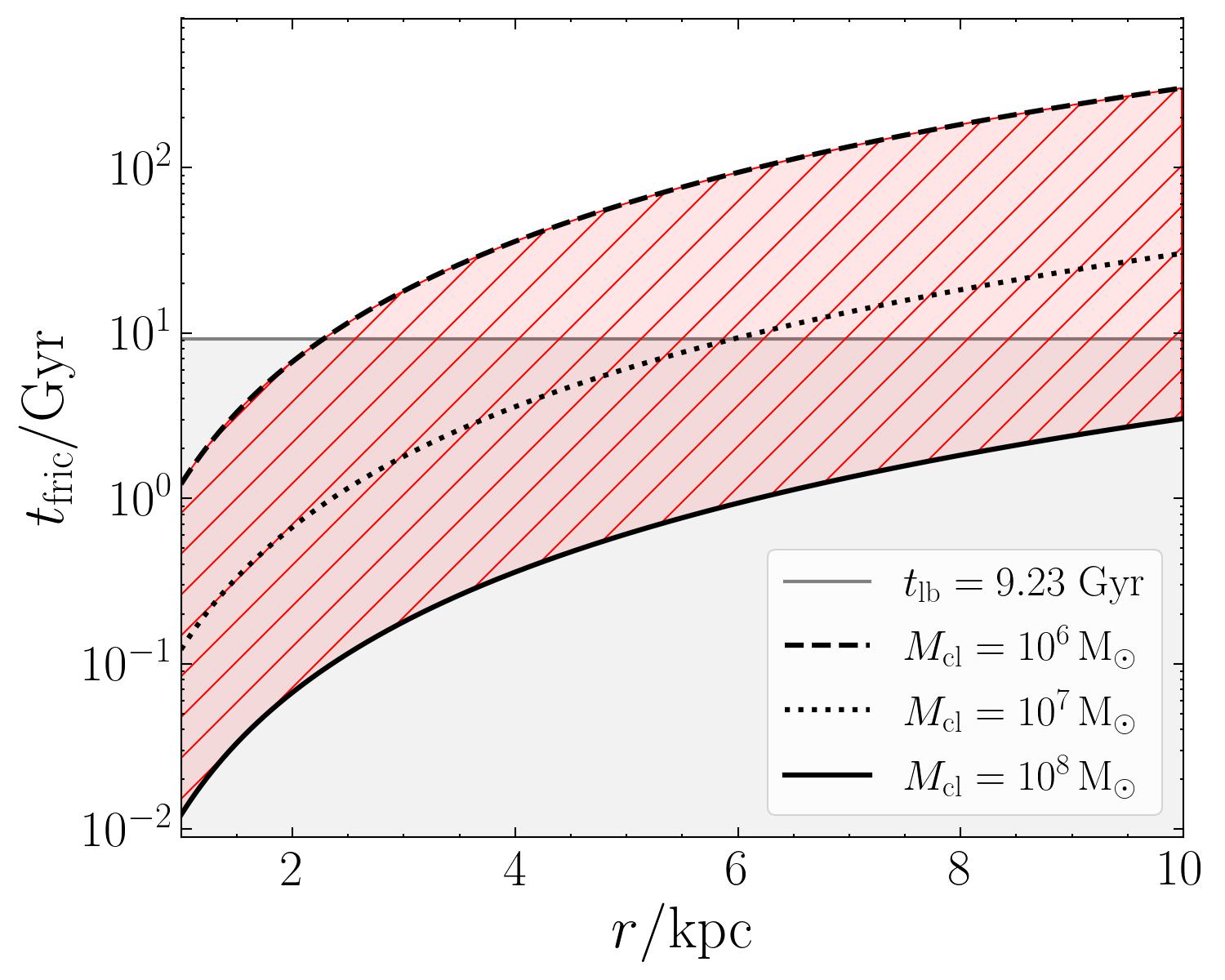}
 \caption{Dynamical-friction timescales for clumps of mass $10^6$, $10^7$ and $10^8\,\mathrm{M_\odot}$ (dashed, dotted and solid lines, respectively) as a function of the distance from the centre of the halo. The red-dashed area covers the range of dynamical friction timescales relative to the interval of clump masses and distances of interest. The grey horizontal line corresponds to the look-back time at $z\approx1.4$ $t_\mathrm{lb}=9.23$ Gyr, which divides the plot between the region in which the dynamical friction is most likely inefficient ($t_\mathrm{fric}>9.23$ Gyr) and efficient ($t_\mathrm{fric}<9.23$ Gyr), the latter marked by the grey area.}
 \label{fig:tfric}
\end{figure}

In Fig. \ref{fig:tfric}, we show the dynamical friction timescales for a halo defined as in Sect. \ref{sec:halo_properties} and for clumps with a mass of $10^6$, $10^7$ and $10^8\,\mathrm{M_\odot}$ and at distances between 1 and 10 kpc. The resulting timescales span a wide range of values, between $0.01$ and $100$ Gyr. The timescale is comparable to or shorter than the look-back time to $z\approx1.4$ (about $9.23$ Gyr) for relevant combinations of distances and masses.
Therefore, numerical simulations are needed to address quantitatively the question of whether the clumps will sink to the centre of the galaxy by $z=0$.

\subsection{Host galaxy model in the simulations}\label{sec:spark_dyn_mass}
In the simulations presented in this section, the clumps are modelled as single particles, with the Sparkler-like halo set according to the parameters derived in \ref{sec:halo_properties} and listed in Table \ref{tab:halo}. The halo is made of $10^7$ particles of mass $5.43\times 10^4\,\msun$ each. \rr{Modelling the halo by means of DM particles allows us to include directly the effects of dynamical friction in the simulation thanks to their mutual interaction with the particles representing the clumps.}

Throughout our work, we model the total mass of the Sparkler system as equal to the aforementioned DM halo mass, unless otherwise stated. The NFW density profile is characterized by a central cusp, such that $\rho_\mathrm{NFW}(r\ll r_{-2})\propto r^{-1}$. A central cusp is consistent with the profiles of the total mass (thus including both baryonic and DM components), computed from rotation curves in disk galaxies (such as the SPARC sample, \citealt{Lelli2016}, with central slope between $-1$ and $-0.5$).
Therefore, we can assume that the NFW profile is a good representation of the total mass density distribution of a Sparkler-like system.

The initial conditions of the halo are defined following the prescription described in \cite{Springel2005}. By fixing $M_{200}$ and $c_{200}$ of the desired NFW potential, the prescription by \cite{Springel2005} determines the \cite{Hernquist1990} model that most closely resembles the original NFW model in the innermost regions, yet having a finite and converging total mass equal to $M_{200}$. The Hernquist density profile is

\begin{equation}\label{eq:hernquist}
    \rho_\mathrm{Hernquist}=\dfrac{M_\mathrm{tot}}{2\pi}\dfrac{a}{r\,(r+a)^3},
\end{equation}

\noindent where $M_\mathrm{tot}$ is the total mass and $a$ is the scale radius. In our case, $M_\mathrm{tot}=M_{200}(z=0)$, with scale radius $a=32.3$ kpc and consequent $r_{200}=151$ kpc.
\rr{The softening of the gravitational force field is realised adopting a Plummer kernel \citep{Aarseth1963} with scale-length} $150$ pc (one third of the average inter-particle distance inside the half-mass radius).

\subsection{Simulations of single clumps in circular orbits}\label{sec:sim_single_clumps}

As a preliminary numerical exploration, we ran a set of 20 simulations in which one clump, initially on a circular orbit, is followed for $9.23$ Gyr in the Sparkler-like halo defined in Sect. \ref{sec:spark_dyn_mass}.
The initial distance from the centre of the host ranges from 2 to 6 kpc -- the range of clump observed distances in the Sparkler galaxy. The explored masses range between $1\times 10^6\,\msun$ and $6\times 10^7\,\msun$, consistent with the observed mass range.

\begin{figure*}[t!]
\centering
 \includegraphics[width=0.99\textwidth]{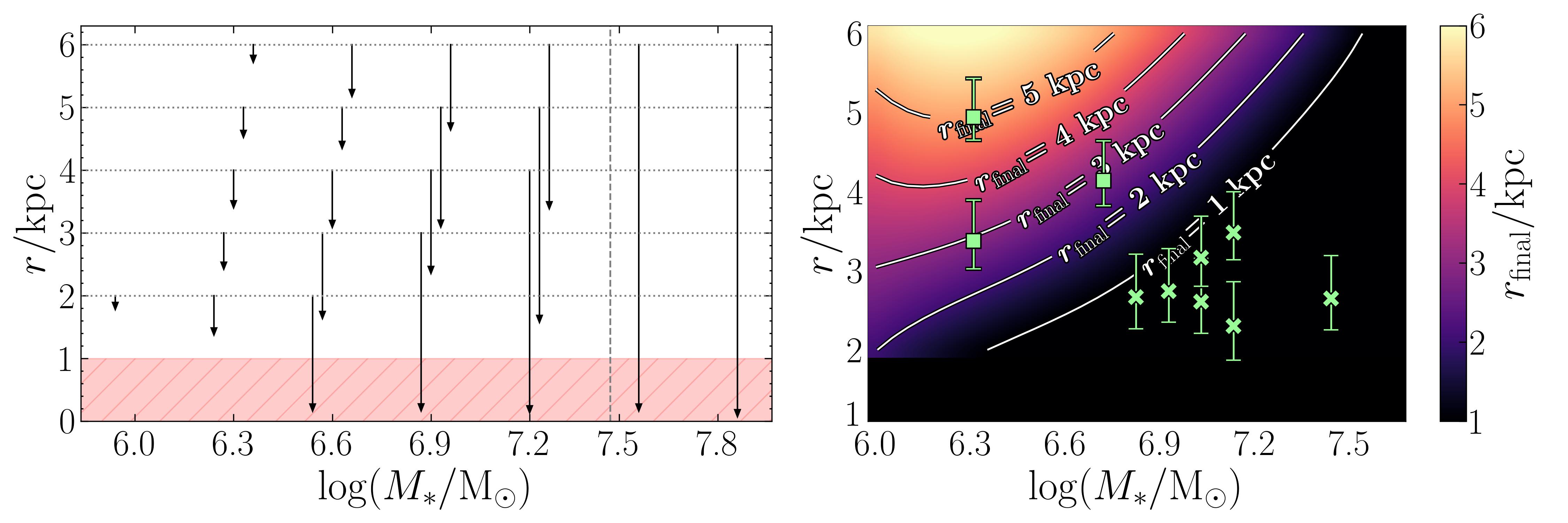}
 \caption{Left panel: evolution over $9.23$ Gyr of the distances from the centre of the point-mass clumps orbiting in a Sparkler-like potential with circular orbit initial conditions. On the $x-$axis, the mass of the clump is reported, and the black arrows point from the initial to the final distance. The final distance is computed as the average distance from the centre of the system, within the last $5\%$ of the simulated time.
 The masses larger than the vertical dotted grey line are also larger than the observed stellar masses of the Sparkler clumps. We classify as survived the clumps with final distance out of the shaded area.
 Right panel: colour map of the final distance of the clumps as a function of their initial distance ($y-$axis) and dynamical mass ($x-$axis), obtained by interpolating the simulations shown in the left panel. The colour bar traces the final distance and goes from 1 kpc (or less) to 6 kpc (the largest simulated initial distance).
 The contours mark the combinations of initial distance and mass that result in specific values of final distance (from 1 to 5 kpc). The green data points are the extraplanar Sparkler clumps \citep{Claeyssens2023,Forbes2023}. For each clump three distances are computed, de-projecting the observed distances depending on the three spatial configurations studied in Sect. \ref{sec:clumps_dist_distribution} (face-on disk, edge-on disk and spherical), as described in Sect. \ref{sec:sim_single_clumps}. The symbols (squares and crosses) indicate the median of the three possible distances, with the error-bars corresponding to the minimum and maximum distance.
 We used squares for the clumps that survive, while we used crosses for the clumps for which the inferred final distance is smaller than 1 kpc.}
 \label{fig:migration_live_halo}
\end{figure*}

The left panel of Fig. \ref{fig:migration_live_halo} shows the migration of clumps from the initial to the final position, after $9.23$ Gyr, for given mass. 
Clumps with total mass $\log(M_*/\msun)\leq6.3$ never reach the centre of the system, even when starting at a distance of 2 kpc. Clumps with higher masses are more effectively brought to the centre of the system: for masses higher than about $3\times 10^7\,\msun$, dynamical friction is extremely efficient across the whole range of initial orbital radii and the clumps reach the centre even when starting at 6 kpc. These results strongly suggest that, if the clumps had a DM halo, and therefore a dynamical mass significantly higher than the stellar mass, it would be very unlikely to observe them in the local Universe, because they would quickly sink to the centre of the system.

To generalize the results of these simulations to any pair $M_*-$initial radius in the relevant range, we interpolated the map of the left panel of Fig. \ref{fig:migration_live_halo} using a smooth bivariate spline\footnote{\url{https://docs.scipy.org/doc/scipy/reference/generated/scipy.interpolate.SmoothBivariateSpline.html}}. The right panel of Fig. \ref{fig:migration_live_halo} shows the map of the final distance (described by the colour scale) as a function of the initial distance and the mass of the clump. Under these hypotheses, the Sparkler clumps (assumed to contain no DM) are shown in the right panel of Fig. \ref{fig:migration_live_halo}. For each clump we computed three estimates of the intrinsic distance, taking into account the possible projection effects of each of the three spatial configurations analysed in Sect. \ref{sec:clumps_dist_distribution}. In the face-on disk scenario, the observed distance of each clump coincides with the intrinsic one by construction, therefore no corrections are needed. In the edge-on disk and in the spherical configurations, the de-projected position of the $i$th clumps is computed as $(R_{\mathrm{obs},i}^2+\widetilde{z}_i^2)^{1/2}$, where
$\widetilde{z}_i$ is the median of its line-of-sight probability distribution $p_{\mathrm{los},i}(|z|)$. In the edge-on disk configuration, $p_{\mathrm{los},i}(z)=p_\mathrm{exp}[(R_{\mathrm{obs},i}^2+z^2)^{1/2}\,;R_\mathrm{e}]$, where $p_\mathrm{exp}$ is the edge-on disk exponential distribution, with effective radius $R_\mathrm{e}$ (listed in the first row of the third column in Table \ref{tab:spatial_df}), and $R_{\mathrm{obs},i}$ is the observed position of the $i$th clump.
In the spherical configuration, $p_{\mathrm{los,}i}(z)=p_\mathrm{Sersic3D}[(R_{\mathrm{obs},i}^2+z^2)^{1/2}\,;R_\mathrm{e},n]$, with $p_\mathrm{Sersic3D}$ from Eq. \ref{eq:sersic_3D}, with effective radius $R_\mathrm{e}$ and Sérsic index $n$ (listed in the first row of the fourth column in Table \ref{tab:spatial_df}).

As one can notice, $7$ clumps out of $10$ fall within $1$ kpc from the centre.
This experiment hints that dynamical friction is able to drastically change the distribution of the clumps surrounding the Sparkler galaxy. If the clumps do not possess a DM halo, the outermost ones (at $3-5$ kpc) are poorly affected by dynamical friction, while in the innermost regions only the low-mass ones (less massive than about $5\times 10^6\,\msun$) do not sink to the centre of the system.

\subsection{Simulations of Sparkler-like clump system}\label{sec:simul_sparkler_like}

In this section, in order to assess the effects of dynamical evolution on the Sparkler clumps we ran a series of simulations that explore three Sparkler-like spatial configurations -- face-on disk, edge-on disk, or spherical -- as defined in Sect. \ref{sec:clumps_dist_distribution}. Each simulation includes $10$ clumps with initial masses sampling the distribution discussed in Sect. \ref{sec:clumps_mass_distribution}.
For each spatial configuration we ran $15$ simulations, in order to obtain statistically relevant results.
The NFW halo, representing the total mass density distribution of the Sparkler galaxy, is built according to the procedure described in Sect. \ref{sec:spark_dyn_mass}.

For each simulation, the clump masses are extracted within $3\sigma$ of the best-fitting Gaussian distribution (see Table \ref{tab:mdf}).
For the initialization of positions and velocities, the steps used in the disk configurations (either face-on and edge-on) are the following:

\begin{enumerate}
    \item \textit{Initial distance}: extracted from the spatial distribution corresponding to the configuration under study. In particular, the initial distances are imposed to be between 1 and 10 kpc (this interval encloses $\approx 85\%$ of the total probability).
    \item \textit{Position angle}: randomly extracted under the condition that the clumps are initialized at least at $300$ pc (about 10 times the smallest scale radius of the Sparkler clumps, see the discussion in Sect. \ref{sec:size_mass}) from each other and that the centre of mass of the clump system is consistent with the centre of the halo within the softening length.
    \item \textit{Velocity}: the clumps are initialized on co-planar circular orbits.
\end{enumerate}

For the spherical configuration, the clump initial positions and velocities are extracted from the ergodic distribution function of the system, generated by putting in equilibrium the clump best-fit Sérsic distribution (Eq. \ref{eq:sersic}) with the Sparkler-like Hernquist gravitational potential, as described in Sect. \ref{sec:spark_dyn_mass}. The initial conditions are generated by means of the software code \textsc{OpOpGadget}\footnote{\url{https://github.com/giulianoiorio/OpOpGadget}}.Since we are modelling the evolution of GC-progenitor candidates, which have ages of a few Gyr at $z\approx1.4$ \citep{Mowla2022,Claeyssens2023,Adamo2023}, we can neglect mass losses by stellar evolution and keep the mass of the clump-particles constant \citep{Calura2014}.

\begin{figure*}[t!]
\centering
 \includegraphics[width=0.98\textwidth]{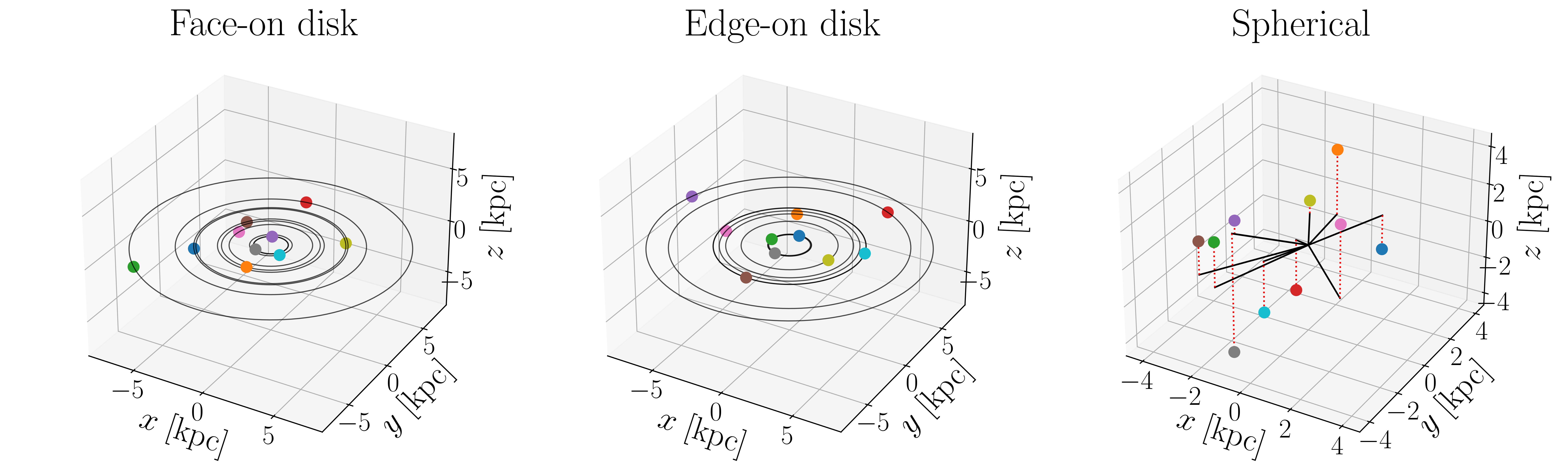}
 \caption{\rr{Initial spatial distribution of the 10 clumps in representative simulations of the three different spatial configurations described in Sect. \ref{sec:clumps_dist_distribution}. The clump initial positions are determined as described in Sect. \ref{sec:simul_sparkler_like}. From left to right: face-on disk, edge-on disk and spherical configurations. Each clump is displayed as a coloured dot. In the first two panels, we plot in grey the trajectory of the circular orbit at which each clump is initialized. This also makes apparent that the orbits of the clumps are co-planar in these two configurations. In the right panel, for each clump we plot the projected distance on the $x-y$ plane as the black solid line, and the $z$ coordinate of the clump as the red dotted line, to show that the clumps are not co-planar.}}
 \label{fig:configurations_plot3d}
\end{figure*}

\rr{To illustrate the differences among the three spatial configurations, we present three spatial distributions of 10 clumps for each configuration in Figure \ref{fig:configurations_plot3d}. It is evident that the clumps in the face-on and edge-on disk configurations are co-planar, while in the spherical configuration, they do not occupy the same orbital plane.}

\subsection{Results}\label{sec:res_configuration}
We present here the outcomes of the simulations described in Sect. \ref{sec:simul_sparkler_like}.
In the following, we will call survived clumps all the clumps with a final distance larger than 1 kpc. The final distance is computed as the average distance in the last $5\%$ of the simulation, which is $\approx 500$ Myr, about 25 times the crossing time at 1 kpc. \rr{We have chosen to evaluate the final average distance since clumps on elliptical orbits may have peri-centre distances below our threshold, yet spending most of their time beyond this critical value.} Such threshold has been chosen firstly because the focus of this work is on testing the hypothesis that the Sparkler extraplanar clumps are progenitors of local halo/disk GCs, therefore distances smaller than 1 kpc would be incompatible with such scenario.
Furthermore, the dynamical evolution of the clumps in the central regions depends on a number of factors, such as the central density profile of the DM halo and the presence of substructures and of a bulge/proto-bulge, which go beyond the scope of this work.

\subsubsection{Face-on disk configuration}\label{sec:res_faceon}
\begin{figure}[t!]
\centering
 \includegraphics[width=0.49\textwidth]{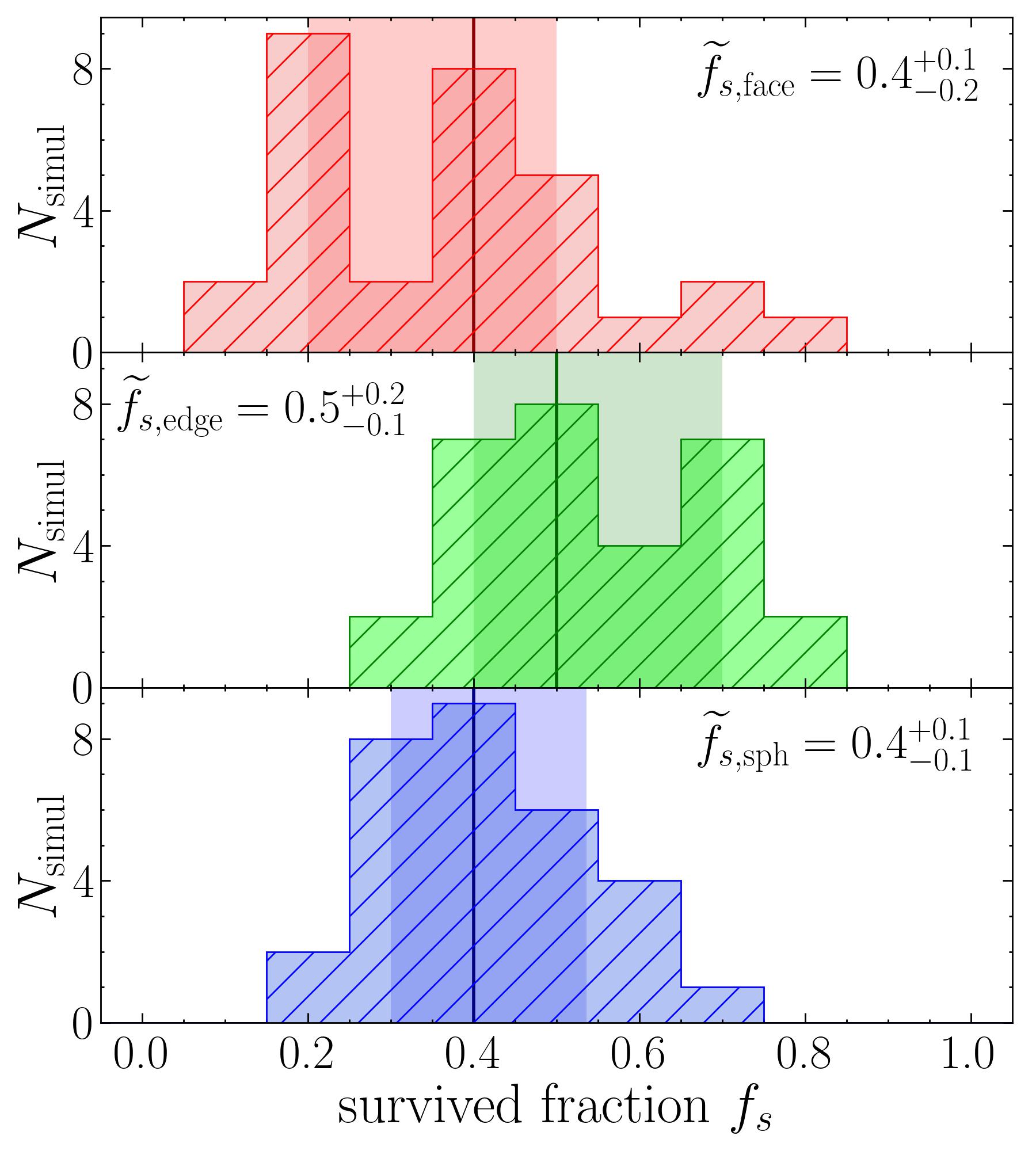}
 \caption{Fraction of survived clumps of each of the 30 simulations ran in the face-on (top panel, in red), edge-on (middle panel, in green) and spherical (bottom panel, in blue) configurations. A clump is flagged as survived if its distance from the centre of the halo at the end of the simulation is larger than 1 kpc. The vertical solid line and the shaded area show the median fraction of survived clumps $\widetilde{f}_s$ and the $1\sigma$ interval, respectively (also reported in the legend).}
 \label{fig:surv_fraction_conf}
\end{figure}

We first focus on the results obtained for the simulations of the face-on configuration.
The top panel in Fig. \ref{fig:surv_fraction_conf} shows the histogram of the fraction of survived clumps considering the 30 simulations. The median value is $0.4^{+0.1}_{-0.2}$.
Interestingly, in all the runs there is at least one survived clump. These results suggest that the Sparkler galaxy is expected to retain $3-4$ clumps in the outskirts, while the rest of them will most likely fall into its centre.

\begin{figure}[t!]
\centering
 \includegraphics[width=0.49\textwidth]{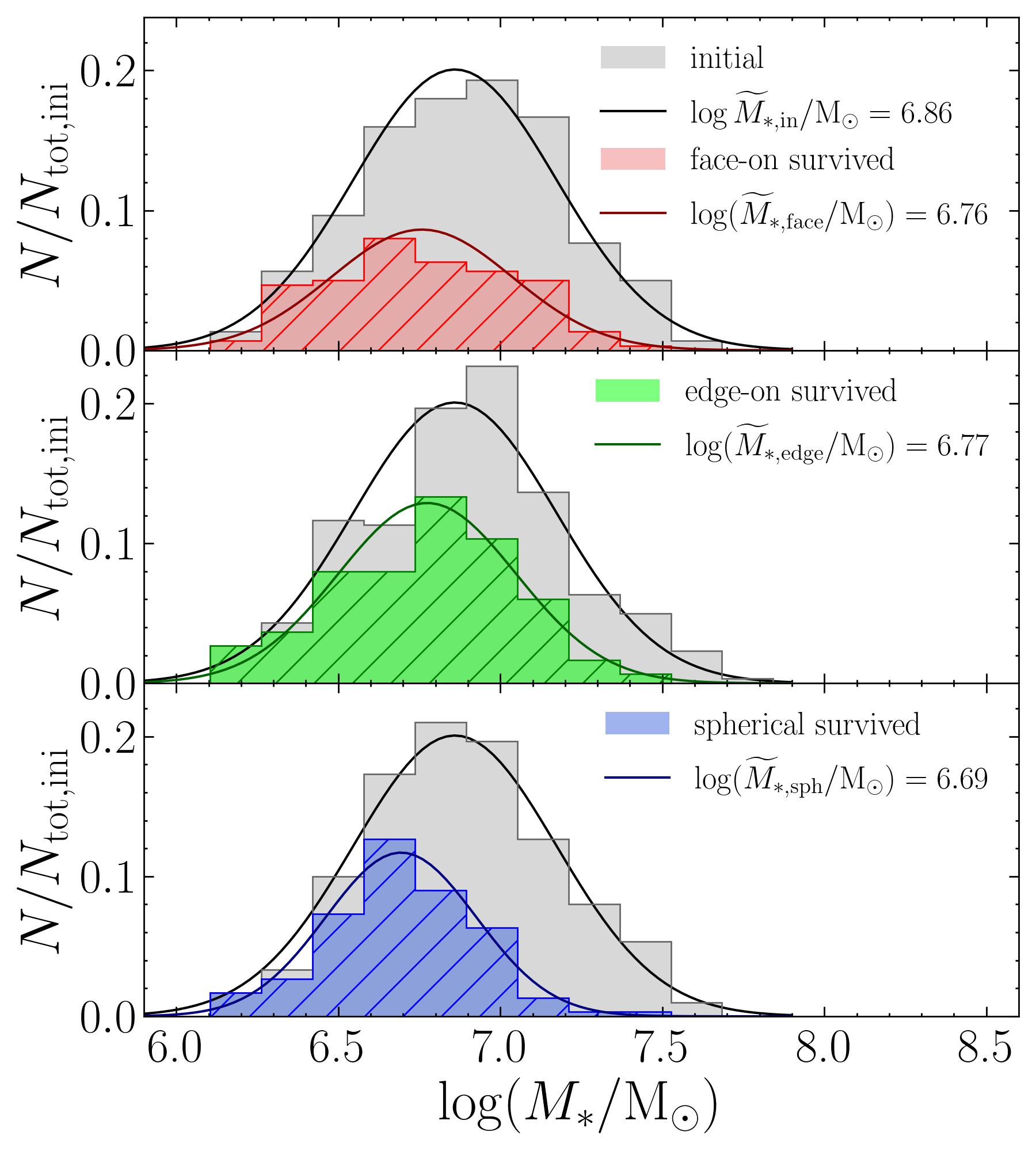}
 \caption{Fraction of initial clumps and survived clumps per mass-bin, shown as grey histograms and coloured histograms (red for the face-on configuration, green for the edge-on configuration and blue for the spherical configuration), respectively. The face-on configuration is shown in the top panel, the edge-on configuration in the middle panel, the spherical configuration in the bottom panel. The histograms are built binning all the clumps (either initial or survived) of the 30 simulations, normalized for the initial total number of clumps (10 for each simulation, therefore 300). The black and coloured lines are the best-fitting Gaussian distributions, of which we report the mean in the legend.}
 \label{fig:mass_distr_conf}
\end{figure}

The top panel of Fig. \ref{fig:mass_distr_conf} shows the comparison between the initial mass distribution and the mass distribution of survived clumps.
The survived distribution gets slightly narrower and peaks at smaller masses than the initial one. Indeed, the mean mass of the survived clumps is $\log (\widetilde{M}_*/\msun)=6.76$, $\approx0.1$ dex smaller than the initial one. Even though a non-negligible fraction of $\sim 10^7\,\msun$ clumps is predicted to survive, the survived clumps are preferentially the low mass ones.

\begin{figure}[t!]
\centering
 \includegraphics[width=0.49\textwidth]{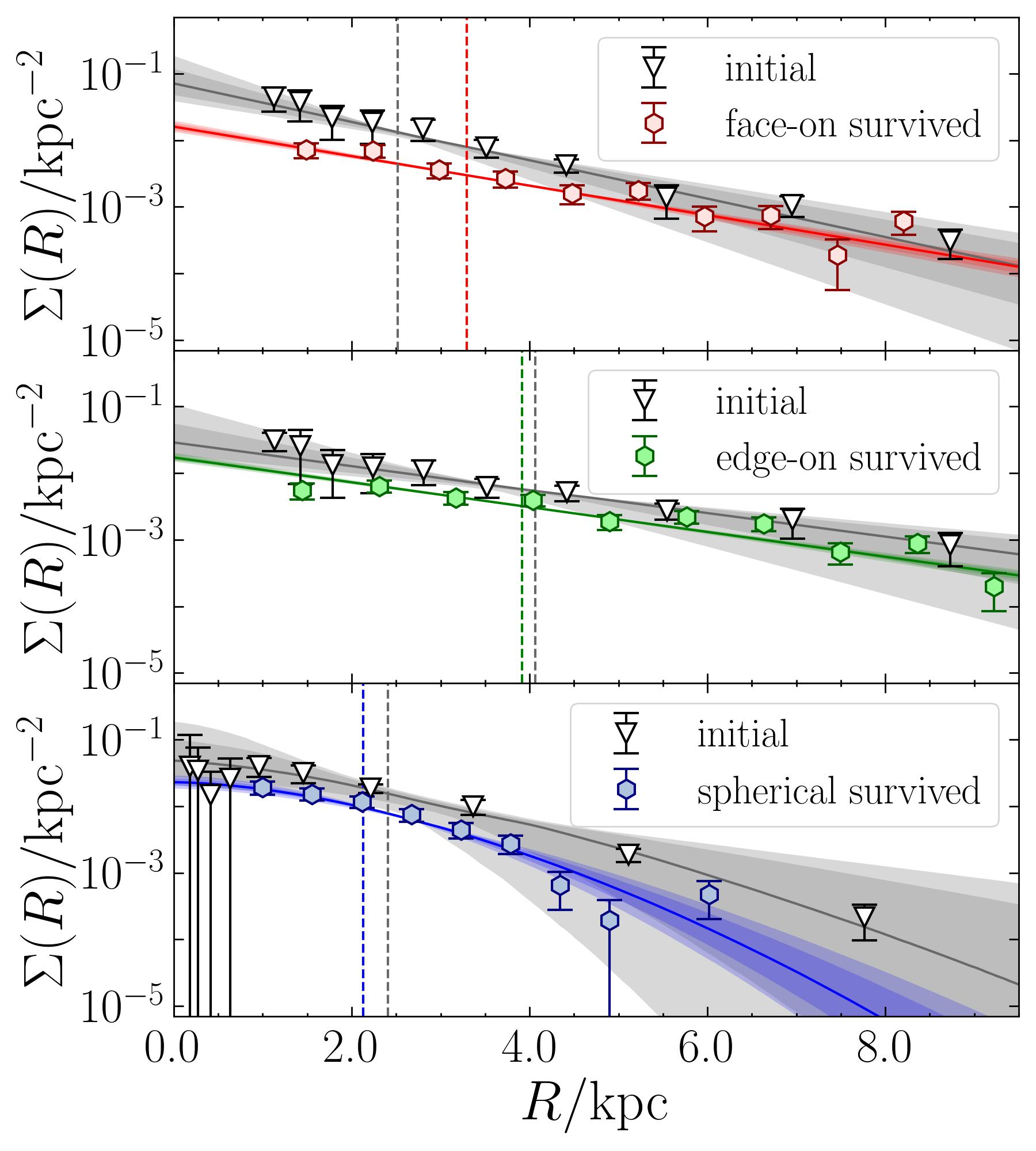}
 \caption{Surface number density profiles of the clumps in the face-on (top panel), edge-on (middle panel) and spherical configuration (bottom panel). In each panel, the dark grey line is the initial distribution, with the grey shaded areas covering the $1\sigma$ and $2\sigma$ intervals, built adopting the corresponding function described in Sect. \ref{sec:clumps_dist_distribution} with the best-fit parameters listed in the first row of Table \ref{tab:spatial_df}. All the profiles are projected along the $z$ axis. The triangles show the profiles of the initial clump positions in the $x-y$ plane, computed stacking together the 10 clumps in all the 30 runs. The coloured hexagons show the surface number density profiles at the end of the simulations, obtained stacking together all the survived clumps. The corresponding coloured solid line is the best-fit profile to the final projected distances of the survived clumps, obtained as described in Sect. \ref{sec:res_configuration}. All the profiles are normalized to the total number of initial clumps, so the integral of the initial profiles is equal to 1, and the integral of the survived profiles is equal to the average surviving fraction (distributions shown in Fig. \ref{fig:surv_fraction_conf}).}
 \label{fig:dist_distr_conf}
\end{figure}

Finally, in the top panel of Fig. \ref{fig:dist_distr_conf} we show the clump initial number density profiles (black triangles), projected along the $z-$axis, compared to the profiles of the survivors at the end of the simulations (red squared). The latter was fitted again with an exponential profile (shown as the red solid line), which still describes remarkably well the distance distribution of the survived clumps.
In particular, the best-fitting scale radius is $3.30^{+0.23}_{-0.21}$ kpc, larger than the initial value of $2.52^{+0.68}_{-0.50}$ kpc (shown as the dark grey solid line), testifying how most of the lost clumps were in the innermost regions of the system. Indeed, the profile of the survivors is flatter than the initial one.

\subsubsection{Edge-on disk configuration}\label{sec:res_edgeon}

In this configuration, the initial intrinsic distance distribution has a larger scale radius $R_\mathrm{e}$ than in the face-on disk (see Table \ref{tab:spatial_df}). As a consequence, the clumps are initialized further from the centre and are less affected by dynamical friction.
This difference results in a larger survived fraction (middle panel of Fig. \ref{fig:surv_fraction_conf}), with a median value of $0.5^{+0.2}_{-0.1}$. Therefore, on average, $5-6$ clumps out of 10 survive down to $z=0$.
As displayed in the middle panel of Fig. \ref{fig:mass_distr_conf}, the survived mass distribution is consistent with the face-on one, with a mean stellar mass $\log (\widetilde{M}_*/\msun)=6.77$.

For what regards the evolution of the spatial distribution, the middle panel of Fig. \ref{fig:dist_distr_conf} shows the clump initial number density profile, projected along the $z-$axis, and the profile of the survived clumps at the end of the simulations.
The survived distribution is fitted to an exponential profile following the same method described in Sect. \ref{sec:clumps_dist_distribution} and shown as the green solid line, to be compared with the initial profile. The final scale radius of the survived clumps is $R_\mathrm{e}=3.92^{+0.22}_{-0.21}$ kpc, indistinguishable from the initial one ($R_\mathrm{e}=4.06^{+1.55}_{-1.02}$).

\subsubsection{Spherical configuration}\label{sec:res_spherical}
When the configuration is spherical, the fraction of survived clumps (bottom panel in Fig. \ref{fig:surv_fraction_conf}) is $\widetilde{f}_s=0.4\pm 0.1$, consistent with the one obtained in the face-on disk case (top panel of Fig. \ref{fig:surv_fraction_conf}). However, the final mass distribution peaks at smaller masses than in both disk configurations, with a mean survived mass $\log (\widetilde{M}_*/\msun)=6.69$ (bottom panel in Fig. \ref{fig:mass_distr_conf}).
The projected number density profile of the survived clumps remains consistent with the initial one, as shown in the bottom panel of Fig. \ref{fig:dist_distr_conf}, similarly to what we have found for the edge-on disk configuration.
In order to quantitatively compare with the initial distribution (see the solid curves in bottom panel of Fig. \ref{fig:dist_distr_conf}), we have fitted the final distances of survived fraction of clusters to a de-projected Sérsic profile (Eq. \ref{eq:sersic_3D}), following the same procedure described in Sect. \ref{sec:clumps_dist_distribution}. The values of the fit parameters (the scale radius and the Sérsic index) are reported in the third column of Table \ref{tab:spatial_df} and are consistent with the initial parameters within $1\sigma$.

\section{Tidal stripping simulations}\label{sec:resolved_clumps}

In this section, we study the mass evolution of the Sparkler clumps under the effects of the tidal stripping exerted by gravitational potential of the host system.
Assuming that the satellite (a clump in our case) is on a circular orbit around the host system (Sparkler) and that both systems are characterized by a spherical mass distribution, the effects of tidal stripping can be evaluated approximately by computing the tidal radius $r_\mathrm{tidal}$.
The tidal radius divides the regions in which the satellite mass is highly affected by the host gravitational force (beyond $r_\mathrm{tidal}$) from that in which the self-gravity of the satellite is strong enough to retain it (within $r_\mathrm{tidal}$). It is defined as the distance from the centre of the satellite at which

\begin{equation}\label{eq:rtidal}
    3\overline{\rho}_\mathrm{h}(r)=\overline{\rho}_\mathrm{cl}(r_\mathrm{tidal}),
\end{equation}

\noindent where $\rho_\mathrm{h}$ and $\rho_\mathrm{cl}$ are the mean halo and clump density profile within $r$ and $r_\mathrm{tidal}$, respectively, and $r$ is the distance between the centre of the system and the centre of the clump \citep[see][their Eq. 8.92]{Binney2008}. The halo density $\rho_\mathrm{h}$ is defined as in Eq. \ref{eq:nfw}, whereas for $\rho_\mathrm{cl}$ we opted for a Plummer density profile \citep{Plummer1911}

\begin{equation}\label{eq:plummer}
    \rho_\mathrm{plummer}(r) = \dfrac{3M_*}{4\pi r_\mathrm{s}^3}\left( 1+\dfrac{r^2}{r_\mathrm{s}^2} \right)^{-5/2},
\end{equation}

\noindent where $M_*$ is the total stellar mass and $r_\mathrm{s}$ is the scale radius. For a Plummer density profile, the projected half-mass radius is $R_\mathrm{eff}=r_\mathrm{s}$.

\begin{figure}[t!]
\centering
 \includegraphics[width=0.49\textwidth]{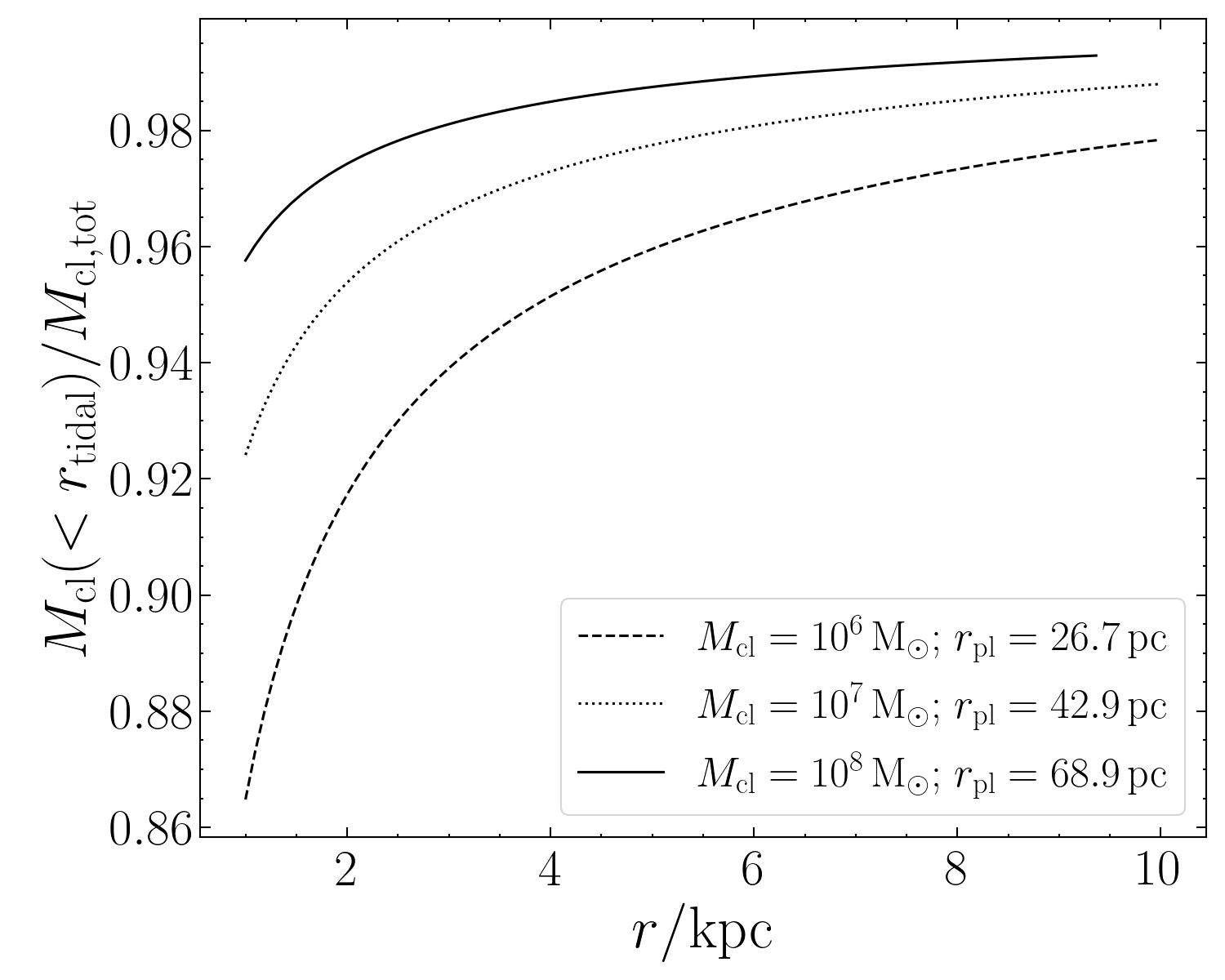}
 \caption{Fraction of the clump mass enclosed within the tidal radius, defined as in Eq. \ref{eq:rtidal}. The masses of the clumps are $10^6$, $10^7$ and $10^8\,\msun$ (dashed, dotted and solid lines, respectively). The radius of the clumps is given by Eq. \ref{eq:size_mass}. Tidal radii are computed for distances between 1 and 10 kpc.}
 \label{fig:rtidal}
\end{figure}

We have computed the fraction of clump mass enclosed within $r_\mathrm{tidal}$ for clumps of different mass and for distances ranging between 1 and 10 kpc. The results are shown in Fig. \ref{fig:rtidal}.
For each clump mass, we compute $R_\mathrm{eff}$ from the size-mass relation (Eq. \ref{eq:size_mass}).
The mass within $r_\mathrm{tidal}$ is always larger than the $86\%$ of the initial total mass. We note that this study is representative only of the sub-sample of seven resolved extraplanar clumps in the Sparkler system. Considering that the upper limits to $R_\mathrm{eff}$ of the unresolved clumps are $11-12$ pc, we conclude they are more compact than the resolved ones, therefore they will tend to retain more mass than the resolved ones, at a given mass.
Even though these results are just an approximate prediction, they suggest that the mass loss due to tidal stripping is expected to have minor effects on our predictions about the effects of dynamical friction shown in Sect. \ref{sec:res_configuration}. However, the mass distribution of the clumps can change between $z\approx1.4$ and $z=0$ due to mass loss, especially at the low-mass end, and therefore it is useful to model tidal stripping by means of numerical simulations. In the following sections, we will describe the simulations that we performed to better study the effects of this process and to test different shapes of the host.

\subsection{Tidal stripping in a spherical host}\label{sec:resolved_clumps_setup}
In these experiments, the Sparkler total mass is included as an external potential (i.e. a static potential), while the clumps are resolved. In order to be directly comparable to the mass profile defined in Sect. \ref{sec:spark_dyn_mass}, the external potential is modelled as a Hernquist model with the same parameters listed in Sect. \ref{sec:halo_properties} and obtained following the procedure described in \cite{Springel2005}.

The clumps are modelled as Plummer spheres (Eq. \ref{eq:plummer}). \rr{Even though the density profile of local GCs is typically well described by the King model \citep{King1966}, we opted for a profile which is still characterized by an inner core, but drops down less steeply in the outskirts. This choice is supported by the fact that the progenitors of local GCs do not necessarily already possess the final density profile, especially considering that, in the specific case of the Sparkler clumps, both masses and sizes are different from those observed in local GCs. Nonetheless, we verified a posteriori that the clump density becomes comparable to the King profile within $\approx 500\,\mathrm{Myr}$, since it gets truncated by the galactic tidal field.}
At a fixed mass $M_*$, the Plummer scale radius $r_\mathrm{s}$ is given by the size-mass relation (Sect. \ref{sec:size_mass}). We tested stellar masses between $\log (M_*/\msun)=6$ and $\log (M_*/\msun)=7.8$ in steps of $0.3$ dex. The clumps are always made of $10^5$ particles with mass $M_*/10^5$, therefore ranging from $10\,\msun$ to $630\,\msun$.
The softening length is $0.6$ pc, about one third of the average inter-particle distance within \rr{the clump half-mass radius ($\approx 1.7r_\mathrm{s}$), for our choice of softening. With this setup, any collisional effect is neglected. Indeed the 2-body relaxation time of these clumps, computed as $t_\mathrm{relax}=0.1(N/\ln N)\,t_\mathrm{cross}$ at the half-mass radius, in the range between $16-33$ Gyr (for $\log (M/\msun)=6$) and $360-720$ Gyr (for $\log (M/\msun)=7.8$), assuming that the stars have masses in the range $0.5-1\,\msun$.}
The clumps are initialized on circular orbits at 1 kpc from the centre of the halo.

\begin{figure}[t!]
\centering
 \includegraphics[width=0.49\textwidth]{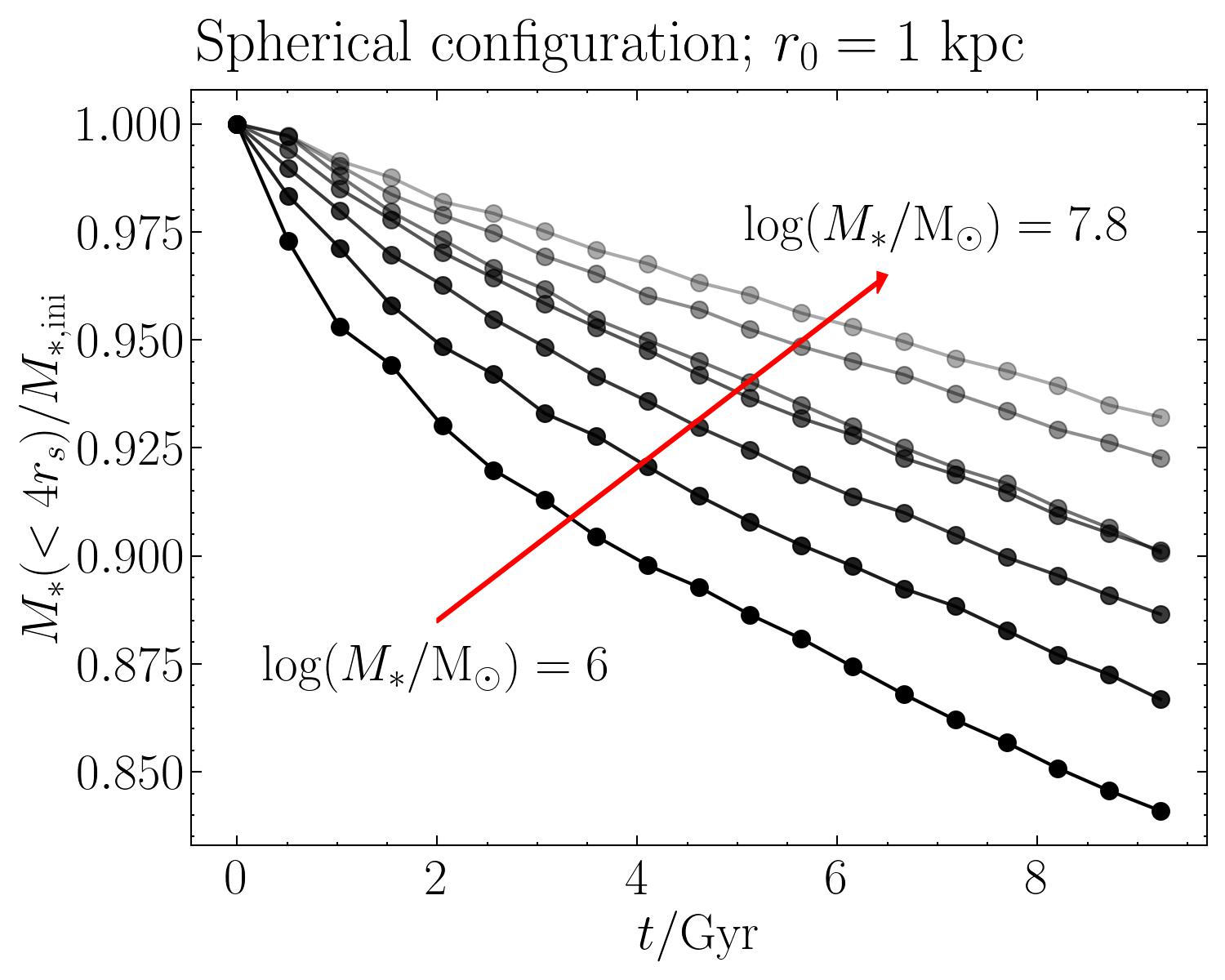}\\
 \includegraphics[width=0.49\textwidth]{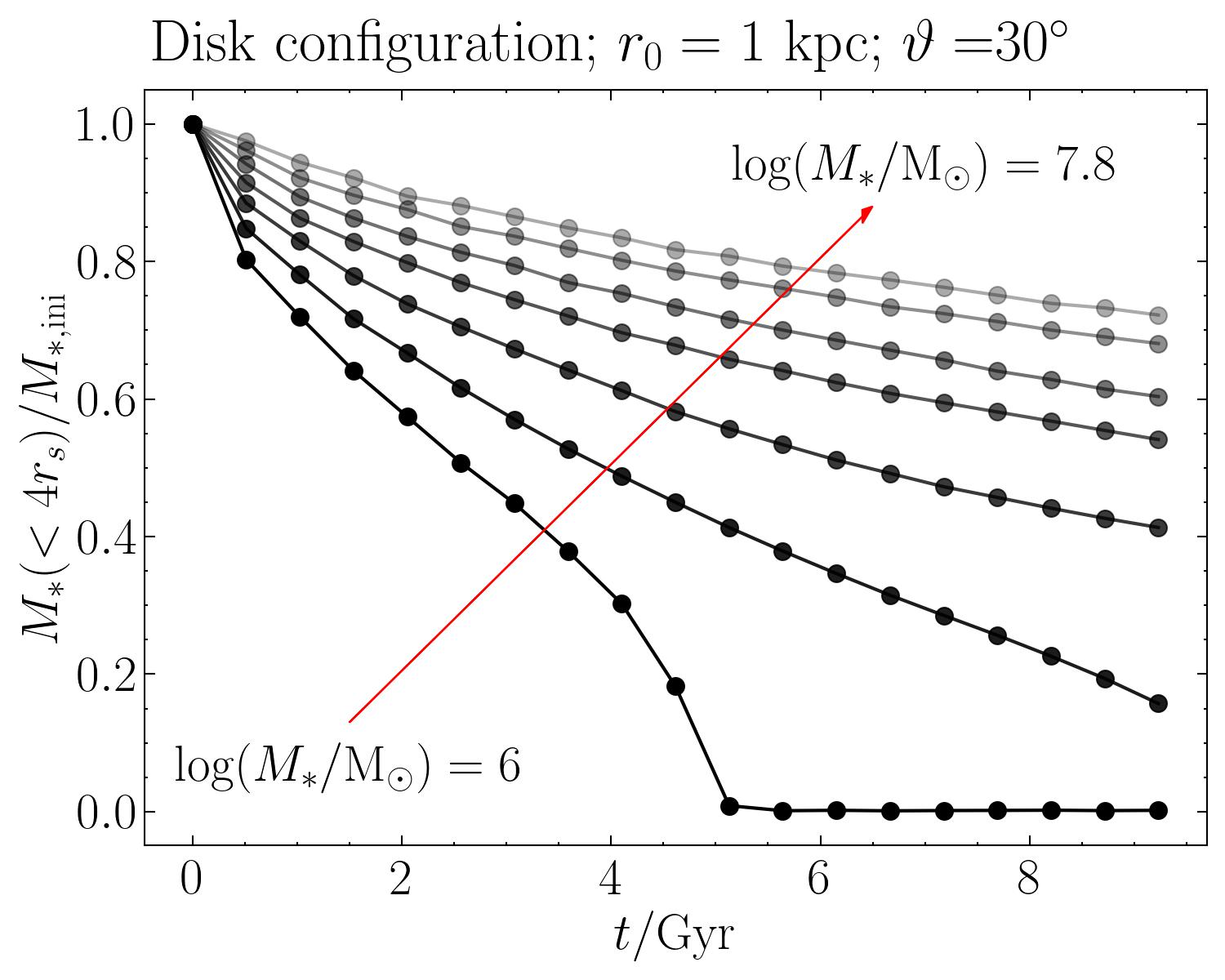}
 \caption{Top panel: retained mass (relative to the initial one) as a function of time, for clumps of different masses in circular orbits at 1 kpc from the centre of the system. The residual mass is computed within $4r_\mathrm{s}$, where $r_\mathrm{s}$ is the initial Plummer scale radius. In this case, the clumps orbit in a Sparkler-like spherical potential representing the total mass, defined in Sect. \ref{sec:spark_dyn_mass}. Masses go from $\log (M_*/\msun)=6$ to $\log (M_*/\msun)=7.8$ with a step of $0.3$ dex (the arrow points from the smallest to the highest mass).
 Bottom panel: same as the left panel, but for clumps in quasi-circular orbit at 1 kpc from the centre of the system including a stellar disk component (modelled as in Sect. \ref{sec:tidal_shock}). The clump orbital plane is inclined of 30 degrees with respect to the plane of the disk.
 }
 \label{fig:tidal_losses}
\end{figure}

In the top panel of Fig. \ref{fig:tidal_losses}, we show the mass retained by the clump relative to the initial one, computed within $4$ initial scale radii, as a function of time. The fraction of retained final mass is lower for low-mass clumps than high-mass clumps. However, even for $\log(M_*/\msun)=6$, less than $\approx15\%$ of the initial mass is lost.
These results are consistent with the predictions shown in Fig. \ref{fig:rtidal} and confirm that, under the assumption of circular orbit in the considered spherical system, the mass loss due to tidal stripping {is negligible}.

\rr{In Appendix \ref{app:tidal_elliptical}, we also explore the case of clumps on elliptical orbits, starting at larger distances ($3-4$ kpc) and reaching the final distance of $\approx1$ kpc under the effects of the dynamical friction deceleration, modelled with the Chandrasekhar formula \citep{Chandrasekhar1943,Binney2008}. However, we obtain that the mass lost increases only for clumps in pronounced radial orbits, even though the lost mass increases only by a factor $2$, still insufficient to remove enough mass from the Sparkler clumps in order to reach GC-like masses.}

\subsection{Tidal stripping in presence of a stellar disk}\label{sec:tidal_shock}
In Sect. \ref{sec:resolved_clumps_setup}, we have analysed the tidal effects modelling the Sparkler as a spherical system. Here, we perform a series of simulations of resolved clumps exploring the case in which the stellar component of the host is a disk. This is consistent with the fact that Sparkler is a star-forming galaxy \citep{Mowla2022}. A spherical DM halo is included as an external Hernquist potential similarly to what has been done in Sect. \ref{sec:resolved_clumps_setup}.

The gravitational field gradient experienced by the clumps as they cross the disk results in the so-called tidal shocks, which can cause the clumps to lose significantly more mass than in the case of spherical symmetry or when the clumps are co-planar with the disk \citep{Gnedin1997,Kruijssen2009,Lamers2006,Elmegreen2010a,Elmegreen2010b,Webb2014b,Kruijssen2015,Miholics2017,Pfeffer2018,Li2019,ReinaCampos2022}.
Aiming at maximizing the tidal effects of the disk, for the whole simulation we fix the disk properties at those that it would possess at $z=0$, after accreting mass for almost $10$ Gyr. 
The disk mass ($8.79\times 10^9\,\msun$) is determined as described in Sect. \ref{sec:halo_properties}, using the halo-to-stellar mass relation at $z=0$ from the \textsc{universemachine} \citep{Behroozi2019}. The effective radius is $R_\mathrm{eff}=4.65$ kpc, determined from the galaxy mass-size relations by \cite{vanderWel2014}, and the disk scale height is $h_z=0.25$ kpc, in accordance with recent studies of disk galaxies of similar mass and redshift \citep{Bershady2010,Bacchini2019,Bacchini2024,Ranaivoharimina2024}. The disk mass density profile is modelled with a double exponential profile \citep{Freeman1970,Binney2008}

\begin{equation}\label{eq:expdisk}
    \rho_\mathrm{d}(R,z)=\rho_0\,e^{-R/R_\mathrm{d}}\,e^{-|z|/h_z},
\end{equation}

\noindent where $\rho_0$ is the central density and $R_\mathrm{d}$ is the scale radius. The scale radius $R_\mathrm{d}=2.77$ kpc is computed from the effective radius $R_\mathrm{eff}$ following Eq. 13 in \cite{LimaNeto1999}, for which we fix the Sérsic index to 1.
We modelled the exponential disk as a sum of three Miyamoto-Nagai (MN) disks \citep{Miyamoto1975} of suitable parameters, determined following the prescriptions and correlations by \cite{Smith2015}. This choice allows us to compute the gravitational acceleration of the disc analytically, as the sum of those of the three MN disks. Each MN disk is defined by three parameters (following Eq. 6 in \citealt{Smith2015}), which are listed in Table \ref{tab:3mn}: the total mass $M_\mathrm{d}$, the scale radius $a$ and the scale height $b$. We point out that the disk MN2 (third row in Table \ref{tab:3mn}) has a negative mass, but the sum of the three MN disks results in a density profile which is always positive.

\begin{table}[t!]
\fontsize{10pt}{10pt}\selectfont
\setlength{\tabcolsep}{4pt}
\renewcommand{\arraystretch}{1.4} % Default value: 1
\centering
\caption{Parameters of the Sparkler-like exponential disk (first row) and of the three Miyamoto-Nagai (MN) disks (second, third and fourth rows) that, summed together, better approximate the potential of the original exponential disk (see \citealt{Smith2015}).}
\begin{tabular}{l|ccc}
\hline\hline
\multirow{2}{*}{Disk} & $M_\mathrm{d}$ & scale radius & scale height\\
& $[10^9\,\msun]$ & $\mathrm{[kpc]}$ & $\mathrm{[kpc]}$\\
\hline
exp & $8.79$ & $2.77$ & $0.25$\\
\hline
MN1 & $1.37$ & $1.57$ & $0.3$\\
MN2 & $-48.82$ & $7.16$ & $0.3$\\
MN2 & $56.87$ & $6.23$ & $0.3$\\
\toprule
\end{tabular}
\tablefoot{The first column is the ID of the disk (exp for the exponential disk; MN1, MN2 and MN3 for the three MN disks), the second column is the total mass of the disk $M_\mathrm{d}$, the third column is the disk scale radius ($R_\mathrm{d}$ for the exponential disk and $a$ for the MN disks) and the fourth column is the disk scale height ($h_z$ for the exponential disk and $b$ for the MN disks).}
\label{tab:3mn}
\end{table}

Similarly to what has been done in the previous section, the mass of the clumps varies from $\log (M_*/\msun)=6$ to $\log (M_*/\msun)=7.8$ (with steps of $0.3$ dex) and the size is determined by the Sparkler size-mass relation (Eq. \ref{eq:size_mass}). The initial conditions of the centre of mass of the clumps are always set in the same way.
The initial position is $(x_0,y_0,z_0)=(r_0\cos \vartheta,0,r_0\sin \vartheta)$, with $\vartheta=\pi/6$. Therefore the clump is initialized at a distance of $r_0=1$ kpc, with the initial position vector inclined by 30 degrees with respect to the plane of the disk (coinciding with the $x-y$ plane).
The initial velocity is $(v_{x,0},v_{y,0},v_{z,0})=(0,v_{\mathrm{c},0},0)$, where $v_{\mathrm{c},0}$ is the speed of an object in circular orbit at $R=r_0$ and $z=0$ co-planar to the disk. Therefore, the initial speed is given by the relation $v_{c,0}^2=G\,M_\mathrm{halo}(r_0)/r_0 + r_0[d\Phi_\mathrm{disk}/dR](r_0,0)$, where $\Phi_\mathrm{disk}(R,z)$ is the disk potential.

\begin{figure}[t!]
\centering
 \includegraphics[width=0.49\textwidth]{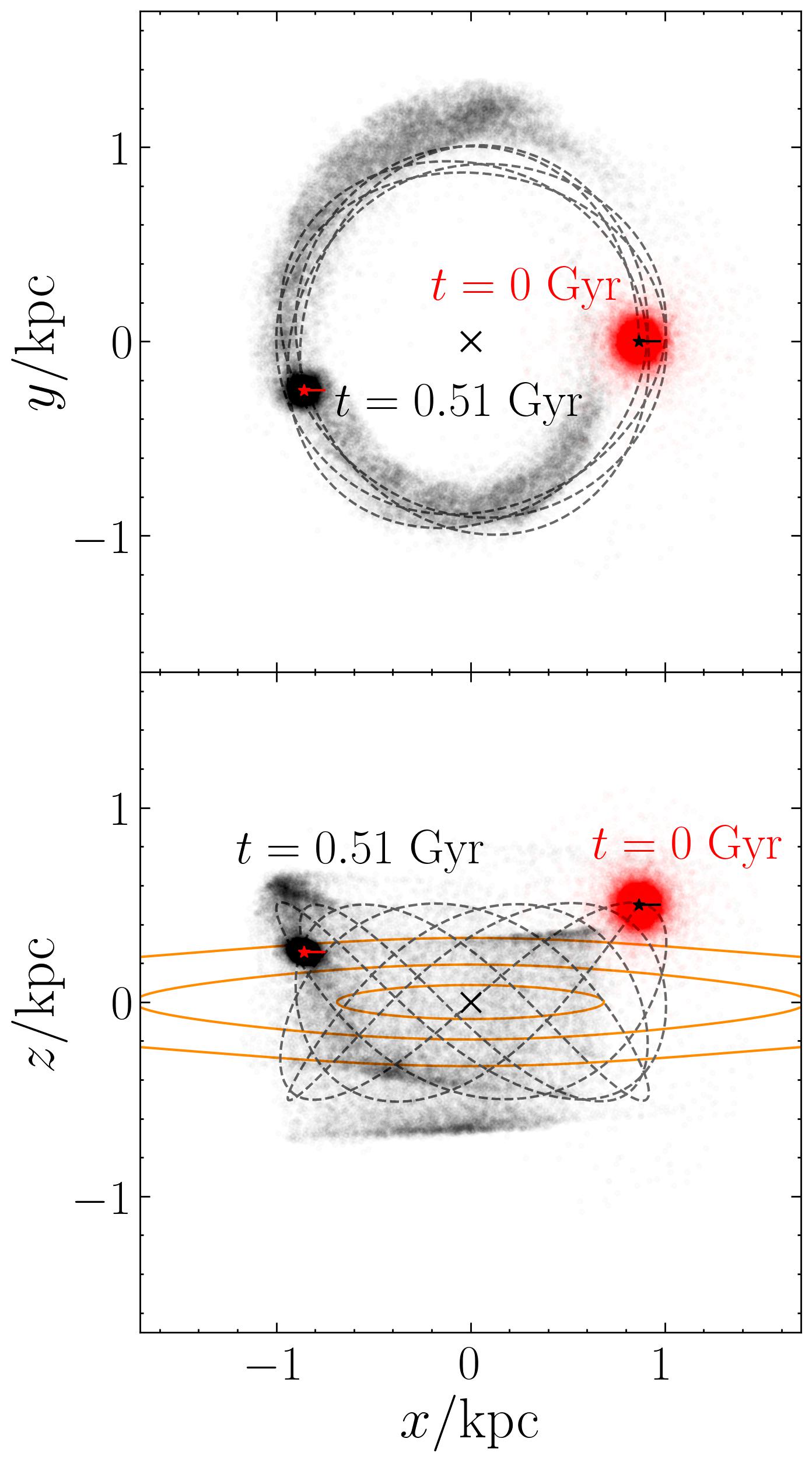}
 \caption{Comparison between the initial conditions and the snapshot at $0.51$ Gyr for the tidal-shock simulation described in Sect. \ref{sec:tidal_shock} in the case of mass $M_*=10^6\,\msun$. The top and bottom panels show the projections on the $x-y$ and $x-z$ planes, respectively. The clump stellar particles are shown as red and black dots for the snapshots at $t=0$ Gyr and $t=0.51$ Gyr, respectively. The centres of clump at $t=0$ Gyr and $t=0.51$ Gyr are shown as black and red star, respectively, with an horizontal line of length $4\,r_{s,i}$, where $r_{s,i}$ is the initial Plummer scale radius. The grey dashed curves represent the trajectory of the clump. The orange curves trace iso-density contours of the Sparkler disk external potential. The centre of the system is marker by a black cross.}
 \label{fig:orbit}
\end{figure}

We tested many values of $\vartheta$ finding that, for $\vartheta>\pi/6$, the orbits are such that the distance of the clump from the centre of the system displays significant variations, experiencing a number of close peri-centric passages, yet showing similar effects in terms of mass loss. Therefore we opted for $\vartheta=\pi/6$, for which the clump keeps an almost-constant distance from the centre of the system, without losing the effects of tidal shocks.
The initial distance is fixed at $r_0=1$ kpc in order to maximize the tidal effects within the interval of observed clump distances.
In Fig. \ref{fig:orbit} we show the first $0.51$ Gyr of the orbit, on the $x-y$ and $x-z$ planes, for the case of a clump with $M_*=10^6\,\msun$. We note that the orbit is limited to a finite volume, represented by a thick ring in the $x-y$ plane and by a box in the $x-z$ plane, even though the distance of the clump from the centre of the system remains close to 1 kpc, with variations as large as $10\%$. Furthermore, the snapshot at $0.51$ Gyr already show the effects of the tidal shocks, with long tails of stellar particles departing from the main body of the clump. Finally, we note that, since in these simulations the Sparkler potential is included as an external potential, the clump is uneffected by dynamical friction, so the clump does not inspiral.

The resulting clump loss is shown in the bottom panel of Fig. \ref{fig:tidal_losses}, which demonstrates how the tidal shock strongly changes the results with respect to the spherical configuration shown in the top panel. A clump of $\log (M_*/\msun)=7.8$ retains about $70-75\%$ of the initial mass (with respect to $>90\%$ in the absence of a disk), while a clump with mass $\log (M_*/\msun)=6$ gets completely disrupted (without a disk it retains about $90\%$ of its initial mass). Again, we point out that these results are relative to the most extreme cases of tidal shocks, since we considered the highest possible mass for a Sparkler-like disk and the clumps are very close to the central regions. Therefore, not all the clumps as small as $10^6\,\msun$ will be disrupted. However, at the same time, even in presence of tidal shocks the massive clumps tend to remain particularly massive if compared with the typical mass of present-day GCs.

\section{Discussion}\label{sec:discussion}
The results presented in Sects. \ref{sec:dyn_fric} and \ref{sec:resolved_clumps} suggest that the fraction of surviving clumps is $\approx 40-60\%$.
Previous $N$-body simulations of MW-like \citep{Pfeffer2018,ReinaCampos2019,Li2019} and dwarf galaxies \citep{MorenoHilario2024} find smaller surviving fraction ($10-30\%$). This difference, yet not particularly pronounced, may arise from many different factors, such as the different mass of the simulated host system. The simulations presented in Sect. \ref{sec:dyn_fric} do not include tidal stripping effects, which, on the one hand, may increase the fraction of lost clumps as a consequence of their disruption but, on the other hand, may also halt the sinking caused by dynamical friction. Furthermore, in this work we have focused on the clumps located far away from the main body of the galaxy, neglecting those observed within it, therefore closer to the centre of the system and more susceptible to dynamical friction. The fraction of survivors among these clumps would most likely be smaller, decreasing the average fraction of total survivors. Finally, the sample of clumps observed in the Sparkler galaxy is most likely biased towards the high-mass end of the clump mass distribution, even though the sample completeness in lensed images is very difficult to study.

The mass distribution of the survivors changes from the initial one only for the loss of the clumps accreted onto the centre of the system. Tidal stripping from a spherical host system, presented in Sect. \ref{sec:resolved_clumps_setup}, has a minor effect on shaping the final mass distribution.
Therefore, in the scenario in which the Sparkler galaxy is better described by a spherical geometry, our study predicts the presence of very massive surviving clumps, more massive than $10^7\,\msun$ and inconsistent with the observed properties of present-day GCs in a Sparkler-like progenitor (see \citealt{Baumgardt2018} for the MW; \citealt{Baumgardt2013} for the Large Magellanic Cloud; \citealt{Hunter2003} for the Small Magellanic Cloud).
Such outliers are very uncommon in the local Universe. Some effects not taken into account in our simulations can be additional channels to remove mass from these clumps or completely disrupt them. For example, the presence of DM in the clumps would increase their mass and enhance the dynamical friction efficiency, bringing them to the centre on shorter timescales. Alternatively, if these clumps have been accreted by a previous merger or are ejected from the disk by a subsequent one, the orbits of the clumps could tend to be eccentric.
The small peri-centric distances of such orbits can favour both dynamical friction and tidal stripping effects, and therefore their fall to the centre of the system or the loss of the stellar mass in excess \citep{Gnedin1997,Gnedin1999,Baumgardt2003,Webb2013,Webb2014a,Brockamp2014}.
Finally, a number of works \citep{Gieles2006,Lamers2006,Webb2019,Webb2024,ReinaCampos2022} have shown that the encounters with giant molecular clouds (GMCs) cause a series of tidal shocks able to significantly strip mass from the clumps. However, we note that these works conclude that a high frequency of encounters with GMCs is needed in order to make the tidal shocks effective. Therefore this process is effective if the clumps have formed and orbit inside the stellar disk; on the other hand,
if the clumps are located outside the stellar disk and cross it only twice per orbit (as modelled in Sect. \ref{sec:tidal_shock}), then the interactions with the GMCs are not frequent enough and therefore negligible with respect to the tidal shocks exerted by the stellar disk.

\subsection{Clump size evolution}\label{sec:clumps_size_evolution}
The half-mass radii of GCs in the local Universe are typically of a few parsecs \citep{Jordan2005,Spitler2006,Baumgardt2023}. Therefore, the size of the Sparkler clumps are expected to evolve to consistent values, down to $z=0$, in order to be promising GC progenitors.

\begin{figure}[t!]
\centering
 \includegraphics[width=0.48\textwidth]{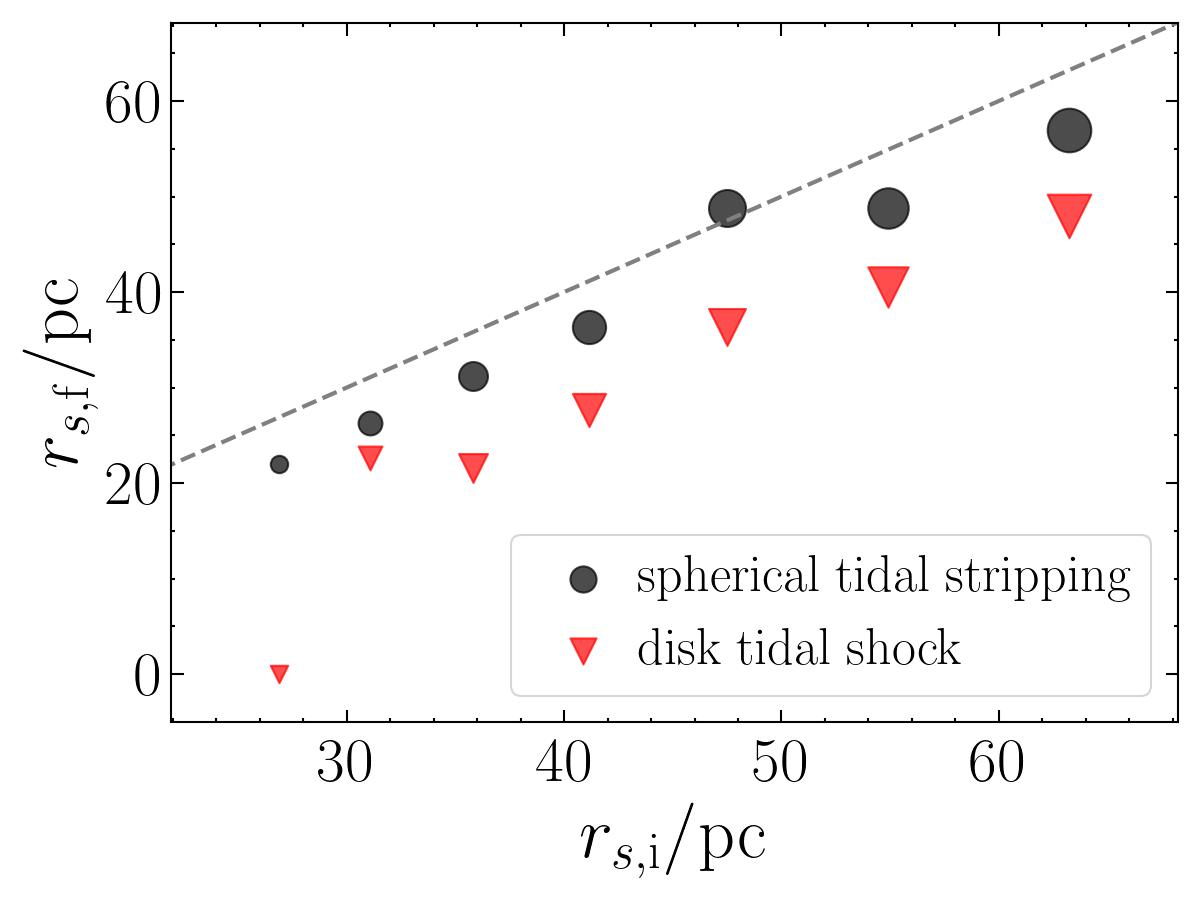}
 \caption{Comparison between the final ($r_{s,\mathrm{f}}$) and initial ($r_{s,\mathrm{i}}$) scale radii of the clumps simulated in Sect. \ref{sec:resolved_clumps}, when modelled as Plummer sphere, as in Eq. \ref{eq:plummer}). The initial scale radii are computed from the size-mass relation of Eq. \ref{sec:size_mass}. The final scale radii are obtained fitting a Plummer density profile to the final snapshots of the simulations of resolved clumps in a spherically symmetric Sparkler-like system (Sect. \ref{sec:resolved_clumps_setup}, circles) and of the simulations including the effects of a disk (Sect. \ref{sec:tidal_shock}, triangles). The dashed line is the $1:1$ relation.
 }
 \label{fig:clumps_size_evolution}
\end{figure}

By removing the outermost stars of the clumps, tidal stripping reduces the half-mass radius. To quantify this effect, we took the final snapshot of each simulation presented in Sects. \ref{sec:resolved_clumps_setup} and \ref{sec:tidal_shock} and fitted the clump to a Plummer sphere (Eq. \ref{eq:plummer}).
The fitting procedure is described in detail in Appendix \ref{app:fit_plummer} and gives us the final total mass $M_\mathrm{*,f}$ and scale radius $r_{s,\mathrm{f}}$.
In Fig. \ref{fig:clumps_size_evolution}, we compare the initial scale radii $r_{s,\mathrm{i}}$ with the final ones, both for the tidal stripping in a spherical system, and for tidal shocks of a Sparkler-like disk.
Such sizes, even considering the effects of tidal stripping and shocks, remain systematically larger than 20 pc down to $z=0$, with the only exception of the clump with mass $10^6\,\msun$ in the tidal shock simulation, which gets completely disrupted. Therefore, the final sizes of these clumps are incompatible with those observed in local GCs.

If one assumed that the observed sizes of these objects are overestimated because of the limited angular resolution of JWST, thus not affecting the mass estimate, and that these clumps possess actual GC-like sizes, then the clumps would become even denser, making the tidal stripping negligible. This would imply that the over-massive surviving outliers discussed in the previous section would become even more difficult to strip of their excess mass.
\rr{Alternatively, if the magnification factors $\mu$ of the clumps were underestimated, then their intrinsic properties would be biased towards large values. The intrinsic mass of a lensed object $m_\mathrm{i}$ is related to the observed and lensed one $m_\mathrm{o}$ by $m_\mathrm{i}=m_\mathrm{o}/\mu$, while the intrinsic length $l_\mathrm{i}$ is related to the observed and lensed one by $l_\mathrm{i}=l_\mathrm{o}/\mu^{1/2}$. If, for instance, the true magnification factor were underestimated by a factor $\approx 10$, then the clumps would have masses consistent with those of massive local GCs and sizes larger by a factor up to $5$. A magnification factor $\mu\approx100$ would be consistent with the one of the critical line derived for the Sparkler galaxy by \cite{Mahler2023}.}

\subsection{Combining dynamical friction and tidal shocks}\label{sec:dyn_fric_tid_shock}
In this section, we combine our dynamical friction and tidal shock results, in order to predict the expected mass distribution of the survived Sparkler clumps at $z=0$.
For this experiment, we consider the spherical configuration described in Sect. \ref{sec:clumps_dist_distribution}. In this scenario, the clump orbits are not co-planar to the stellar disk of the Sparkler system, therefore they undergo a series of tidal shocks in correspondence to the crossings of the galactic disk, as modelled in Sect. \ref{sec:tidal_shock}.

\begin{figure*}[t!]
\centering
 \includegraphics[width=0.49\textwidth]{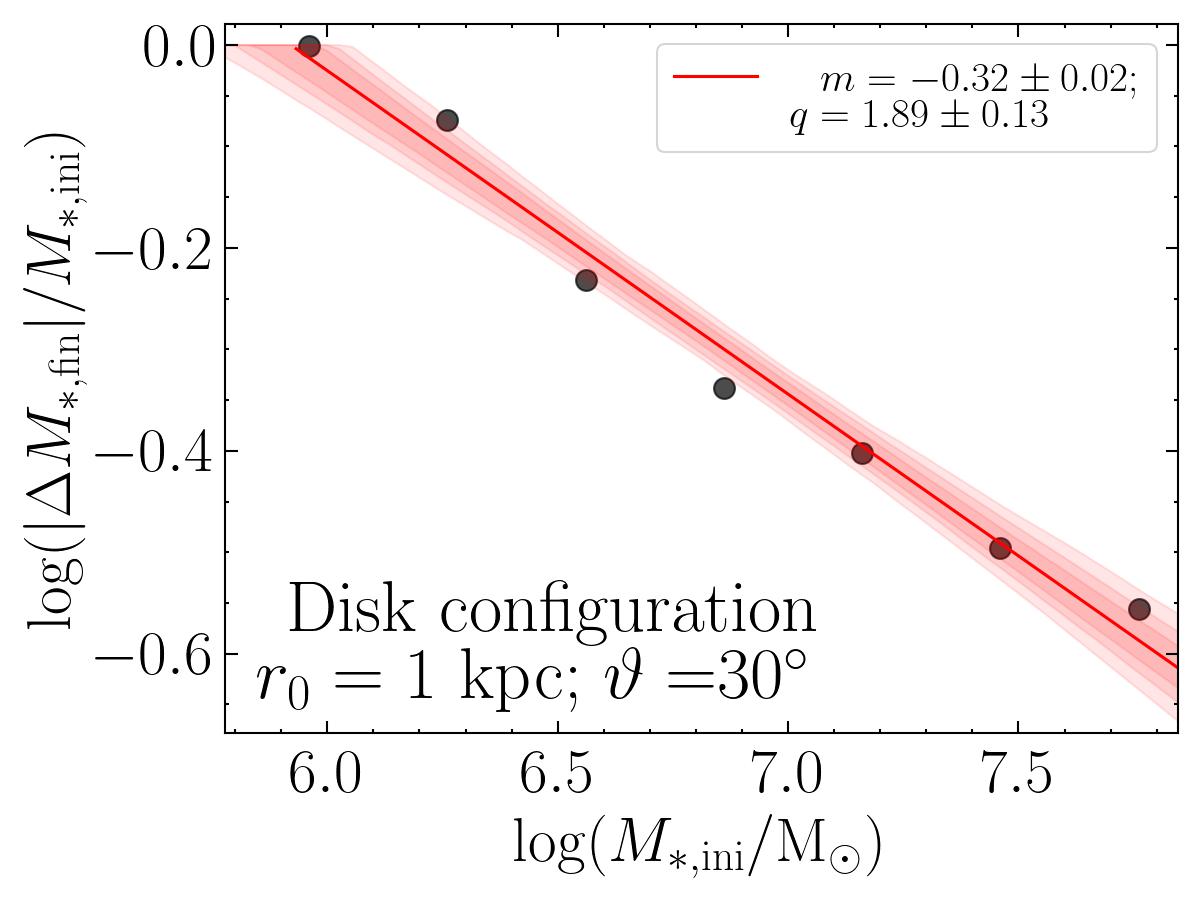}
 \includegraphics[width=0.49\textwidth]{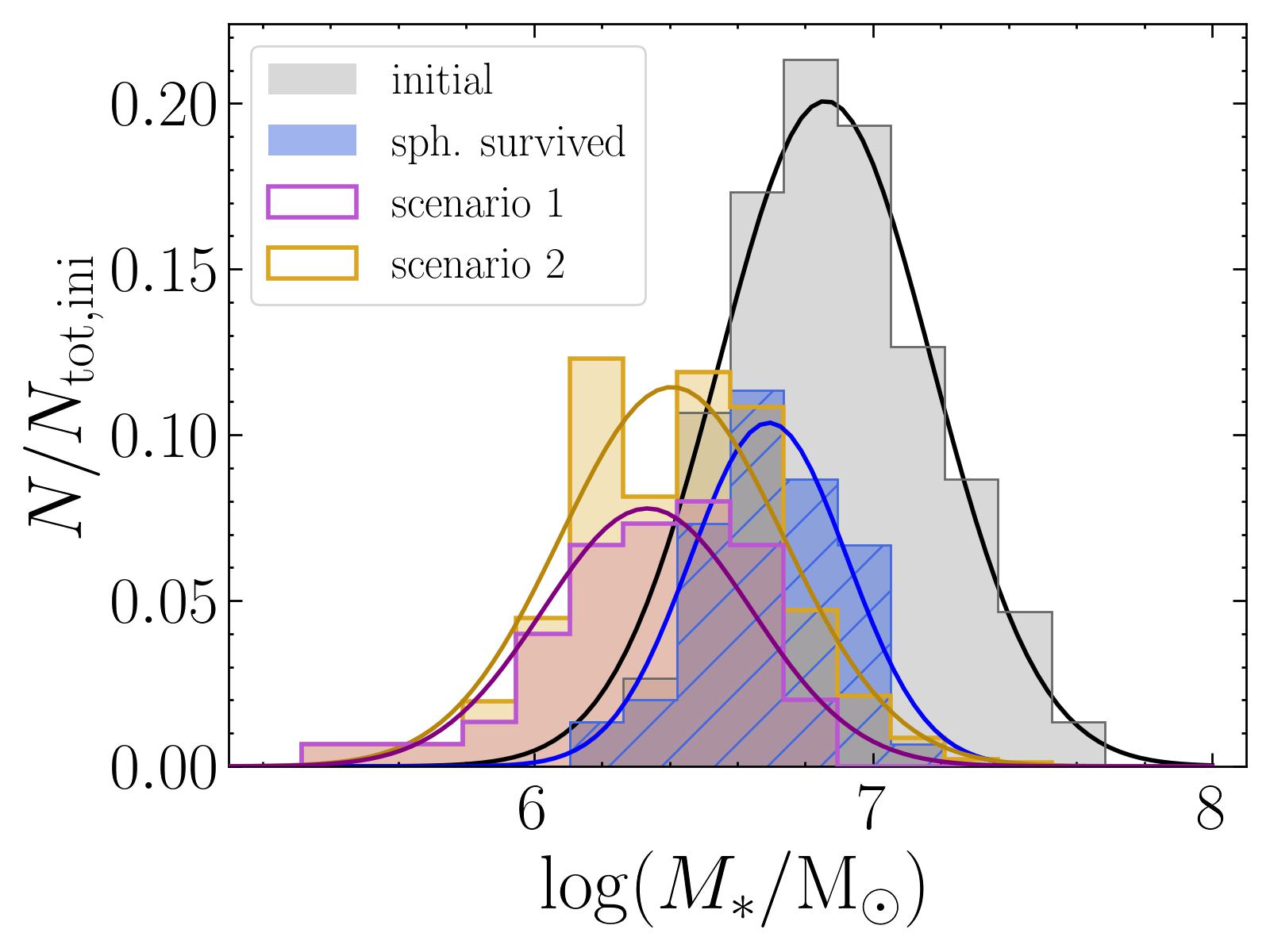}
 \caption{Left panel: fraction of the clump mass lost after $9.23$ Gyr $\log(|\Delta M_{*,\mathrm{fin}}|/M_{*,\mathrm{ini}})$ as a function of the initial mass $\log(M_{*,\mathrm{ini}})$. Here $\Delta M_{*,\mathrm{fin}}=M_\mathrm{*,fin}(4r_\mathrm{s})-M_\mathrm{*,ini}$, where $M_\mathrm{*,fin}(4r_\mathrm{s})$ is the final mass within four initial Plummer scale radii, computed as in Sect. \ref{sec:tidal_shock}. The circles are the results from the simulations of resolved clumps described in Sect. \ref{sec:tidal_shock}, while the line is the best fit (Eq. \ref{eq:mass_lost}) and the shaded area is the $1\sigma$ interval.
 Right panel: distributions of the number of clumps per mass bin, normalized to the total initial number of clumps. The grey and blue histograms are the analogues of those shown in the bottom panel of Fig. \ref{fig:mass_distr_conf}, and they show the initial and surviving number of clumps, respectively, in the spherical configuration case considering the effects of dynamical friction only. The purple histogram shows the mass distribution of the surviving clumps of scenario 1 (short dynamical-friction timescale; Sect. \ref{sec:tidal_shock}), while the golden histogram shows the mass distribution of the surviving clumps of scenario 2 (short tidal-stripping timescale; Sect. \ref{sec:tidal_shock}).
 A Gaussian function is fitted to each histogram and shown with the corresponding colour.
 }
 \label{fig:tidal_shock_fit}
\end{figure*}

The mass losses triggered by the tidal shocks can affect the orbital evolution of the clump (see the bottom panel of Fig. \ref{fig:mass_distr_conf}) since a lower mass results in a less efficient dynamical friction.
In order to quantify these tidal effects, we modelled the relation between the final fraction of mass lost $|\Delta M_{*,\mathrm{fin}}|/M_{*,\mathrm{in}}$ and the initial mass $M_{*,\mathrm{in}}$ as

\begin{equation}\label{eq:mass_lost}
    \log(|\Delta M_{*,\mathrm{fin}}|/M_{*,\mathrm{ini}})=\mathrm{min}[m\,\log (M_{*,\mathrm{ini}}/\msun) + q,0],
\end{equation}

\noindent which we report in the left panel of Fig. \ref{fig:tidal_shock_fit}. The mass losses are quantified as described in Sect. \ref{sec:tidal_shock}, as the difference between the initial and final mass within four initial Plummer scale radii.
According to this model, there is a minimum mass $M_{*,\mathrm{min}}$ below which the clumps are always destroyed. Since our results show that a clump of mass $10^6\,\msun$ is completely disrupted, $M_{*,\mathrm{min}}$ must be close to such value. The best-fitting parameters obtained by means of the non-linear least squares method, implemented in the \textsc{numpy}\footnote{\url{https://numpy.org/doc/stable/index.html}} function \textsc{curve\_fit}, are $m=-0.35\pm 0.02$ and $q=2.06\pm 0.15$, which roughly correspond to $M_{*,\mathrm{min}}=8\times 10^5\,\msun$.

In order to understand how the final clump mass distribution changes, we consider two different scenarios: 1) the dynamical-friction timescale is shorter than the tidal-stripping one and 2) the tidal-stripping timescales is shorter than the dynamical-friction one. The resulting fractions of survived clumps per mass bin are shown in the right panel of Fig. \ref{fig:tidal_shock_fit} as the purple and golden histograms, respectively, together with the initial distribution in grey and the results from Sect. \ref{sec:res_spherical} in blue.
The mass distributions are computed as follows:

\begin{itemize}
    \item \textit{Scenario 1}: starting from the mass distribution of survived clumps discussed in Sect. \ref{sec:res_spherical}, shown in the right panel of Fig. \ref{fig:tidal_shock_fit} with the blue histogram, the final masses are shifted at lower masses because of tidal stripping, according to the model of Eq. \ref{eq:mass_lost} (the purple histogram in the right panel in Fig. \ref{fig:tidal_shock_fit}). Low-mass clumps are shifted below $10^6\,\msun$, in a regime which is absent in the initial distribution, while the massive survived clumps do not change significantly their mass;
    \item \textit{Scenario 2}: as a first step, we shift the initial mass distribution according to the model of Eq. \ref{eq:mass_lost}. Then, we compute the fraction of survived clumps per mass bin as the ratio between the number of survived clumps in the spherical dynamical friction-only scenario and the number of total clumps (blue and grey histogram in the right plot of Fig. \ref{fig:tidal_shock_fit}, respectively).
    The final result is shown as the golden histogram: similarly to scenario 1, a good fraction of clumps is shifted to masses comparable with those of massive local GCs, yet retaining $\approx16\%$ of clumps with masses above the mass of $\omega$ Centauri ($5\times 10^6\,\msun$). Furthermore, in this scenario the fraction of survived clumps increases from $0.42$ of the spherical dynamical friction-only case to $0.60$.
\end{itemize}

If we compare scenario 1 and scenario 2, we note that the low-mass end is very similar, while the high-mass end in scenario 2 reaches higher clump masses than in scenario 1. This is a consequence of the fact that the tidal shocks reduces the mass of very massive clumps, affecting the efficiency of the dynamical friction.
Nonetheless, the two final mass distributions are very similar: this is a remarkable result since it implies that in the Sparkler system the relative efficiency of dynamical friction and tidal shock mainly affects the number of survivors, but not their mass distribution. The mean masses for scenario 1 and scenario 2 are indeed $\log (\widetilde{M}_*/\mathrm{\msun})=6.33_{-0.04}^{+0.05}$ and $\log(\widetilde{M}_*/\mathrm{\msun})=6.41\pm 0.1$, respectively. Both values are smaller than the one in absence of tidal effects, but consistent within $1\sigma$.
The fraction of survived over-massive clusters ($M_*>5\times 10^6\,\msun$, the mass of $\omega$ Centauri) is between $0.07$ and $0.16$ (scenario 1 and 2, respectively).

However, we note once again that our assumptions aimed at maximizing the tidal effects and therefore the shift applied to the distribution of survival clumps in the spherical configuration. This would imply an over-estimate of the fraction of low-mass clumps and an under-estimated of the fraction of high-mass clumps.

\subsection{The fate of the accreted clumps}\label{sec:fate_clumps}
The accretion of massive clumps, which are more frequent at high redshift and more affected by dynamical friction, has been proposed as an efficient channel for the formation and mass growth of galactic bulges \citep{Scannapieco2003,Hopkins2010}. Such scenario has been supported in the past couple of decades by the observation of massive GCs, namely Terzan 5 and Liller 1, within 1 kpc from the centre of the MW. These GCs, also called bulge fossil fragments (BFFs), have present stellar masses of a few $10^6\,\msun$ \citep{Lanzoni2010,Saracino2015,Baumgardt2023} and are supposed to be among the few survived stellar clumps accreted by the Galactic bulge and not completely disrupted. Recent studies based on their metallicity and star formation history have concluded that the initial stellar mass of the progenitors of these BFFs ranges from $4\times 10^7\,\msun$ \citep{Romano2023,Crociati2024} up to $10^9\,\msun$ \citep{Ferraro2021}.
The masses of the clumps accreted by the centre of the Sparkler galaxy (Sect. \ref{sec:dyn_fric}) are consistent with the values found by \cite{Romano2023} for the progenitors of the BFFs, favouring a scenario in which a fraction of the clumps observed around the Sparkler galaxy are BFF progenitors.
We point out that the half-mass radii of Terzan 5 and Liller 1 are $4.03$ and $2.42$ pc, respectively, a factor $\sim 10$ smaller than the sizes of the Sparkler resolved clumps, as discussed in Sect. \ref{sec:clumps_size_evolution}. However, we remind the presence of three unresolved clumps among those of our sample. In particular, C4 and C9 have masses larger than $10^7\,\msun$ and upper limits on their effective radii of $11-12$ pc (Table \ref{tab:clumps}), making them promising BFF progenitors also in terms of size.

Another possible evolutionary path of the massive clumps of Sparkler is the one leading to the formation of a nuclear star cluster (NSC; see e.g.\ \citealt{Neumayer2020}). The NSCs are located by definition at the dynamical
centre of the host galaxy and are usually detected from the light excess above the extrapolation of the host galaxy's surface brightness profile into its innermost regions ($\lesssim 10^2$ pc).
NSCs are found in both early- and late-type galaxies and are usually the brightest clusters of their hosts, possibly as a consequence of the merging of the clumps and GCs fallen into the centre of the system, dragged by dynamical friction \citep{Tremaine1975,Capuzzo1993,Oh2000,Agarwal2011,Neumayer2011}.
Their effective radii are usually of a few parsecs, but a non-negligible fraction of NSCs with sizes larger than 10 pc exists \citep{Georgiev2016} and would be compatible with those of the Sparkler clumps. Their masses are typically higher than those of GCs and consistent with those of the Sparkler clumps, peaking at $\sim10^{6-7}\,\msun$ and reaching also values as large as $\sim10^{8-9}\,\msun$ \citep{Georgiev2016}.

Finally, we note that the simulations presented in this work are not capable to capture the complex processes that a clump undergoes in the centre of the Sparkler galaxy, therefore at the moment we are unable to quantify the fraction of accreted clumps that will completely disrupt or will survive as BFFs or, alternatively, if the clumps will sink and merge to the very centre of Sparkler forming a NSC.

\section{Conclusions}\label{sec:summary}
In this work we have studied the dynamical evolution of the stellar clump system observed in the Sparkler gravitationally lensed galaxy at $z\approx1.4$. We have focused our study on the clumps observed far away from the galactic stellar component, among which recent JWST photometric observation have hinted the presence of candidate progenitors of the GCs observed in the local Universe \citep{Mowla2022,Claeyssens2023,Adamo2023}. These objects, consistent with GCs in terms of formation redshift and metallicity, possess very large masses ($\log (M_*/\msun)=6.3-7.4$) and sizes (half-mass radii $R_\mathrm{eff}=30-52$ pc) and are located at projected distances from the centre of the galaxy of $2-5$ kpc. Their dynamical evolution from $z\approx1.4$ to $z=0$ (about $9.23$ Gyr) can affect their position in the galactic system, but also their final mass and size.

We first focused on their evolution by taking into account the effect of the dynamical friction exerted by the Sparkler system, modelled as a spherical system.
Since little is known about the morphology of the galaxy and of the clump system, we studied three possible configurations: face-on disk, edge-on disk and spherical. The disk configurations (Sects. \ref{sec:res_faceon} and \ref{sec:res_edgeon}) are interesting in the context of GC formation. According to previous works, GCs may form in the galactic disks in an epoch of bursty star formation (favoured in the star-forming sites of disk galaxies) to then be ejected far away from the galactic plane by subsequent major mergers \citep{Hopkins2009,Kruijssen2012,Kruijssen2015}.
In the first two configurations, the clumps are initialized on circular orbits co-planar to each other, while the latter one assumes that the clump system is spherically symmetric and isotropic.
For each scenario, we performed multiple simulations of Sparkler-like clump systems, computing the number of survived clumps (which ended the simulation at a distance larger than 1 kpc from the centre of the system) and the mass and distance distributions. The main results of these simulations can be summarized as follows:

\begin{itemize}
    \item on average, about 4 clumps per simulation survive, depending on the scenario, while the others are accreted onto the centre of the system;
    \item the distance distribution remains essentially unchanged, with the only exception of the face-on disk case in which the final distance distribution is more extended than the initial one;
    \item the final mass distribution slightly shifts to smaller masses, but the peak value $\log (\widetilde{M}_*/\msun)$ remains larger than $\sim 6.7$. Therefore, a significant fraction of survived clumps are more massive than a few $10^6\,\msun$, which is unexpected for a galaxy such as Sparkler, which is thought to reach a stellar mass of $10^{9-10}\,\msun$ at $z=0$.
\end{itemize}

In addition to the aforementioned simulations, we also studied the clump mass loss under the effects of the tidal field of the host galaxy, by means of a different simulation setup in which the Sparkler system is implemented as an external gravitational potential and the clumps as resolved Plummer spheres made of $10^5$ particles.
We tested both the case in which the Sparkler is a spherical system and in which the Sparkler stellar component is an exponential disk. In presence of a disk, the clumps are initialized at 1 kpc outside the plane of the disks, experiencing strong tidal shocks every time that they cross such a plane as a consequence of the large gravitational field gradient.
The main outcome of these simulations are as follows:

\begin{itemize}
    \item in the spherical case, the fraction of mass lost is smaller than $\approx15\%$, at any clump stellar mass;
    \item in presence of the stellar disk, tidal shocks can completely disrupt low-mass clumps, while they have minor effects for very massive clumps;
    \item even when a large fraction of mass is removed, the sizes of the clumps remain between 20 and 50 pc, about a factor of 10 larger than the sizes typically observed in local GCs.
\end{itemize}

When combining the aforementioned results of dynamical friction and tidal shock from a disk, we bracketed a realistic behaviour. If we consider first the mass lost by tidal shocks, the diminished mass of the clumps partially halts the effects of dynamical friction. Consequently, the fraction of survivors increases to $60\%$ of the initial sample. On the other hand, if we consider first the evolution of the mass distribution under the effects of dynamical friction, the mass loss caused by the tidal shocks shift the mass distribution of the survived clumps at smaller values. In both cases, the peak mass is $\log(\widetilde{M}_*/\msun)\approx 6.4$. Such values are marginally consistent with the high-mass end of the mass distribution of MW GCs \citep{Baumgardt2018}, even though we note the strong difference between the sizes of the Sparkler clumps and those of local GCs.

\rr{Apart from the possible underestimate of the lensing magnification factor of the clumps}, further effects not considered in this work, such as preferentially radial orbits and encounters with giant molecular clouds, may increase the fraction of stripped mass, explaining these outliers. Radial orbits make the clump explore the innermost and densest regions of the system, experiencing a larger tidal field \citep{Gnedin1997,Gnedin1999,Baumgardt2003,Webb2013,Webb2014a,Brockamp2014}. Close encounters with dense and massive molecular clouds can trigger a number of tidal shocks \citep{Gieles2006,Lamers2006,Webb2019,Webb2024,ReinaCampos2022}.
Alternatively, these objects may possess DM mini-halos that would justify the observed sizes, more consistent with those of dwarf galaxies than GCs \citep{Tolstoy2009}. In such case, they would sink more efficiently to the centre of the system, owing to their higher mass, explaining the absence of these over-massive outliers at low redshift.
\rr{On top of that, a major merger occurring between $z\approx 1.4$ and $z=0$ could strongly affect the spatial distribution and orbital properties of the clumps, possibly ejecting them into the galactic halo \citep{Kruijssen2012}. However, the clump ejection would limit further dynamical process, making it hard to strip the excess mass \citep{Pfeffer2018,Li2019}. Finally, it is not excluded that the Sparkler galaxy could actually be an uncommon system and have no counterparts in the Local Universe.}

Our study shows how the morphology of the Sparkler galaxy and of its clump system affect the dynamical evolution of the clumps and suggests that most of the observed Sparkler clumps are likely to sink to the centre of the system, rather than getting disrupted by tidal stripping. Once they reach the central regions, these clumps can eventually contribute to the formation of the bulge, or will survive as compact stellar system similar to those observed in recent studies of the MW bulge or will eventually merge forming a nuclear star cluster. The survived clumps tend to have masses and sizes larger than the expected values observed in the GCs of the MW.

Future data constraining the kinematics and morphology of the Sparkler and its clump system will help shed light on the dynamical evolution of candidate proto-GCs, by making it possible to build more realistic models of this galaxy.

\begin{acknowledgements}
    \rr{We thank the anonymous referee for their useful comments that helped improve this manuscript.}
    The research activities described in this paper have been co-funded by the European Union – NextGenerationEU within PRIN 2022 project n.20229YBSAN – Globular clusters in cosmological simulations and in lensed fields: from their birth to the present epoch.
    This research made use of Astropy, a community developed core Python package for Astronomy by the Astropy Collaboration (\citeyear{Astropy2018}).
\end{acknowledgements}

%%% BIBLIOGRAPHY %%%
\bibliography{biblio_sparkler}{}

\begin{thebibliography}{106}
\expandafter\ifx\csname natexlab\endcsname\relax\def\natexlab#1{#1}\fi

\bibitem[{{Aarseth}(1963)}]{Aarseth1963}
{Aarseth}, S.~J. 1963, \mnras, 126, 223

\bibitem[{{Adamo} {et~al.}(2023){Adamo}, {Usher}, {Pfeffer}, \&
  {Claeyssens}}]{Adamo2023}
{Adamo}, A., {Usher}, C., {Pfeffer}, J., \& {Claeyssens}, A. 2023, \mnras, 525,
  L6

\bibitem[{{Agarwal} \& {Milosavljevi{\'c}}(2011)}]{Agarwal2011}
{Agarwal}, M. \& {Milosavljevi{\'c}}, M. 2011, \apj, 729, 35

\bibitem[{and A.~M. Price-Whelan {et~al.}(2018)and A.~M. Price-Whelan,
  Sip{\H{o}}cz, Günther, Lim, Crawford, Conseil, Shupe, Craig, Dencheva,
  Ginsburg, VanderPlas, Bradley, P{\'{e}}rez-Su{\'{a}}rez, de~Val-Borro,
  Aldcroft, Cruz, Robitaille, Tollerud, Ardelean, Babej, Bach, Bachetti,
  Bakanov, Bamford, Barentsen, Barmby, Baumbach, Berry, Biscani, Boquien,
  Bostroem, Bouma, Brammer, Bray, Breytenbach, Buddelmeijer, Burke, Calderone,
  Rodr{\'{\i}}guez, Cara, Cardoso, Cheedella, Copin, Corrales, Crichton,
  D'Avella, Deil, Depagne, Dietrich, Donath, Droettboom, Earl, Erben, Fabbro,
  Ferreira, Finethy, Fox, Garrison, Gibbons, Goldstein, Gommers, Greco,
  Greenfield, Groener, Grollier, Hagen, Hirst, Homeier, Horton, Hosseinzadeh,
  Hu, Hunkeler, Ivezi{\'{c}}, Jain, Jenness, Kanarek, Kendrew, Kern,
  Kerzendorf, Khvalko, King, Kirkby, Kulkarni, Kumar, Lee, Lenz, Littlefair,
  Ma, Macleod, Mastropietro, McCully, Montagnac, Morris, Mueller, Mumford,
  Muna, Murphy, Nelson, Nguyen, Ninan, Nöthe, Ogaz, Oh, Parejko, Parley,
  Pascual, Patil, Patil, Plunkett, Prochaska, Rastogi, Janga, Sabater,
  Sakurikar, Seifert, Sherbert, Sherwood-Taylor, Shih, Sick, Silbiger,
  Singanamalla, Singer, Sladen, Sooley, Sornarajah, Streicher, Teuben, Thomas,
  Tremblay, Turner, Terr{\'{o}}n, van Kerkwijk, de~la Vega, Watkins, Weaver,
  Whitmore, Woillez, Zabalza, , \& and}]{Astropy2018}
and A.~M. Price-Whelan, Sip{\H{o}}cz, B.~M., Günther, H.~M., {et~al.} 2018,
  The Astronomical Journal, 156, 123

\bibitem[{{Bacchini} {et~al.}(2019){Bacchini}, {Fraternali}, {Iorio}, \&
  {Pezzulli}}]{Bacchini2019}
{Bacchini}, C., {Fraternali}, F., {Iorio}, G., \& {Pezzulli}, G. 2019, \aap,
  622, A64

\bibitem[{{Bacchini} {et~al.}(2024){Bacchini}, {Nipoti}, {Iorio},
  {Roman-Oliveira}, {Rizzo}, {Mancera Pi{\~n}a}, {Marasco}, {Zanella}, \&
  {Lelli}}]{Bacchini2024}
{Bacchini}, C., {Nipoti}, C., {Iorio}, G., {et~al.} 2024, \aap, 687, A115

\bibitem[{{Bagla}(2002)}]{Bagla2002}
{Bagla}, J.~S. 2002, Journal of Astrophysics and Astronomy, 23, 185

\bibitem[{{Barnes} \& {Hut}(1986)}]{Barnes1986}
{Barnes}, J. \& {Hut}, P. 1986, \nat, 324, 446

\bibitem[{{Baumgardt} {et~al.}(2023){Baumgardt}, {H{\'e}nault-Brunet},
  {Dickson}, \& {Sollima}}]{Baumgardt2023}
{Baumgardt}, H., {H{\'e}nault-Brunet}, V., {Dickson}, N., \& {Sollima}, A.
  2023, \mnras, 521, 3991

\bibitem[{{Baumgardt} \& {Hilker}(2018)}]{Baumgardt2018}
{Baumgardt}, H. \& {Hilker}, M. 2018, \mnras, 478, 1520

\bibitem[{{Baumgardt} \& {Makino}(2003)}]{Baumgardt2003}
{Baumgardt}, H. \& {Makino}, J. 2003, \mnras, 340, 227

\bibitem[{{Baumgardt} {et~al.}(2013){Baumgardt}, {Parmentier}, {Anders}, \&
  {Grebel}}]{Baumgardt2013}
{Baumgardt}, H., {Parmentier}, G., {Anders}, P., \& {Grebel}, E.~K. 2013,
  \mnras, 430, 676

\bibitem[{{Behroozi} {et~al.}(2019){Behroozi}, {Wechsler}, {Hearin}, \&
  {Conroy}}]{Behroozi2019}
{Behroozi}, P., {Wechsler}, R.~H., {Hearin}, A.~P., \& {Conroy}, C. 2019,
  \mnras, 488, 3143

\bibitem[{{Bekki} \& {Freeman}(2003)}]{Bekki2003}
{Bekki}, K. \& {Freeman}, K.~C. 2003, \mnras, 346, L11

\bibitem[{{Bekki} \& {Tsujimoto}(2019)}]{Bekki2019}
{Bekki}, K. \& {Tsujimoto}, T. 2019, \apj, 886, 121

\bibitem[{{Bershady} {et~al.}(2010){Bershady}, {Verheijen}, {Westfall},
  {Andersen}, {Swaters}, \& {Martinsson}}]{Bershady2010}
{Bershady}, M.~A., {Verheijen}, M. A.~W., {Westfall}, K.~B., {et~al.} 2010,
  \apj, 716, 234

\bibitem[{{Binney} \& {Tremaine}(2008)}]{Binney2008}
{Binney}, J. \& {Tremaine}, S. 2008, {Galactic Dynamics: Second Edition}
  (Princeton University Press)

\bibitem[{{Binney} \& {Wong}(2017)}]{Binney2017}
{Binney}, J. \& {Wong}, L.~K. 2017, \mnras, 467, 2446

\bibitem[{{Brockamp} {et~al.}(2014){Brockamp}, {K{\"u}pper}, {Thies},
  {Baumgardt}, \& {Kroupa}}]{Brockamp2014}
{Brockamp}, M., {K{\"u}pper}, A.~H.~W., {Thies}, I., {Baumgardt}, H., \&
  {Kroupa}, P. 2014, \mnras, 441, 150

\bibitem[{{Brown} \& {Gnedin}(2021)}]{Brown2021}
{Brown}, G. \& {Gnedin}, O.~Y. 2021, \mnras, 508, 5935

\bibitem[{Bryan \& Norman(1998)}]{Bryan1998}
Bryan, G.~L. \& Norman, M.~L. 1998, The Astrophysical Journal, 495, 80

\bibitem[{{Calura} {et~al.}(2014){Calura}, {Ciotti}, \& {Nipoti}}]{Calura2014}
{Calura}, F., {Ciotti}, L., \& {Nipoti}, C. 2014, \mnras, 440, 3341

\bibitem[{{Calura} {et~al.}(2019){Calura}, {D'Ercole}, {Vesperini}, {Vanzella},
  \& {Sollima}}]{Calura2019}
{Calura}, F., {D'Ercole}, A., {Vesperini}, E., {Vanzella}, E., \& {Sollima}, A.
  2019, \mnras, 489, 3269

\bibitem[{{Calura} {et~al.}(2022){Calura}, {Lupi}, {Rosdahl}, {Vanzella},
  {Meneghetti}, {Rosati}, {Vesperini}, {Lacchin}, {Pascale}, \&
  {Gilli}}]{Calura2022}
{Calura}, F., {Lupi}, A., {Rosdahl}, J., {et~al.} 2022, \mnras, 516, 5914

\bibitem[{{Calura} {et~al.}(2024){Calura}, {Pascale}, {Agertz}, {Andersson},
  {Lacchin}, {Lupi}, {Meneghetti}, {Nipoti}, {Ragagnin}, {Rosdahl}, {Vanzella},
  {Vesperini}, \& {Zanella}}]{Calura2024}
{Calura}, F., {Pascale}, R., {Agertz}, O., {et~al.} 2024, arXiv e-prints,
  arXiv:2411.02502

\bibitem[{{Capuzzo-Dolcetta}(1993)}]{Capuzzo1993}
{Capuzzo-Dolcetta}, R. 1993, \apj, 415, 616

\bibitem[{{Chandrasekhar}(1943)}]{Chandrasekhar1943}
{Chandrasekhar}, S. 1943, \apj, 97, 255

\bibitem[{{Ciotti} \& {Bertin}(1999)}]{Ciotti1999}
{Ciotti}, L. \& {Bertin}, G. 1999, \aap, 352, 447

\bibitem[{{Claeyssens} {et~al.}(2025){Claeyssens}, {Adamo}, {Messa},
  {Dessauges-Zavadsky}, {Richard}, {Kramarenko}, {Matthee}, \&
  {Naidu}}]{Claeyssens2025}
{Claeyssens}, A., {Adamo}, A., {Messa}, M., {et~al.} 2025, \mnras
  [\eprint[arXiv]{2410.10974}]

\bibitem[{{Claeyssens} {et~al.}(2023){Claeyssens}, {Adamo}, {Richard},
  {Mahler}, {Messa}, \& {Dessauges-Zavadsky}}]{Claeyssens2023}
{Claeyssens}, A., {Adamo}, A., {Richard}, J., {et~al.} 2023, \mnras, 520, 2180

\bibitem[{{Correa} {et~al.}(2015){Correa}, {Wyithe}, {Schaye}, \&
  {Duffy}}]{Correa2015}
{Correa}, C.~A., {Wyithe}, J. S.~B., {Schaye}, J., \& {Duffy}, A.~R. 2015,
  \mnras, 452, 1217

\bibitem[{{Crain} {et~al.}(2015){Crain}, {Schaye}, {Bower}, {Furlong},
  {Schaller}, {Theuns}, {Dalla Vecchia}, {Frenk}, {McCarthy}, {Helly},
  {Jenkins}, {Rosas-Guevara}, {White}, \& {Trayford}}]{Crain2015}
{Crain}, R.~A., {Schaye}, J., {Bower}, R.~G., {et~al.} 2015, \mnras, 450, 1937

\bibitem[{{Crociati} {et~al.}(2024){Crociati}, {Cignoni}, {Dalessandro},
  {Pallanca}, {Massari}, {Ferraro}, {Lanzoni}, {Origlia}, \&
  {Valenti}}]{Crociati2024}
{Crociati}, C., {Cignoni}, M., {Dalessandro}, E., {et~al.} 2024, \aap, 691,
  A311

\bibitem[{{Di Cintio} {et~al.}(2024){Di Cintio}, {Iorio}, {Calura}, {Nipoti},
  \& {Cantari}}]{DiCintio2024}
{Di Cintio}, P., {Iorio}, G., {Calura}, F., {Nipoti}, C., \& {Cantari}, M.
  2024, \aap, 690, A61

\bibitem[{{Diemer}(2018)}]{Diemer2018}
{Diemer}, B. 2018, \apjs, 239, 35

\bibitem[{{Dutton} \& {Macci{\`o}}(2014)}]{Dutton2014}
{Dutton}, A.~A. \& {Macci{\`o}}, A.~V. 2014, \mnras, 441, 3359

\bibitem[{{Elmegreen}(2010)}]{Elmegreen2010b}
{Elmegreen}, B.~G. 2010, \apjl, 712, L184

\bibitem[{{Elmegreen} \& {Hunter}(2010)}]{Elmegreen2010a}
{Elmegreen}, B.~G. \& {Hunter}, D.~A. 2010, \apj, 712, 604

\bibitem[{{Ferraro} {et~al.}(2021){Ferraro}, {Pallanca}, {Lanzoni}, {Crociati},
  {Dalessandro}, {Origlia}, {Rich}, {Saracino}, {Mucciarelli}, {Valenti},
  {Geisler}, {Mauro}, {Villanova}, {Moni Bidin}, \& {Beccari}}]{Ferraro2021}
{Ferraro}, F.~R., {Pallanca}, C., {Lanzoni}, B., {et~al.} 2021, Nature
  Astronomy, 5, 311

\bibitem[{{Forbes} \& {Romanowsky}(2023)}]{Forbes2023}
{Forbes}, D.~A. \& {Romanowsky}, A.~J. 2023, \mnras, 520, L58

\bibitem[{{Freeman}(1970)}]{Freeman1970}
{Freeman}, K.~C. 1970, \apj, 160, 811

\bibitem[{{Georgiev} {et~al.}(2016){Georgiev}, {B{\"o}ker}, {Leigh},
  {L{\"u}tzgendorf}, \& {Neumayer}}]{Georgiev2016}
{Georgiev}, I.~Y., {B{\"o}ker}, T., {Leigh}, N., {L{\"u}tzgendorf}, N., \&
  {Neumayer}, N. 2016, \mnras, 457, 2122

\bibitem[{Gieles {et~al.}(2006)Gieles, Zwart, Baumgardt, Athanassoula, Lamers,
  Sipior, \& Leenaarts}]{Gieles2006}
Gieles, M., Zwart, S. F.~P., Baumgardt, H., {et~al.} 2006, Monthly Notices of
  the Royal Astronomical Society, 371, 793

\bibitem[{{Gnedin} {et~al.}(1999){Gnedin}, {Hernquist}, \&
  {Ostriker}}]{Gnedin1999}
{Gnedin}, O.~Y., {Hernquist}, L., \& {Ostriker}, J.~P. 1999, \apj, 514, 109

\bibitem[{{Gnedin} \& {Ostriker}(1997)}]{Gnedin1997}
{Gnedin}, O.~Y. \& {Ostriker}, J.~P. 1997, \apj, 474, 223

\bibitem[{{Hernquist}(1990)}]{Hernquist1990}
{Hernquist}, L. 1990, \apj, 356, 359

\bibitem[{{Hopkins} {et~al.}(2010){Hopkins}, {Bundy}, {Croton}, {Hernquist},
  {Keres}, {Khochfar}, {Stewart}, {Wetzel}, \& {Younger}}]{Hopkins2010}
{Hopkins}, P.~F., {Bundy}, K., {Croton}, D., {et~al.} 2010, \apj, 715, 202

\bibitem[{{Hopkins} {et~al.}(2009){Hopkins}, {Cox}, {Younger}, \&
  {Hernquist}}]{Hopkins2009}
{Hopkins}, P.~F., {Cox}, T.~J., {Younger}, J.~D., \& {Hernquist}, L. 2009,
  \apj, 691, 1168

\bibitem[{{Horta} {et~al.}(2021){Horta}, {Hughes}, {Pfeffer}, {Bastian},
  {Kruijssen}, {Reina-Campos}, \& {Crain}}]{Horta2021}
{Horta}, D., {Hughes}, M.~E., {Pfeffer}, J.~L., {et~al.} 2021, \mnras, 500,
  4768

\bibitem[{{Hunter} {et~al.}(2003){Hunter}, {Elmegreen}, {Dupuy}, \&
  {Mortonson}}]{Hunter2003}
{Hunter}, D.~A., {Elmegreen}, B.~G., {Dupuy}, T.~J., \& {Mortonson}, M. 2003,
  \aj, 126, 1836

\bibitem[{Jordán {et~al.}(2005)Jordán, Côté, Blakeslee, Ferrarese,
  McLaughlin, Mei, Peng, Tonry, Merritt, Milosavljević, Sarazin, Sivakoff, \&
  West}]{Jordan2005}
Jordán, A., Côté, P., Blakeslee, J.~P., {et~al.} 2005, The Astrophysical
  Journal, 634, 1002

\bibitem[{{King}(1966)}]{King1966}
{King}, I.~R. 1966, \aj, 71, 64

\bibitem[{{Kravtsov} \& {Gnedin}(2005)}]{Kravtsov2005}
{Kravtsov}, A.~V. \& {Gnedin}, O.~Y. 2005, \apj, 623, 650

\bibitem[{Kruijssen(2015)}]{Kruijssen2015}
Kruijssen, J. M.~D. 2015, Monthly Notices of the Royal Astronomical Society,
  454, 1658

\bibitem[{{Kruijssen} \& {Mieske}(2009)}]{Kruijssen2009}
{Kruijssen}, J.~M.~D. \& {Mieske}, S. 2009, \aap, 500, 785

\bibitem[{{Kruijssen} {et~al.}(2012){Kruijssen}, {Pelupessy}, {Lamers},
  {Portegies Zwart}, {Bastian}, \& {Icke}}]{Kruijssen2012}
{Kruijssen}, J.~M.~D., {Pelupessy}, F.~I., {Lamers}, H. J.~G.~L.~M., {et~al.}
  2012, \mnras, 421, 1927

\bibitem[{{Lamers} \& {Gieles}(2006)}]{Lamers2006}
{Lamers}, H.~J.~G.~L.~M. \& {Gieles}, M. 2006, \aap, 455, L17

\bibitem[{{Lanzoni} {et~al.}(2010){Lanzoni}, {Ferraro}, {Dalessandro},
  {Mucciarelli}, {Beccari}, {Miocchi}, {Bellazzini}, {Rich}, {Origlia},
  {Valenti}, {Rood}, \& {Ransom}}]{Lanzoni2010}
{Lanzoni}, B., {Ferraro}, F.~R., {Dalessandro}, E., {et~al.} 2010, \apj, 717,
  653

\bibitem[{{Lee} {et~al.}(1999){Lee}, {Joo}, {Sohn}, {Rey}, {Lee}, \&
  {Walker}}]{Lee1999}
{Lee}, Y.~W., {Joo}, J.~M., {Sohn}, Y.~J., {et~al.} 1999, \nat, 402, 55

\bibitem[{{Lelli} {et~al.}(2016){Lelli}, {McGaugh}, \& {Schombert}}]{Lelli2016}
{Lelli}, F., {McGaugh}, S.~S., \& {Schombert}, J.~M. 2016, \aj, 152, 157

\bibitem[{Li \& Gnedin(2019)}]{Li2019}
Li, H. \& Gnedin, O.~Y. 2019, Monthly Notices of the Royal Astronomical
  Society, 486, 4030

\bibitem[{{Lima Neto} {et~al.}(1999){Lima Neto}, {Gerbal}, \&
  {M{\'a}rquez}}]{LimaNeto1999}
{Lima Neto}, G.~B., {Gerbal}, D., \& {M{\'a}rquez}, I. 1999, \mnras, 309, 481

\bibitem[{{Mackey} \& {Gilmore}(2004)}]{Mackey2004}
{Mackey}, A.~D. \& {Gilmore}, G.~F. 2004, \mnras, 355, 504

\bibitem[{{Mahler} {et~al.}(2023){Mahler}, {Jauzac}, {Richard}, {Beauchesne},
  {Ebeling}, {Lagattuta}, {Natarajan}, {Sharon}, {Atek}, {Claeyssens},
  {Cl{\'e}ment}, {Eckert}, {Edge}, {Kneib}, \& {Niemiec}}]{Mahler2023}
{Mahler}, G., {Jauzac}, M., {Richard}, J., {et~al.} 2023, \apj, 945, 49

\bibitem[{{Me{\v{s}}tri{\'c}} {et~al.}(2022){Me{\v{s}}tri{\'c}}, {Vanzella},
  {Zanella}, {Castellano}, {Calura}, {Rosati}, {Bergamini}, {Mercurio},
  {Meneghetti}, {Grillo}, {Caminha}, {Nonino}, {Merlin}, {Cupani}, \&
  {Sani}}]{Mestric2022}
{Me{\v{s}}tri{\'c}}, U., {Vanzella}, E., {Zanella}, A., {et~al.} 2022, \mnras,
  516, 3532

\bibitem[{Miholics {et~al.}(2017)Miholics, Kruijssen, \& Sills}]{Miholics2017}
Miholics, M., Kruijssen, J. M.~D., \& Sills, A. 2017, Monthly Notices of the
  Royal Astronomical Society, 470, 1421

\bibitem[{{Miyamoto} \& {Nagai}(1975)}]{Miyamoto1975}
{Miyamoto}, M. \& {Nagai}, R. 1975, \pasj, 27, 533

\bibitem[{{Moreno-Hilario} {et~al.}(2024){Moreno-Hilario}, {Martinez-Medina},
  {Li}, {Souza}, \& {P{\'e}rez-Villegas}}]{MorenoHilario2024}
{Moreno-Hilario}, E., {Martinez-Medina}, L.~A., {Li}, H., {Souza}, S.~O., \&
  {P{\'e}rez-Villegas}, A. 2024, \mnras, 527, 2765

\bibitem[{{Mowla} {et~al.}(2022){Mowla}, {Iyer}, {Desprez},
  {Estrada-Carpenter}, {Martis}, {Noirot}, {Sarrouh}, {Strait}, {Asada},
  {Abraham}, {Brammer}, {Sawicki}, {Willott}, {Bradac}, {Doyon}, {Muzzin},
  {Pacifici}, {Ravindranath}, \& {Zabl}}]{Mowla2022}
{Mowla}, L., {Iyer}, K.~G., {Desprez}, G., {et~al.} 2022, \apjl, 937, L35

\bibitem[{{Navarro} {et~al.}(1996){Navarro}, {Frenk}, \& {White}}]{Navarro1996}
{Navarro}, J.~F., {Frenk}, C.~S., \& {White}, S. D.~M. 1996, \apj, 462, 563

\bibitem[{{Neumayer} {et~al.}(2020){Neumayer}, {Seth}, \&
  {B{\"o}ker}}]{Neumayer2020}
{Neumayer}, N., {Seth}, A., \& {B{\"o}ker}, T. 2020, \aapr, 28, 4

\bibitem[{{Neumayer} {et~al.}(2011){Neumayer}, {Walcher}, {Andersen},
  {S{\'a}nchez}, {B{\"o}ker}, \& {Rix}}]{Neumayer2011}
{Neumayer}, N., {Walcher}, C.~J., {Andersen}, D., {et~al.} 2011, \mnras, 413,
  1875

\bibitem[{{Oh} \& {Lin}(2000)}]{Oh2000}
{Oh}, K.~S. \& {Lin}, D.~N.~C. 2000, \apj, 543, 620

\bibitem[{{Pallottini} {et~al.}(2022){Pallottini}, {Ferrara}, {Gallerani},
  {Behrens}, {Kohandel}, {Carniani}, {Vallini}, {Salvadori}, {Gelli},
  {Sommovigo}, {D'Odorico}, {Di Mascia}, \& {Pizzati}}]{Pallottini2022}
{Pallottini}, A., {Ferrara}, A., {Gallerani}, S., {et~al.} 2022, \mnras, 513,
  5621

\bibitem[{{Pascale} {et~al.}(2025){Pascale}, {Calura}, {Vesperini}, {Rosdahl},
  {Nipoti}, {Giunchi}, {Lacchin}, {Lupi}, {Messa}, {Meneghetti}, {Ragagnin},
  {Vanzella}, \& {Zanella}}]{Pascale2025}
{Pascale}, R., {Calura}, F., {Vesperini}, E., {et~al.} 2025, arXiv e-prints,
  arXiv:2505.06346

\bibitem[{{Petts} {et~al.}(2016){Petts}, {Read}, \& {Gualandris}}]{Petts2016}
{Petts}, J.~A., {Read}, J.~I., \& {Gualandris}, A. 2016, \mnras, 463, 858

\bibitem[{{Pfeffer} {et~al.}(2018){Pfeffer}, {Kruijssen}, {Crain}, \&
  {Bastian}}]{Pfeffer2018}
{Pfeffer}, J., {Kruijssen}, J.~M.~D., {Crain}, R.~A., \& {Bastian}, N. 2018,
  \mnras, 475, 4309

\bibitem[{{Plummer}(1911)}]{Plummer1911}
{Plummer}, H.~C. 1911, \mnras, 71, 460

\bibitem[{{Pontoppidan} {et~al.}(2022){Pontoppidan}, {Barrientes}, {Blome},
  {Braun}, {Brown}, {Carruthers}, {Coe}, {DePasquale}, {Espinoza}, {Marin},
  {Gordon}, {Henry}, {Hustak}, {James}, {Jenkins}, {Koekemoer}, {LaMassa},
  {Law}, {Lockwood}, {Moro-Martin}, {Mullally}, {Pagan}, {Player}, {Proffitt},
  {Pulliam}, {Ramsay}, {Ravindranath}, {Reid}, {Robberto}, {Sabbi}, {Ubeda},
  {Balogh}, {Flanagan}, {Gardner}, {Hasan}, {Meinke}, \&
  {Nota}}]{Pontoppidan2022}
{Pontoppidan}, K.~M., {Barrientes}, J., {Blome}, C., {et~al.} 2022, \apjl, 936,
  L14

\bibitem[{{Posti} \& {Helmi}(2019)}]{Posti2019}
{Posti}, L. \& {Helmi}, A. 2019, \aap, 621, A56

\bibitem[{{Power} {et~al.}(2003){Power}, {Navarro}, {Jenkins}, {Frenk},
  {White}, {Springel}, {Stadel}, \& {Quinn}}]{Power2003}
{Power}, C., {Navarro}, J.~F., {Jenkins}, A., {et~al.} 2003, \mnras, 338, 14

\bibitem[{{Ranaivoharimina} {et~al.}(2024){Ranaivoharimina},
  {Randriamampandry}, {Wang}, {Men{\'e}ndez-Delmestre}, \&
  {Gon{\c{c}}alves}}]{Ranaivoharimina2024}
{Ranaivoharimina}, N., {Randriamampandry}, T.~H., {Wang}, J.,
  {Men{\'e}ndez-Delmestre}, K., \& {Gon{\c{c}}alves}, T.~S. 2024, arXiv
  e-prints, arXiv:2410.09762

\bibitem[{{Reina-Campos} {et~al.}(2022){Reina-Campos}, {Keller}, {Kruijssen},
  {Gensior}, {Trujillo-Gomez}, {Jeffreson}, {Pfeffer}, \&
  {Sills}}]{ReinaCampos2022}
{Reina-Campos}, M., {Keller}, B.~W., {Kruijssen}, J.~M.~D., {et~al.} 2022,
  \mnras, 517, 3144

\bibitem[{{Reina-Campos} {et~al.}(2019){Reina-Campos}, {Kruijssen}, {Pfeffer},
  {Bastian}, \& {Crain}}]{ReinaCampos2019}
{Reina-Campos}, M., {Kruijssen}, J.~M.~D., {Pfeffer}, J.~L., {Bastian}, N., \&
  {Crain}, R.~A. 2019, \mnras, 486, 5838

\bibitem[{{Romano} {et~al.}(2023){Romano}, {Ferraro}, {Origlia}, {Portegies
  Zwart}, {Lanzoni}, {Crociati}, {Massari}, {Dalessandro}, {Mucciarelli},
  {Rich}, {Calura}, \& {Matteucci}}]{Romano2023}
{Romano}, D., {Ferraro}, F.~R., {Origlia}, L., {et~al.} 2023, \apj, 951, 85

\bibitem[{{Saracino} {et~al.}(2015){Saracino}, {Dalessandro}, {Ferraro},
  {Lanzoni}, {Geisler}, {Mauro}, {Villanova}, {Moni Bidin}, {Miocchi}, \&
  {Massari}}]{Saracino2015}
{Saracino}, S., {Dalessandro}, E., {Ferraro}, F.~R., {et~al.} 2015, \apj, 806,
  152

\bibitem[{{Scannapieco} \& {Tissera}(2003)}]{Scannapieco2003}
{Scannapieco}, C. \& {Tissera}, P.~B. 2003, \mnras, 338, 880

\bibitem[{{Schaye} {et~al.}(2015){Schaye}, {Crain}, {Bower}, {Furlong},
  {Schaller}, {Theuns}, {Dalla Vecchia}, {Frenk}, {McCarthy}, {Helly},
  {Jenkins}, {Rosas-Guevara}, {White}, {Baes}, {Booth}, {Camps}, {Navarro},
  {Qu}, {Rahmati}, {Sawala}, {Thomas}, \& {Trayford}}]{Schaye2015}
{Schaye}, J., {Crain}, R.~A., {Bower}, R.~G., {et~al.} 2015, \mnras, 446, 521

\bibitem[{{Sersic}(1968)}]{Sersic1968}
{Sersic}, J.~L. 1968, {Atlas de Galaxias Australes} (Observatorio Astronomico,
  Universidad Nacional de Cordoba)

\bibitem[{{Smith} {et~al.}(2015){Smith}, {Flynn}, {Candlish}, {Fellhauer}, \&
  {Gibson}}]{Smith2015}
{Smith}, R., {Flynn}, C., {Candlish}, G.~N., {Fellhauer}, M., \& {Gibson},
  B.~K. 2015, \mnras, 448, 2934

\bibitem[{Spitler {et~al.}(2006)Spitler, Larsen, Strader, Brodie, Forbes, \&
  Beasley}]{Spitler2006}
Spitler, L.~R., Larsen, S.~S., Strader, J., {et~al.} 2006, The Astronomical
  Journal, 132, 1593

\bibitem[{{Springel} {et~al.}(2005{\natexlab{a}}){Springel}, {Di Matteo}, \&
  {Hernquist}}]{Springel2005}
{Springel}, V., {Di Matteo}, T., \& {Hernquist}, L. 2005{\natexlab{a}}, \mnras,
  361, 776

\bibitem[{{Springel} {et~al.}(2005{\natexlab{b}}){Springel}, {White},
  {Jenkins}, {Frenk}, {Yoshida}, {Gao}, {Navarro}, {Thacker}, {Croton},
  {Helly}, {Peacock}, {Cole}, {Thomas}, {Couchman}, {Evrard}, {Colberg}, \&
  {Pearce}}]{Springel2005b}
{Springel}, V., {White}, S. D.~M., {Jenkins}, A., {et~al.} 2005{\natexlab{b}},
  \nat, 435, 629

\bibitem[{{Tolstoy} {et~al.}(2009){Tolstoy}, {Hill}, \& {Tosi}}]{Tolstoy2009}
{Tolstoy}, E., {Hill}, V., \& {Tosi}, M. 2009, \araa, 47, 371

\bibitem[{{Tomasetti} {et~al.}(2024){Tomasetti}, {Moresco}, {Lardo}, {Courbin},
  {Jimenez}, {Verde}, {Millon}, \& {Cimatti}}]{Tomasetti2024}
{Tomasetti}, E., {Moresco}, M., {Lardo}, C., {et~al.} 2024, arXiv e-prints,
  arXiv:2412.06903

\bibitem[{{Tremaine} {et~al.}(1975){Tremaine}, {Ostriker}, \&
  {Spitzer}}]{Tremaine1975}
{Tremaine}, S.~D., {Ostriker}, J.~P., \& {Spitzer}, Jr., L. 1975, \apj, 196,
  407

\bibitem[{{van der Wel} {et~al.}(2014){van der Wel}, {Franx}, {van Dokkum},
  {Skelton}, {Momcheva}, {Whitaker}, {Brammer}, {Bell}, {Rix}, {Wuyts},
  {Ferguson}, {Holden}, {Barro}, {Koekemoer}, {Chang}, {McGrath},
  {H{\"a}ussler}, {Dekel}, {Behroozi}, {Fumagalli}, {Leja}, {Lundgren},
  {Maseda}, {Nelson}, {Wake}, {Patel}, {Labb{\'e}}, {Faber}, {Grogin}, \&
  {Kocevski}}]{vanderWel2014}
{van der Wel}, A., {Franx}, M., {van Dokkum}, P.~G., {et~al.} 2014, \apj, 788,
  28

\bibitem[{{Vanzella} {et~al.}(2017){Vanzella}, {Calura}, {Meneghetti},
  {Mercurio}, {Castellano}, {Caminha}, {Balestra}, {Rosati}, {Tozzi}, {De
  Barros}, {Grazian}, {D'Ercole}, {Ciotti}, {Caputi}, {Grillo}, {Merlin},
  {Pentericci}, {Fontana}, {Cristiani}, \& {Coe}}]{Vanzella2017}
{Vanzella}, E., {Calura}, F., {Meneghetti}, M., {et~al.} 2017, \mnras, 467,
  4304

\bibitem[{{Vesperini}(1997)}]{Vesperini1997}
{Vesperini}, E. 1997, \mnras, 287, 915

\bibitem[{{Webb} {et~al.}(2013){Webb}, {Harris}, {Sills}, \&
  {Hurley}}]{Webb2013}
{Webb}, J.~J., {Harris}, W.~E., {Sills}, A., \& {Hurley}, J.~R. 2013, \apj,
  764, 124

\bibitem[{{Webb} {et~al.}(2014){Webb}, {Leigh}, {Sills}, {Harris}, \&
  {Hurley}}]{Webb2014a}
{Webb}, J.~J., {Leigh}, N., {Sills}, A., {Harris}, W.~E., \& {Hurley}, J.~R.
  2014, \mnras, 442, 1569

\bibitem[{{Webb} {et~al.}(2019){Webb}, {Reina-Campos}, \&
  {Kruijssen}}]{Webb2019}
{Webb}, J.~J., {Reina-Campos}, M., \& {Kruijssen}, J.~M.~D. 2019, \mnras, 486,
  5879

\bibitem[{{Webb} {et~al.}(2024){Webb}, {Reina-Campos}, \&
  {Kruijssen}}]{Webb2024}
{Webb}, J.~J., {Reina-Campos}, M., \& {Kruijssen}, J.~M.~D. 2024, \apj, 975,
  242

\bibitem[{Webb {et~al.}(2014)Webb, Sills, Harris, \& Hurley}]{Webb2014b}
Webb, J.~J., Sills, A., Harris, W.~E., \& Hurley, J.~R. 2014, Monthly Notices
  of the Royal Astronomical Society, 445, 1048

\bibitem[{{Weinberger} {et~al.}(2020){Weinberger}, {Springel}, \&
  {Pakmor}}]{Weinberger2020}
{Weinberger}, R., {Springel}, V., \& {Pakmor}, R. 2020, \apjs, 248, 32

\bibitem[{{Whitaker} {et~al.}(2025){Whitaker}, {Cutler}, {Chandar}, {Pan},
  {Setton}, {Furtak}, {Bezanson}, {Labb{\'e}}, {Leja}, {Suess}, {Wang},
  {Weaver}, {Atek}, {Brammer}, {Feldmann}, {F{\"o}rster Schreiber},
  {Glazebrook}, {de Graaff}, {Greene}, {Khullar}, {Marchesini}, {Maseda},
  {Miller}, {Mo}, {Mowla}, {Nanayakkara}, {Nelson}, {Price}, {Rizzo}, {van
  Dokkum}, {Williams}, {Zhang}, {Zhang}, \& {Zitrin}}]{Whitaker2025}
{Whitaker}, K.~E., {Cutler}, S.~E., {Chandar}, R., {et~al.} 2025, arXiv
  e-prints, arXiv:2501.07627

\end{thebibliography}
\bibliographystyle{aa}

\begin{appendix}
\onecolumn

\section{Fit of resolved clump density profiles}\label{app:fit_plummer}
In order to get the final sizes of the resolved clumps shown in Fig. \ref{fig:clumps_size_evolution}, the clumps density profiles were fitted to the one of a Plummer sphere.
The first step is the identification of the centre of the clump, which is computed iteratively by using the shrinking sphere method \citep{Power2003}. At the first iteration, the algorithm computes the average position of a fraction $f_\mathrm{max}$ of the stellar particles closest to the centre of the simulation box, therefore $(0,0,0)$, and this is assumed as the new guess as centre of the clump.
Then, at the $k$th iteration the new guess is computed as the average position of the $f_\mathrm{max}-k\Delta f$ fraction of particles closest to the $(k-1)$th guess centre of the system. The algorithm is iterated down to a minimum fraction $f_\mathrm{min}$. In our case, we set $f_\mathrm{max}=0.950$, $f_\mathrm{min}=0.025$, $\Delta f=0.025$.

\begin{figure*}[t!]
\centering
 \includegraphics[width=0.95\textwidth]{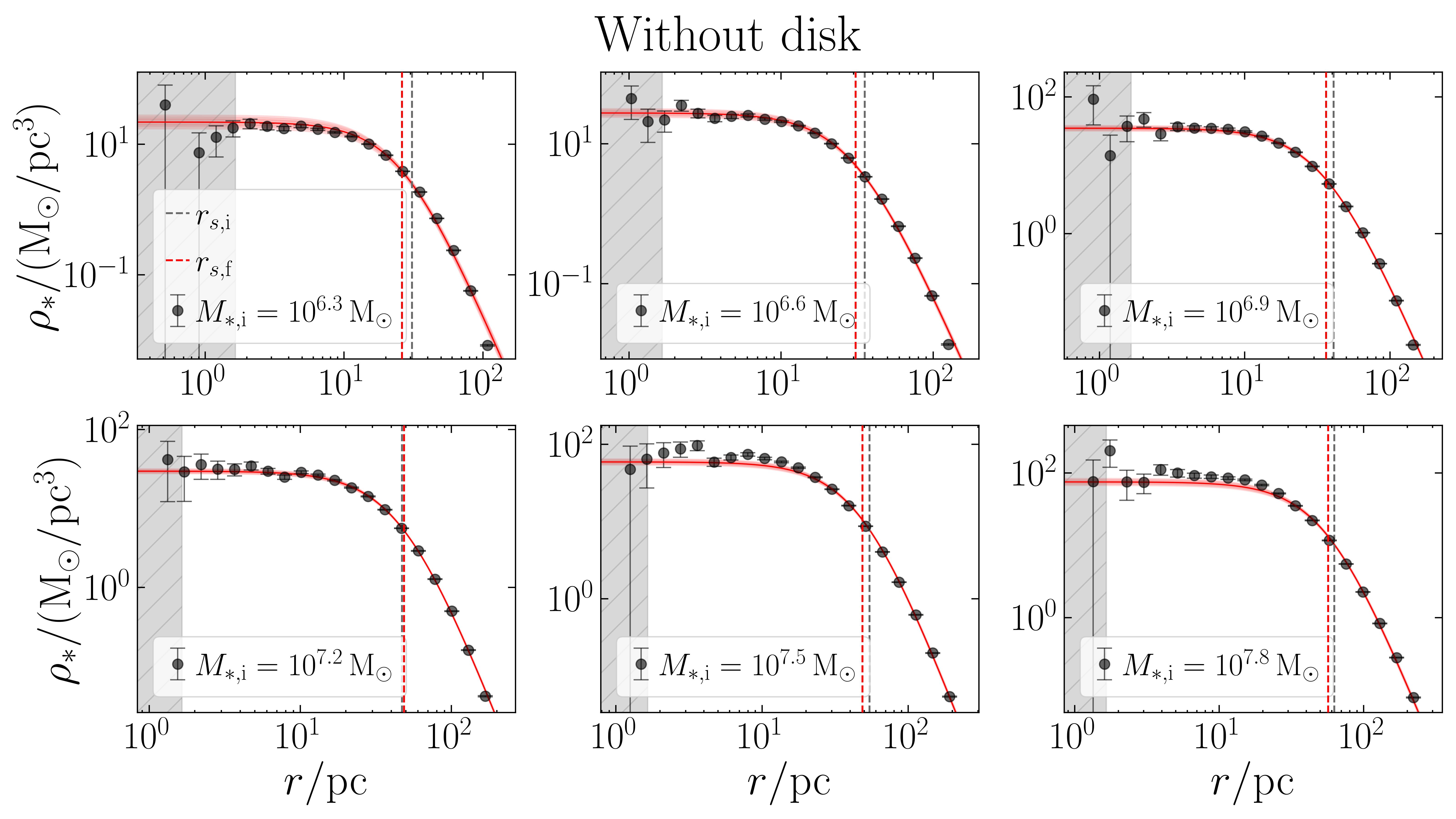}\\
 \includegraphics[width=0.95\textwidth]{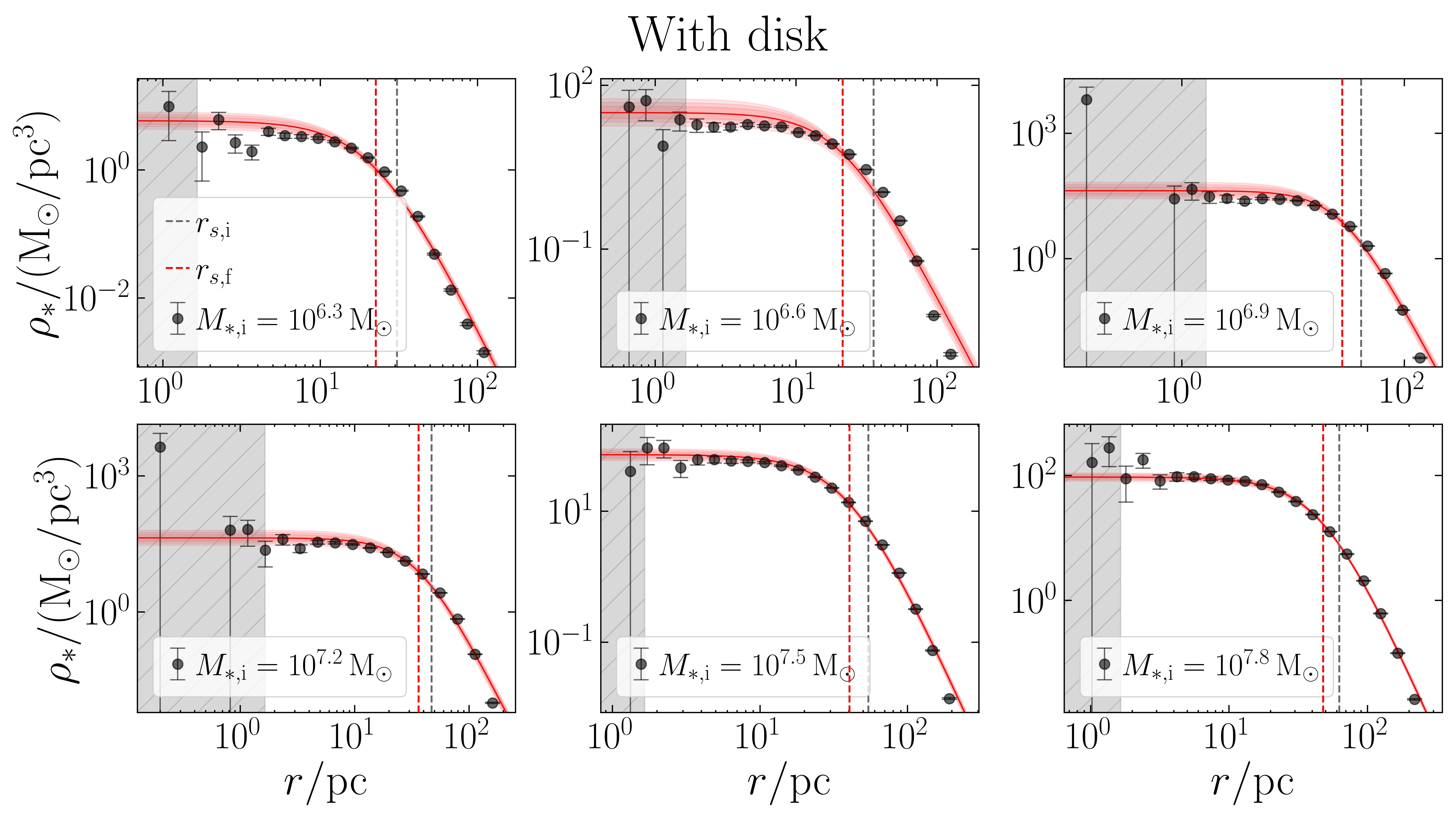}
 \caption{Final 3D density profiles of the clumps in the spherical-host simulations presented in Sect. \ref{sec:resolved_clumps_setup} (first two rows) and for the case in which the external potential includes also a Sparkler-like stellar disk as described in Sect. \ref{sec:tidal_shock} (last two rows). First and third row: clumps with mass $10^{6.3}\,\msun$, $10^{6.6}\,\msun$ and $10^{6.9}\,\msun$ (left, central and right panels, respectively); second and fourth row:  clumps with mass $10^{7.2}\,\msun$, $10^{7.5}\,\msun$ and $10^{7.8}\,\msun$ (left, central and right panels, respectively).
 The black dots are the $N$-body density profiles, while the red solid line is the best-fit Plummer density profile (with the shaded are marking the $1\sigma$, $2\sigma$ and $3\sigma$ confidence levels), as described in Appendix \ref{app:fit_plummer}. The grey shaded area highlights the region within the softening length, which is excluded from the fit. The grey dashed line is the initial Plummer scale radius $r_{s,\mathrm{i}}$, the red dashed line is the best-fit final scale radius $r_{s,\mathrm{f}}$.}
 \label{fig:spherical_plummer}
\end{figure*}

Once the centre of the clump is found, the distances of the particles are computed with respect to this centre.
Given $r_{s,\mathrm{i}}$ the initial Plummer scale radii, computed from the clump size-mass relation (Sect. \ref{sec:size_mass}), we evaluate the clump density profile within $4\,r_{s,\mathrm{i}}$, dividing the particles in $20$ bins equally spaced in the logarithmic space, with limits $\{r_0,...,r_{21}\}$.
The density of the $i$th bin $\rho_i$ is computed as

\begin{equation}\label{eq:density_bin}
    \rho_i=\dfrac{m_pN_i}{\dfrac{4}{3}\pi\left(r_{i+1}^3-r_i^3\right)},
\end{equation}

\noindent where $m_p$ is the particle mass (defined in Sect. \ref{sec:resolved_clumps_setup}) and $N_i$ is the number of particles with distances from the centre of the clump within $r_i$ and $r_{i+1}$. The uncertainty $\delta\rho_i$ on the number of particles is assumed to be Poissonian, therefore $\delta\rho_i=\rho_i/N_i^{1/2}$.
The resulting profile is fitted by a Plummer sphere by means of the non-linear least squares method, implemented in the \textsc{numpy}\footnote{\url{https://numpy.org/doc/stable/index.html}} function \textsc{curve\_fit}, with the final total mass $M_\mathrm{*,f}$ and scale radius $r_{s,\mathrm{f}}$ as free parameters. We note that we excluded from the fit the bins falling within $1.8$ pc (three times the softening length; Sect. \ref{sec:res_configuration}).
In Fig. \ref{fig:spherical_plummer} we plot the best-fit model compared to the binned profile, for the cases of tidal stripping without and with a disk (first two rows and last two rows, respectively), respectively (Sects. \ref{sec:resolved_clumps_setup} and \ref{sec:tidal_shock}). We excluded the case $\log (M_*/\msun)=6$ since in the case of tidal stripping with a disk the clump gets completely disrupted. In each panel we also show the initial ($r_{s,\mathrm{i}}$) and the best-fit final ($r_{s,\mathrm{f}}$) Plummer scale radius.
The Plummer profile describes remarkably well the binned profile especially in the cases with little stripping.

\FloatBarrier

\section{Tidal stripping in elliptical orbits}\label{app:tidal_elliptical}
\rr{In Sect. \ref{sec:resolved_clumps_setup} we have shown the tidal stripping history of clumps of different mass in circular orbit at 1 kpc from the centre of the Sparkler-like spherical halo, modelled with an external gravitational potential. This setup implies that the clumps are not affected by dynamical friction and do not deviate from their initial orbit. We have chosen this setup in order to maximize the tidal stripping effects, initializing the clumps as close as possible to the threshold that, according to our definition, divides surviving and lost clumps.
The results presented in Sect. \ref{sec:resolved_clumps_setup} show that the fraction of tidal mass losses are not sufficient to make the mass of Sparkler-like clumps comparable to local GCs. That is why in Sect. \ref{sec:tidal_shock} we introduced the Sparkler stellar disk as an external potential, in order to trigger tidal shocks and substantially enhance the amount of stripped stars.}

\begin{figure*}[t!]
\centering
 \includegraphics[width=0.85\textwidth]{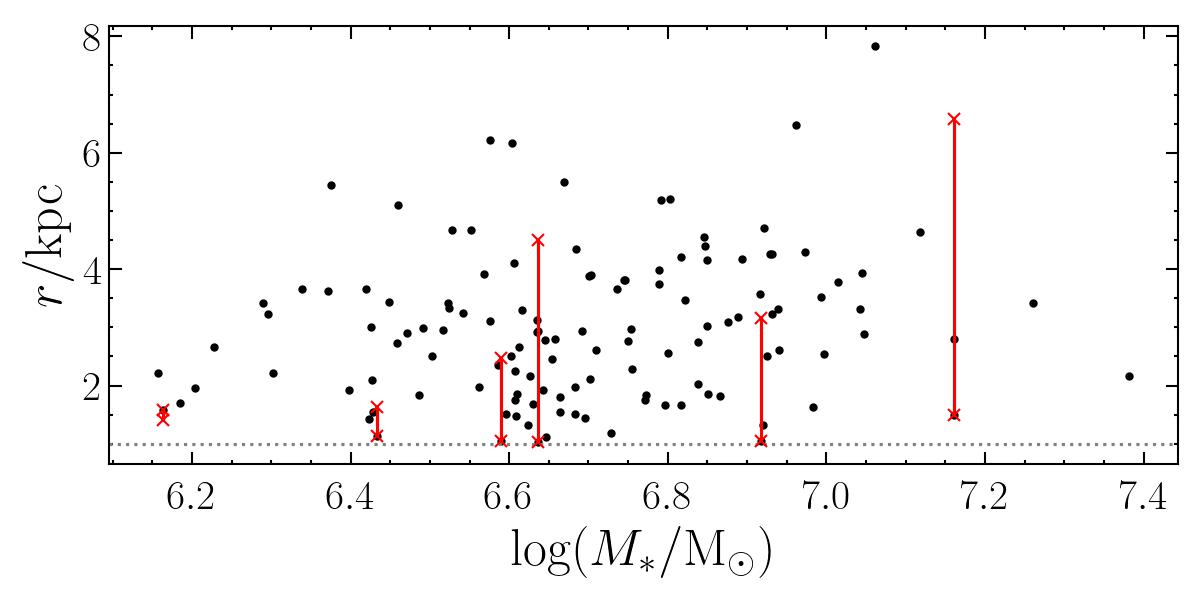}
 \caption{\rr{Final distance of the surviving clumps (black dots) for all the simulations with point-mass clumps in a halo made of particles, in the case of the spherical configuration (Sect. \ref{sec:res_spherical}). The red crosses and lines show the initial and final distance of the six clumps studied in this section. The grey dotted line is at 1 kpc, the threshold that divides the surviving and the lost clumps.}}
 \label{fig:spherical_migration}
\end{figure*}

\rr{In this Appendix we explore the differences arising in the setup of Sect. \ref{sec:resolved_clumps_setup} when we consider non-circular orbits and the dragging effect of dynamical friction in the spherical halo.
To choose the orbital parameters, we selected 6 clumps from the 30 runs discussed in Sect. \ref{sec:res_spherical}, choosing surviving clumps of different masses that at the end of the simulation are at a distance slightly larger than 1 kpc. In Table \ref{tab:tidal_elliptical} we list their initial conditions, while in Fig. \ref{fig:spherical_migration} we compare their initial and final positions with those of the other surviving clumps.}

\rr{The dynamical friction is natively present in the simulations of Sect. \ref{sec:res_spherical}, while in a setup like the one of Sect. \ref{sec:resolved_clumps_setup} it must be implemented via semi-analytical models. Therefore, we modelled the dynamical friction deceleration of the clump centre of mass using the Chandrasekhar formula \citep{Chandrasekhar1943,Binney2008}}

\begin{equation}\label{eq:chandra}
    \pmb{a}_\mathrm{DF}(\pmb{x})=-2\pi G^2 M_*\,\rho_\mathrm{halo}(\pmb{x}_\mathrm{CM})\,\ln\Lambda\left[\mathrm{erf}(X)-\frac{2X}{\sqrt{\pi}}\exp(-X^2)\right]\frac{\pmb{v}_\mathrm{CM}}{|\pmb{v}_\mathrm{CM}|^3}.
\end{equation}

\rr{\noindent In Eq. \ref{eq:chandra} $M_*$ is the clump stellar mass, $\rho_\mathrm{halo}$ is the halo density, $\pmb{x}_\mathrm{CM}=\sum_{i=i}^{N}\pmb{x}_i/N$ and $\pmb{v}_\mathrm{CM}=\sum_{i=i}^{N}\pmb{v}_i/N$ are the position and the velocity of the clump centre of mass, respectively. $X=|\pmb{v}_\mathrm{CM}|/[\sigma_r(|\pmb{x}_\mathrm{CM}|)\sqrt{2}]$, where $\sigma_r$ is the radial velocity dispersion of the background (which is analytical for a Hernquist isotropic system, \citealt{Hernquist1990}).
$\Lambda$ is a constant that goes into the Coulomb logarithm and it can computed as}

\begin{equation}\label{eq:lambda}
    \Lambda=\frac{|\pmb{x}_\mathrm{CM}|/\gamma}{\mathrm{max}(r_\mathrm{hm},GM_*/|\pmb{v}_\mathrm{CM}|^2)},
\end{equation}

\begin{figure*}[t!]
\centering
 \includegraphics[width=0.95\textwidth]{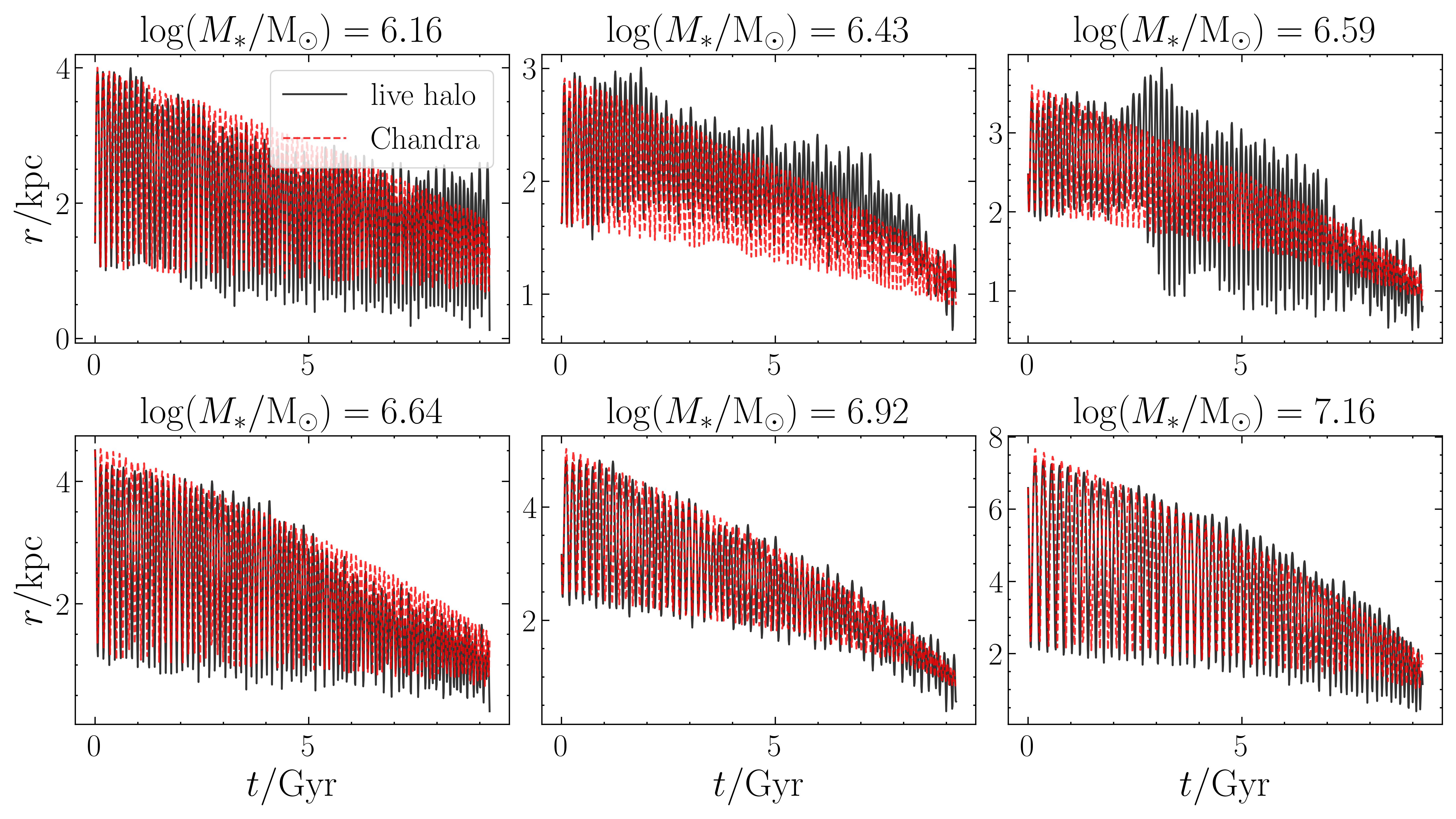}
 \caption{\rr{For each panel: the black solid line is the orbit of the selected clump (parameters in Table \ref{tab:tidal_elliptical}) in the simulation of point-mass clumps in a spherical halo made of particles, for the spherical configuration (Sect. \ref{sec:res_spherical}); the red dashed line is the orbit of the same clump in the same halo, but implemented as an external potential and with the addition of the Chandrasekhar formula, as described in Appendix \ref{app:tidal_elliptical}.}}
 \label{fig:chosen_orbits}
\end{figure*}

\rr{\noindent with $r_\mathrm{hm}$ being the clump half-mass radius and $\gamma=|\mathrm{d}\ln\rho_\mathrm{halo}/\mathrm{d}\ln r$ or fixed arbitrarily. At each time step, this acceleration is added to each particle of the resolved clump.
We fix the clump mass to its initial value and we tuned the value of $\ln\Lambda$ for each clump in order to qualitatively reproduce the orbits of the chosen clumps in the simulations of Sect. \ref{sec:res_spherical}. The values of $\ln\Lambda$ are listed in Table \ref{tab:tidal_elliptical}, while in Fig. \ref{fig:chosen_orbits} we show the comparisons between the orbits of the chosen point-mass clumps as derived from Sect. \ref{sec:res_spherical} and the one obtained with the Chandrasekhar formula.}

\begin{table*}[t!]
\fontsize{10pt}{10pt}\selectfont
\setlength{\tabcolsep}{4pt}
\renewcommand{\arraystretch}{1.4} % Default value: 1
\centering
\caption{\rr{Parameters of the six clumps chosen to test the effects of orbit eccentricity and dynamical friction on  tidal stripping.}}
\begin{tabular}{cccccccc}
\bottomrule\bottomrule
ID & $\log (M_{*,\mathrm{ini}}/\msun)$ & $\pmb{x}_0/$kpc  & $\pmb{v}_0/(\mathrm{km/s})$ & $\ln\Lambda$ & $M_{*,\mathrm{fin}}/M_{*,\mathrm{ini}}$ & $M_{*,\mathrm{fin-sph}}/M_{*,\mathrm{ini}}$ & $M_{*,\mathrm{fin-disk}}/M_{*,\mathrm{ini}}$\\
$(0)$ & $(1)$ & $(2)$ & $(3)$ & $(4)$ & $(5)$ & $(6)$ & $(7)$\\
\hline
1 & $6.16$ & $(-0.41,0.79,-1.11)$ & $(21.33,98.87,-21.95)$ & $10.5$ & $0.72$ & $0.84-0.87$ & $0-0.16$\\
2 & $6.39$ & $(-1.51,-0.60,0.07)$ & $(-19.77,82.63,12.90)$ & $6$ & $0.87$ & $0.87-0.89$ & $0.16-0.41$\\
3 & $6.55$ & $(-0.66,-1.12,2.11)$ & $(-37.77,73.24,-14.16)$ & $7$ & $0.92$ & $0.87-0.89$ & $0.16-0.41$\\
4 & $6.60$ & $(-0.71,-4.14,1.64)$ & $(30.21,14.25,2.13)$ & $5$ & $0.82$ & $0.87-0.89$ & $0.16-0.41$\\
5 & $6.88$ & $(2.28,2.00,-0.88)$ & $(20.59,-94.79,11.97)$ & $6$ & $0.95$ & $0.89-0.90$ & $0.41-0.54$\\
6 & $7.12$ & $(-3.40,-4.61,3.26)$ & $(-8.72,43.81,-56.13)$ & $5$ & $0.94$ & $0.90$ & $0.54-0.60$\\
\toprule
\end{tabular}
\tablefoot{\rr{From left to right: ID of the clump (0), clump initial mass (1), initial position (2), initial velocity (3), Coulomb logarithm (4), fraction of residual mass in the simulations described in this section (5), fraction of residual mass of clumps of similar mass from the spherical case of Sect. \ref{sec:resolved_clumps_setup} (6), fraction of residual mass of clumps of similar mass from the case with a disk of Sect. \ref{sec:tidal_shock} (7).}}
\label{tab:tidal_elliptical}
\end{table*}

\rr{In Table \ref{tab:tidal_elliptical} we also list the fraction of residual mass of the six chosen clumps of this section, compared with the fraction of residual mass of clumps of similar mass in the cases without and with a disk (Sects. \ref{sec:resolved_clumps_setup} and \ref{sec:tidal_shock}). Comparing the results with those obtained previously, it turns out that in 4 out of 6 cases the tidal stripping is consistent with or less efficient than the one derived for the 1-kpc circular orbits in the spherical case. This is most likely due to the fact that the clumps do not start the simulation in the innermost regions of the system, reaching them only by the end of the run because of the sinking effect of dynamical friction.
The only exceptions are clumps 1 and 4, which retain $72\%$ and $82\%$ of their initial mass, while the retained mass fractions in the case of circular 1-kpc orbit are $84\%$ and $87\%$, respectively. We find that the orbits of these two clumps have the largest eccentricity among the six explored, with $e=(a-b)/(a+b)\approx 0.6$, where $a$ and $b$ are the apocentre and pericentre of the orbit, respectively. Therefore, it is possible that the clump, spanning a wider range of distances during its orbit, may experience a larger gravitational field gradient, resulting in an increased tidal stripping.
However, such increase ($\approx 10\%$ of additional lost mass at best) is negligible if compared with the effects of tidal shocks. Indeed, the predicted residual mass in presence of a disk is smaller than the one obtained for the six clumps studied in this section. This comparison demonstrates that the eccentricity of the orbits and the dynamical friction are second-order effects compared to the presence of the disk, especially in our experiments which are aimed at maximizing the tidal losses of the over-massive Sparkler clumps.}

\end{appendix}
\end{document}